\documentclass[12pt]{article}
\usepackage[dvipdfmx]{graphicx}
\usepackage{amsmath,amsthm,amsfonts,amssymb}
\usepackage{color}
\usepackage[normalem]{ulem}
\usepackage{arydshln, mathrsfs}
\usepackage{wrapfig}
\usepackage{dutchcal}
\makeatletter
\@addtoreset{equation}{section}

\makeatother

\begin{document}

	\begin{titlepage}
		\null
		\begin{flushright}
			%-/-
			%\\
			arXiv:2212.11800
			\\
			December, 2022
		\end{flushright}
		
		\vskip 1cm
		\begin{center}
			
{\LARGE \bf Solitons in Open N=2 String Theory}
				
			\vskip 1cm
			\normalsize
			
			{\large 
				Masashi Hamanaka$^{a,}$\footnote{E-mail:hamanaka@math.nagoya-u.ac.jp}, 
				Shan-Chi Huang$^{a,}$\footnote{E-mail:x18003x@math.nagoya-u.ac.jp}, and 
				Hiroaki Kanno$^{a,b,}$\footnote{E-mail:kanno@math.nagoya-u.ac.jp}
			}
			
			\vskip 0.7cm
			
			{$^a$Graduate School of Mathematics, Nagoya University,\\
				Nagoya, 464-8602, JAPAN}
			
			\vskip 0.5cm
			
			{$^b$KMI, Nagoya University, Nagoya, 464-8602, Japan\\
				Nagoya, 464-8602, JAPAN}
			
			\vskip 1.3cm
			
			{\bf \large Abstract}
			\vskip 0.5cm

		\end{center}

The open N=2 string theory is defined on the four-dimensional space-time with the split signature $(+,+,-,-)$. The string field theory action of the open N=2 string theory is described by the four-dimensional Wess-Zumino-Witten (${\mbox{WZW}}_4$) model. 
Equation of motion of the ${\mbox{WZW}}_4$ model is the Yang equation which is equivalent to the anti-self-dual Yang-Mills equation. 
In this paper, we study soliton-type classical solutions of the ${\mbox{WZW}}_4$ model in the split signature by calculating the action density of the ${\mbox{WZW}}_4$ model. We find that the action density of the one-soliton solutions is localized on a three-dimensional hyperplane. This shows that there would be codimension-one-solitonic objects, or equivalently, some kind of three-branes in the open N=2 string theory. We also prove that in the asymptotic region of the space-time, the action density of the $n$-soliton solutions is a ``nonlinear superposition'' of $n$ one-solitons. This suggests the existence of intersecting $n$ three-branes in the N=2 strings. 
Finally we make a reduction to a $(1+2)$-dimensional real space-time to calculate energy densities of the soliton solutions. We can successfully evaluate the energy distribution for the two-soliton solutions and find that there is no singularity in the interacting region. This implies the existence of smooth intersecting codimension-one branes in the whole region.  
Soliton solutions in the Euclidean signature are also discussed. 

	\end{titlepage}
	
	\clearpage
	\baselineskip 6mm
	
	%\make{index
\tableofcontents

\newpage
	
\section{Introduction}

The Anti-Self-Dual Yang-Mills %(ASDYM) 
equation in four dimensions has been of great interest in elementary particle physics and mathematics. In the Euclidean signature, it has quite important soliton solutions, instantons which are crucial to reveal non-perturbative aspects of quantum field theory. In the split signature $(+,+,-,-)$, it has a close relationship to integrable systems. It is well known that by imposing appropriate reduction conditions for the gauge fields, the anti-self-dual Yang-Mills equations in the split signature can be reduced to various lower-dimensional integrable equations, such as the KdV equation, the Non-Linear Schr\"odinger equation, the Toda equations and so on \cite{Ward, MaWo}. The integrability of these equations is well understood in the geometrical framework of twistor theory \cite{MaWo}. Soliton solutions are mostly of codimension-one in the sense that the energy density of the one-soliton solutions is localized on a codimension-one hyperplane in the space-time (See e.g. \cite{Kodama, MaSa}). 

The anti-self-dual Yang-Mills equation is realized in string theories which are classified according to the number N of the world-sheet supersymmetry. %{$N=0$ string theories are called bosonic string theories and defined in $(25+1)$ dimensional space-time. $N=1$ string theories are called (ordinary) superstring theories and defined in $(9+1)$ dimensional space-time.} 
Under the condition that the critical dimension of the target space is positive and that the string world-sheet theory has an appropriate conformal symmetry, the maximal number is found to be N=2 (\cite{GSW} $\S$4.5 and references therein). In the case of the N=2 string theories, the condition of conformal anomaly cancellation determines the critical dimension to be four, and the Virasoro constraints forbid any excited physical states except for massless scalars \cite{GSW}. Hence, the string field theory can be reduced to the conventional field theory. The world-sheet N=2 supersymmetry induces a complex structure on the four-dimensional target space and hence the Minkowski target space is forbidden \cite{OoVa}. This is the reason why non-trivial string field theories are realized only when the metric has the split signature. (In the Euclidean case, momentum of the massless scalar fields becomes identically zero.) Therefore the N=2 string theory is closely related to the Ward conjecture and integrable systems. 

The space-time action of the open N=2 string theory is described by the four dimensional Wess-Zumino-Witten (${\mbox{WZW}}_4$) model \cite{OoVa,LMNS,IKU,IKUX} (see also \cite{Marcus,Nair,NaSc,Parkes}). 
Equation of motion of this model is the Yang equation which is equivalent to the anti-self-dual Yang-Mills equation \cite{Huang}. Hence solutions of the Yang equation are classical solutions of the open string field theory action of the N=2 strings. It is surprising that the action of the string field theory is explicitly written down in terms of massless scalar fields only. Exact analysis of the classical solutions leads to exact analysis of classical aspects of the string field theory. 

Recently a new type of soliton solutions of the Yang equation has been constructed by using the Darboux transformation \cite{GNO,GHHN} in four-dimensional flat spaces with all kinds of signatures, that is, the Euclidean, the Minkowski and the split signatures \cite{HaHu}. These soliton solutions have localized Yang-Mills action densities on three-dimensional hyperplanes, and hence can be interpreted as codimension-one solitons. Furthermore, asymptotic analysis has been also made in \cite{HaHu2, Huang, Huang2}, which suggests the existence of intersecting three-dimensional branes. In the case of the split signature, these solutions are supposed to be relevant to the open N=2 string theory. 
Therefore, to analyze the solitonic behavior (including the interacting region) for the ${\mbox{WZW}}_4$ action is much more appropriate than the Yang-Mills action. 
%Then we should analyze not the Yang-Mills action but the ${\mbox{WZW}}_4$ action and clarify the soliton interactions hopefully including the interacting region.

In this paper, we study the classical soliton solutions in the WZW$_4$ model. The ${\mbox{WZW}}_4$ action (\ref{WZW4}) consists of the non-linear sigma model (NL$\sigma$M) term and the Wess-Zumino (WZ) term \cite{Donaldson}. We calculate the action densities of the NL$\sigma$M model term and the Wess-Zumino term for the soliton solutions.\footnote{In \cite{Parkes}, Parkes discussed similar problems by using the SL$(2,\mathbb{C})$ non-linear plane wave solutions \cite{deVega} without calculating any action density.} 
For the one-soliton solutions, we find that the ${\mbox{WZW}}_4$ action density is localized on a three-dimensional hyperplane. This suggests that there would be a codimension-one solitonic object, or equivalently, some kind of three-brane in the open N=2 string theory. For the multi-soliton solutions, we clarify the asymptotic behavior and conclude that the $n$-soliton solution possesses $n$ isolated and localized lumps of the action density, and can be interpreted as intersecting $n$ soliton walls. More precisely, each lump of the action density is essentially the same as a one-soliton because it preserves its shape and ``velocity'' with a possible position shift (called the phase shift) of the peak in the scattering process. We evaluate the distribution of the NL$\sigma$M term for the two-soliton solutions successfully and find that there is no singularity in the interacting region. This is consistent with the existence of smooth intersecting three-branes in the whole region. %As for the Wess-Zumino term, we show the action densities for the one-soliton solutions are identically zero. This implies that in the asymptotic region, the action densities are identically zero as well.  
Finally, we make a reduction to a $(1+2)$-dimensional real space-time to calculate energy densities of the soliton solutions. In $(2+2)$-dimensions, 
the concept of energy is ambiguous because of the existence of two time directions. 
%there is no definition of the energy because there are two time directions. 
This is the reason why in this paper we discuss action density instead of the energy density. %However, the physical meaning of the action density is not necessarily clear { in the $(2+2)$-dimensions}{($\leftarrow$. 
We compute the energy densities of the one-soliton and two-soliton solutions in the $(1+2)$ dimensions to confirm that they are localized on the same hyperplanes as those the action densities are localized. This suggests that the locus where the action density is localized could be considered as existence of a physical object. 

This paper is organized as follows. In section 2, the WZW$_4$ model is introduced and our conventions are set up. In section 3, soliton solutions of the Yang equation are reviewed and some properties of the solutions, such as the flip symmetry, singularities and an asymptotic behavior are discussed. In section 4, the action density for the one- and two-soliton solutions is calculated. In section 5, asymptotic analysis of the $n$-soliton solution is given. In section 6, we reduce the ${\mbox{WZW}}_4$ model from $(2+2)$-dimensions to $(1+2)$-dimensions and calculate the Hamiltonian density for the one and two-soliton solutions. Section 7 is devoted to conclusion and discussion. Appendix A is a brief review of the quasideterminant. In Appendix B, a statement in footnote 8 is proved (See section 5). Appendix C is a proof of unitarity of the $n$-soliton solutions on the Euclidean space. Appendix D includes miscellaneous formulas and detailed calculations.

\section{Four-Dimensional Wess-Zumino-Witten Model}

%\subsection{Convension and Settings}
%To facilitate the discussion, we set the gauge group to be $G=GL(N, \mathbb{C})$ (or subgroup of $GL(N,\mathbb{C})$). 

In this section, we review the four-dimensional Wess-Zumino-Witten (${\mbox{WZW}}_4$) model.
%soliton solutions in four-dimensional flat real spaces. % with three kinds of metric. 
%, that is the Euclidean space $\mathbb{E}$, the Minkowski space $\mathbb{M}$, and the Ultrahyperbolic space $U$ whose signatures are $(+,+,+,+)$, $+,+,-,-$, and $(+,+,-,-)$, respectively. 
In order to treat it in a unified way, we introduce a four-dimensional space with complex coordinates $(z,\widetilde z, w, \widetilde w)$ and the flat metric:
\begin{eqnarray}
\mathrm{d}s^2&=&g_{mn}\mathrm{d}z^{m}\mathrm{d}z^{n}=2(\mathrm{d}z \mathrm{d}\widetilde z -\mathrm{d}w \mathrm{d}\widetilde w),
~~~m,n=1,2,3,4. \nonumber\\
&&{\mbox{where}}~~
\label{com_metric}
g_{mn}:=\left(
\begin{array}{cccc}
0&1&0&0\\
1&0&0&0\\ 
0&0&0&-1\\
0&0&-1&0
\end{array}
\right),
~(z^{1},z^{2},z^{3},z^{4}):=(z,\widetilde z, w, \widetilde w).
\end{eqnarray}
% $ds^2=2(dz d\widetilde z -dw d\widetilde w)$. 
The space $\mathbb{C}^4$ can be reduced to the three kinds of real spaces by imposing suitable reality conditions on $(z,\widetilde z, w, \widetilde w)$. For example, the Euclidean real space $\mathbb{E}$ is given by $\widetilde z=\overline z,~\widetilde w= -\overline w$, %the Minkowski real space $\mathbb{M}$ is by $z, \widetilde z \in \mathbb{R},~\widetilde w= \overline w$, 
and the Ultrahyperbolic real space $\mathbb{U}$ by (1) $z, \widetilde z,w, \widetilde w\in \mathbb{R}$ or (2) $\widetilde z=\overline z, ~\widetilde w= \overline w$, which are denoted respectively by $\mathbb{U}_1$ and $\mathbb{U}_2$. 
Our choices are shown in terms of real coordinates $x^\mu~(\mu=1,2,3,4)$ as follows:
%between $z,\widetilde z, w, \widetilde w$ and $x^\mu$ 
\begin{eqnarray}
&(\mathbb{E})&	
\label{Reality condition_E}
	\left(\begin{array}{cc}
	z & w \\
	\widetilde{w} & \widetilde{z}
	\end{array}\right)
	=
	\frac{1}{\sqrt{2}}
	\left(\begin{array}{cc}
	x^{1}+ix^{2} & x^{3}+ix^{4} \\
	-(x^{3}-ix^{4}) & x^{1}-ix^{2}
	\end{array}\right),\\
       &&~~~\mathrm{d}s^2=(\mathrm{d}x^{1})^2+(\mathrm{d}x^{2})^2+(\mathrm{d}x^{3})^2+(\mathrm{d}x^{4})^2.\label{E}\\
\label{Reality condition_U}
        &(\mathbb{U}_1)&
	%\begin{array}{l}
	 \left(
	\begin{array}{cc}
	z & w \\
	\widetilde{w} & \widetilde{z}
	\end{array}\right)
	=
	\displaystyle{\frac{1}{\sqrt{2}}}
	\left(\begin{array}{cc}
	x^{1}+x^{3} & x^{2}+x^{4} \\
	-(x^{2}-x^{4}) & x^{1}-x^{3} 
	\end{array}\right),%  ~~\mbox{for}~~\mathbb{U}_1.	
	\\
%\label{Reality condition_U_2}
%&(\mathbb{U}_2)&\left(\begin{array}{cc}
%	z & w \\
%	\widetilde{w} & \widetilde{z}
%	\end{array}\right)
%	=
%	\displaystyle{\frac{1}{\sqrt{2}}}
%	\left(\begin{array}{cc}
%	x^{1}+ix^{2} & x^{3}-ix^{4} \\
%	x^{3}+ix^{4} & x^{1}-ix^{2}
%	\end{array}\right), % ~~~~~\!\mbox{for}~~\mathbb{U}_2.	
	%\end{array}
%	\\
        &&~~ds^2=(\mathrm{d}x^{1})^2+(\mathrm{d}x^{2})^2-(\mathrm{d}x^{3})^2-(\mathrm{d}x^{4})^2.\label{U}
\end{eqnarray}
In this paper, we mainly consider the case of the Ultrahyperbolic space $\mathbb{U}_1$.\footnote{The case of $\mathbb{U}_2$ is not considered in this paper because unitarity condition of $\sigma$ leads to trivial action densities as we will see in Appendix \ref{unitarity}. The case of the Euclidean space is discussed at the end of each (sub)section.}
%because the theory can be realized in a unitary group as we will see in Sec. 3.2. 

%\subsection{Four-dimensional Wess-Zumino-Witten Model}

Let $M_4$ be a four-dimensional flat space 
%which is either of $\mathbb{E}$, $\mathbb{M}$ or $\mathbb{U}$ 
and $\sigma$ be a map from $M_4$ 
to $G=GL(N,\mathbb{C})$ or its subgroup. 
%(In the case of $\mathbb{U}_1$, $G=U(2)$ is realized)
%In the context of the open N=2 strings, the massless scalar 
The action of the ${\mbox{WZW}}_4$ model consists of 
two parts as follows:
%is given by 
\begin{eqnarray}
\label{WZW4_full}
S_{\scriptsize {\mbox{WZW}}_4}&:=&
S_{\scriptsize{\sigma}}+S_{\scriptsize {\mbox{WZ}}},
\\
S_{\scriptsize{\sigma}}&:=&
\frac{i}{4\pi}\int_{M_4}\omega \wedge 
\mathrm{Tr}
\left[
\left(\partial \sigma\right)\sigma^{-1} \wedge (\widetilde{\partial} \sigma)\sigma^{-1}
\right], \\
S_{\scriptsize {\mbox{WZ}}}
&:=&
-\frac{i}{12\pi}\int_{M_5} \omega \wedge
\mathrm{Tr}\left[
\left(\mathrm{d}\widehat{\sigma}\right)\widehat{\sigma}^{-1} 
\right]^3,
\end{eqnarray}
where $M_5 := M_4 \times {[0,1]}$ and 
$\widehat{\sigma}(z, \widetilde{z}, w, \widetilde{w}, t)$, $t \in {[0, 1]}$ 
is a homotopy such that $\widehat{\sigma}(z, \widetilde{z}, w, \widetilde{w}, 0)=\mathrm{Id}$ and $\widehat{\sigma}(z, \widetilde{z}, w, \widetilde{w}, 1)=\sigma(z, \widetilde{z}, w, \widetilde{w})$, 
and $\omega$ is the K\"ahler two-form on $M_4$ given by
\begin{eqnarray}
\omega =\frac{i}{2}\left( \mathrm{d}z \wedge \mathrm{d}\widetilde{z} - \mathrm{d}w \wedge \mathrm{d}\widetilde{w}\right).
\end{eqnarray}
The exterior derivatives are defined as follows:
\begin{eqnarray}
\mathrm{d}:=\partial + \widetilde{\partial}
+ \mathrm{d}t \partial_t,~~~
\partial:=\mathrm{d}w\partial_w + \mathrm{d}z\partial_z,~~~
\widetilde{\partial}:=\mathrm{d}\widetilde{w}\partial_{\widetilde{w}} + \mathrm{d}\widetilde{z}\partial_{\widetilde{z}}.
\end{eqnarray}
The first part $S_{\scriptsize{\sigma}}$ is called the non-linear sigma model (NL$\sigma$M) term 
%, or simply the sigma model term, 
and the second part %$S_{\scriptsize {\mbox{WZ}}}$ 
is called the Wess-Zumino (WZ) term.
In the Wess-Zumino term, we use an abbreviated notation: 
$A^3:=A\wedge A\wedge A$ for a differential one-form $A$. 

The equation of motion is 
\begin{eqnarray}
\label{Yang}
\widetilde{\partial}\left(
\omega \wedge \left(\partial \sigma \right)\sigma^{-1} \right)=0.
%\end{eqnarray}
%~\mbox{or equivalently}, ~
%\begin{eqnarray}
%\partial(\omega \wedge (\widetilde{\partial} \sigma)\sigma^{-1})=0.
\end{eqnarray}
This is derived as follows. Let us consider an infinitesimal variation of the dynamical variable $\sigma$ such that $\delta \sigma |_{\partial M_4}=0$ and $\mathrm{d} \delta=\delta \mathrm{d}$. 
Then, 
%\begin{eqnarray*}
% \delta ((\mathrm{d}\sigma)\sigma^{-1})=d((\delta\sigma)\sigma^{-1})+(\mathrm{d}\sigma)\sigma^{-1}(\delta\sigma)\sigma^{-1}-(\delta\sigma)\sigma^{-1}(\mathrm{d}\sigma)\sigma^{-1}.
%\end{eqnarray*}
\begin{eqnarray*}
	\delta ((\mathrm{d}\sigma)\sigma^{-1})=\mathrm{d}((\delta\sigma)\sigma^{-1})
	-(\mathrm{d}\sigma)\sigma^{-1}(\delta\sigma)\sigma^{-1}
	+(\delta\sigma)\sigma^{-1}(\mathrm{d}\sigma)\sigma^{-1}. 
\end{eqnarray*}
Note that $(\delta\sigma)\sigma^{-1}$ is a $\mathfrak{g}$-valued zero-form 
while $(\mathrm{d}\sigma)\sigma^{-1}$ is a $\mathfrak{g}$-valued one-form, 
where $\mathfrak{g}$ is the Lie algebra of $G$. 
The cyclic symmetry of trace implies
\begin{eqnarray*}
\delta {\mbox{Tr}} \left[(\mathrm{d}\widehat{\sigma})\widehat{\sigma}^{-1} \right]^3 
=3\mathrm{d} {\mbox{Tr}}\left[(\delta\widehat{\sigma})\widehat{\sigma}^{-1} \left((\mathrm{d}\widehat{\sigma})\widehat{\sigma}^{-1}\right)^2\right].
\end{eqnarray*} 
Since $d\omega=0$, we have 
\begin{eqnarray*}
\delta S_{\scriptsize {\mbox{WZ}}}
=
-\frac{i}{4\pi}\int_{M_5} \omega \wedge
\delta\mathrm{Tr}\left[
\left(\mathrm{d}\widehat{\sigma}\right)\widehat{\sigma}^{-1} 
\right]^3
=
-
\frac{i}{4\pi}\int_{M_4} \omega \wedge
\mathrm{Tr}\left[
(\delta\sigma)\sigma^{-1} \left((\mathrm{d}\sigma)\sigma^{-1}\right)^2\right].
\end{eqnarray*}
The variation of the sigma model term is
\begin{eqnarray*}
 \delta S_{\scriptsize{\sigma}}
&=&
\frac{i}{4\pi}\int_{M_4} \omega \wedge
\delta\mathrm{Tr}
\left[
\left(\partial \sigma\right)\sigma^{-1} \wedge (\widetilde{\partial} \sigma)\sigma^{-1}
\right] \\
&=&
\frac{i}{4\pi}\int_{M_4} \omega \wedge
\mathrm{Tr}
\left[
\partial\left(
(\delta\sigma)\sigma^{-1}
(\widetilde{\partial} \sigma)\sigma^{-1}
\right)
-\widetilde{\partial}\left(
(\delta\sigma)\sigma^{-1}
\left(\partial \sigma\right)\sigma^{-1} 
\right)
\right.
\\
&&~~~~~~~~~~~~~~~~~~~~~~~+
(\delta\sigma)\sigma^{-1}
\left(
\widetilde{\partial}\left(\left( \partial \sigma\right)\sigma^{-1} \right)
-
\partial\left((\widetilde{\partial} \sigma)\sigma^{-1} \right)
\right],
\end{eqnarray*}
where we use $\partial\widetilde{\partial}+\widetilde{\partial}\partial=0$
due to 
$\partial^2=0, \widetilde{\partial}^2=0$ and 
$\mathrm{d}=\partial+\widetilde{\partial}$.
The first and second terms become a surface integration 
due to $d\omega=0$ and the fact that:
\begin{eqnarray*}
&&\mathrm{Tr}\left[
\partial \left((\delta\sigma)\sigma^{-1}
(\widetilde{\partial} \sigma)\sigma^{-1}\right)
-\widetilde{\partial}\left(
(\delta\sigma)\sigma^{-1}\left(\partial \sigma\right)\sigma^{-1} 
\right)\right]\\
&&=
\mathrm{Tr}\left[
\mathrm{d}\left(
(\delta\sigma)\sigma^{-1}
(\widetilde{\partial} \sigma)\sigma^{-1}
\right)
-\mathrm{d}\left(
(\delta\sigma)\sigma^{-1}
\left(\partial \sigma\right)\sigma^{-1} 
\right)\right]. 
\end{eqnarray*}
Therefore we get the final form of the total action variation 
and the equation of motion is obtained:
%complete the proof:
%Proceeding the calculation, we finally get 
\begin{eqnarray*}
\delta S_{\scriptsize {\mbox{WZW}}_4}
=
\frac{i}{2\pi}\int_{M_4} \omega \wedge
\mathrm{Tr}\left[
(\delta\sigma)\sigma^{-1}
\widetilde{\partial}\left(\left( \partial \sigma\right)\sigma^{-1}\right)
\right].
\end{eqnarray*}

Finally we rewrite the ${\mbox{WZW}}_4$ action 
without integration over $M_5$. 
By the cyclic property of the trace we have
\begin{eqnarray}
\mathrm{d}
\mathrm{Tr}\left[
\left(\mathrm{d}\widehat{\sigma}\right)\widehat{\sigma}^{-1} 
\right]^3
=
-\mathrm{Tr}\left[
\left(\mathrm{d}\widehat{\sigma}\right)\widehat{\sigma}^{-1} 
\right]^4 =0. 
\end{eqnarray}
The K\"ahler two-form $\omega$ is closed and 
$H^2(M_4,\mathbb{R})=0$ and hence there exists 
a one-form $A$ on the flat space-time such that 
\begin{eqnarray}
\omega = \mathrm{d}A.
\end{eqnarray}
Note that $A$ is not uniquely determined and has ambiguity 
with respect to the following degree of freedom: $A\mapsto A+\mathrm{d}\kappa$, 
where $\kappa$ is an arbitrary zero-form. 

The Wess-Zumino term is written as 
\begin{eqnarray}
%&&\!\!\!\!
\int_{M_5}\mathrm{d}
\left(
A \wedge \mathrm{Tr}\left[
\left(\mathrm{d}\widehat{\sigma}\right)\widehat{\sigma}^{-1} 
\right]^3\right)=
%\nonumber \\&\!\!\!\!=\!\!\!\!&
\int_{M_4}
A \wedge \mathrm{Tr}\left[
\left(\mathrm{d}\widehat{\sigma}\right)\widehat{\sigma}^{-1} 
\right]^3 \bigg|_{t = 1}
-
\int_{M_4}
A \wedge \mathrm{Tr}\left[
\left(\mathrm{d}\widehat{\sigma}\right)\widehat{\sigma}^{-1} 
\right]^3 \bigg|_{t = 0}.
\end{eqnarray}
If there exists a homotopy such that $\widehat{\sigma}(t=0)=\mathrm{Id}$ and $\widehat{\sigma}(t=1)=\sigma$, the second term vanishes and we obtain
\begin{eqnarray}
\label{WZW4}
S_{\scriptsize{\mbox{WZW}}_4}=
\frac{i}{4\pi}\int_{M_4}\omega \wedge 
\mathrm{Tr}\left[
\left(\partial \sigma\right)\sigma^{-1} \wedge (\widetilde{\partial} \sigma)\sigma^{-1}
\right]
-
\frac{i}{12\pi}\int_{M_4} A \wedge
\mathrm{Tr}\left[
\left(\mathrm{d}\sigma\right)\sigma^{-1} 
\right]^3. ~~
\end{eqnarray}
{}From now on, we use this form of action. 
Since our soliton solutions allow such a homotopy as above, 
we can use (\ref{WZW4}) for computing the action density. 

\subsection{Component Representation of WZW$_4$ Action Density}

Let us write down explicit representations of the WZW$_4$ action density (\ref{WZW4}) in the flat four-dimensional real spaces. 

In terms of the local complex coordinates (\ref{com_metric}) $\sim$
(\ref{Reality condition_U}), 
the NL$\sigma$M action is described as follows:
\begin{eqnarray}
S_{\sigma}
&=&
\frac{i}{4\pi}
\int_{M_4}\omega \wedge 
\mbox{Tr} \left[
\left( \partial \sigma \right)\sigma^{-1} \wedge ( \widetilde{\partial} \sigma)\sigma^{-1}\right] \nonumber\\
&=&-\frac{1}{16\pi}
\int_{M_4}%\frac{1}{2}
\mbox{Tr}\left[
\left( \partial_m \sigma \right)\sigma^{-1}
\left( \partial^{m} \sigma\right)\sigma^{-1}\right]
\mathrm{d}z \wedge \mathrm{d}\widetilde{z} \wedge \mathrm{d}w \wedge \mathrm{d}\widetilde{w}, 
\label{Tr(A_m A_n)_differential form}
\end{eqnarray}
where $\partial^m := g^{mn}\partial_n$ and the metric is given by 
 (\ref{com_metric}). This can be represented explicitly in terms of real coordinates on $\mathbb{U, E}$:
\begin{eqnarray}
S_{\sigma} \label{S_sigma}
&=& 
%\left\{
%\begin{array}{l}
-\frac{1}{16\pi}
\int_{\mathbb{U}\scriptsize{\mbox{or}}\mathbb{E}}
%\displaystyle{\frac{1}{2}}
\mbox{Tr}\left[
\left( \partial_{\mu} \sigma \right)\sigma^{-1}
\left( \partial^{\mu} \sigma \right)\sigma^{-1}\right]
\mathrm{d}x^{1}\wedge\mathrm{d}x^{2}\wedge\mathrm{d}x^{3}\wedge\mathrm{d}x^{4},
%\medskip \\
%\displaystyle{\frac{i}{2}}\mbox{Tr}
%\left\{\left(
%\left( \partial_{\mu} \sigma\right)\sigma^{-1}\right)
%\left(\left( \partial^{\mu} \sigma\right)\sigma^{-1}\right)
%\right\}\mathrm{d}x^{0}\wedge\mathrm{d}x^{1}\wedge\mathrm{d}x^{2}\wedge\mathrm{d}x^{3} ~~\mbox{on}~\mathbb{M}
%\end{array}
%\right., 
\end{eqnarray}
where the real space metrics are given in (\ref{E}) and (\ref{U}).
The NL$\sigma$M action density is read from the integrand as 
 ${\cal{L}}_\sigma:=-(1/16\pi)\mbox{Tr}\left[
\left( \partial_{\mu} \sigma \right)\sigma^{-1}
\left( \partial^{\mu} \sigma \right)\sigma^{-1}\right]$. 

%Here we introduce an informal notation $=_{R}$ to denote the cases when the coordinates are real which can be obtained by imposing some suitable conditions (See appendix \ref{Reality Conditions}) on the complex coordinates $z, \widetilde{z}, w, \widetilde{w}$. 
%\subsection{The Wess-Zumino Term for the 4D Flat Spaces}

Similarly the Wess-Zumino action is described as follows:
\begin{eqnarray}
\label{WZ term on 4D flat complex space_1}
S_{\scriptsize{\mbox{WZ}}}
&=& 
-\frac{i}{12\pi}
\int_{M_4} A \wedge 
\mbox{Tr}\left[
\left(\mathrm{d}\sigma\right)\sigma^{-1} \wedge \left(\mathrm{d}\sigma\right)\sigma^{-1} \wedge \left(\mathrm{d}\sigma \right)\sigma^{-1}
\right]   \smallskip \nonumber\\
&\!\!\!\!= \!\!\!\!&
\frac{1}{16\pi}
%\displaystyle{\frac{3}{2}}
\displaystyle{\int_{M_4}}\left\{
\begin{array}{l}
~~\mathrm{Tr}
\left( \theta_{w} \theta_{z}\theta_{\widetilde{z}} - \theta_{w}\theta_{\widetilde{z}} \theta_{z}
\right)w \smallskip \\
+\mathrm{Tr}
\left( \theta_{\widetilde{w}} \theta_{z}\theta_{\widetilde{z}} - \theta_{\widetilde{w}}\theta_{\widetilde{z}} \theta_{z}
\right)\widetilde{w} \smallskip \\
-
\mathrm{Tr} 
\left( \theta_{z} \theta_{w}\theta_{\widetilde{w}} - \theta_{z}\theta_{\widetilde{w}} \theta_{w}
\right)z \smallskip \\
-
\mathrm{Tr} 
\left( \theta_{\widetilde{z}} \theta_{w}\theta_{\widetilde{w}} - \theta_{\widetilde{z}}\theta_{\widetilde{w}} \theta_{w}
\right)\widetilde{z}
\end{array}
\!\!\right\}
\mathrm{d}z \wedge \mathrm{d}\widetilde{z} \wedge \mathrm{d}w \wedge \mathrm{d}\widetilde{w}, 
\end{eqnarray}
where $\theta_{m}:=\left(\partial_{m} \sigma \right)\sigma^{-1}$. 
Here we choose the potential one-form 
$A$ as $A=(i/4)\left(
 z\mathrm{d}\widetilde{z}\right.$ $\left.-\widetilde{z}\mathrm{d}z - w\mathrm{d}\widetilde{w} +\widetilde{w}\mathrm{d}w\right)$. %The other equivalent choices can be referred to Appendix \ref{WZ term_equivalent expressions}. 
This can be reduced to the three kinds of real spaces. 
For example, in the Ultrahyperbolic space $\mathbb{U}_1$, it is 
%by putting the reality conditions. The reduced action density is 
%written in terms of real coordinates:
\begin{eqnarray}
S_{\scriptsize{\mbox{WZ}}} 
=
-\frac{1}{16\pi}
\displaystyle{\int_{\mathbb{U}_1}}
\left\{
\begin{array}{l}
~~\mathrm{Tr}\left( \theta_1 \theta_2 \theta_4 - \theta_1 \theta_4 \theta_2\right)x^{1}
\smallskip \\
+\mathrm{Tr}\left( \theta_2 \theta_1 \theta_3 - \theta_2 \theta_3 \theta_1\right)x^{2}
\smallskip \\
+\mathrm{Tr}\left( \theta_3 \theta_2 \theta_4 - \theta_3 \theta_4 \theta_2\right)x^{3}
\smallskip \\
+\mathrm{Tr}\left( \theta_4 \theta_1 \theta_3 - \theta_4 \theta_3 \theta_1\right)x^{4}
\end{array}
\right\} \mathrm{d}x^{1}\wedge\mathrm{d}x^{2}\wedge\mathrm{d}x^{3}\wedge\mathrm{d}x^{4}.
%\mbox{on}~\mathbb{U}_1
\end{eqnarray}
where $\theta_{\mu}:=\left(\partial_{\mu} \sigma \right)\sigma^{-1}$. 
% ~$m= z,~\widetilde{z},~ w, ~\widetilde{w}$ ~for complex coordinates, and $m=1, 2, 3, 4~(\mbox{or} ~0,1,2,3)$ ~for real coordinates.
The Wess-Zumino action density ${\cal L}_{\scriptsize{\mbox{WZ}}}$ can be read from the integrand. 

\subsection{Useful Formulas for $G=GL(2,\mathbb{C})$}

In this subsection, we focus on the case of $G=GL(2,\mathbb{C})$. 
Additionally we impose 
the condition $\partial_m|\sigma|=0$ on $\sigma$ 
because our soliton solutions $\sigma$ 
obtained in section 3.2 satisfy this condition. Then 
the WZW action density becomes a concise determinant form 
as follows.
By Jacobi's Formula:
$\mbox{Tr}\left[(\partial_{m}\sigma)\sigma^{-1}\right]=\partial_{m}|\sigma|/|\sigma|
=\partial_m \log|\sigma|$, we find that the condition: $\partial_{m}|\sigma|=0$ 
is equivalent to the condition: $\mathrm{Tr}\left[(\partial_{m}\sigma)\sigma^{-1}\right]=0$
%due to Jacobi's Formula:
%$\mbox{Tr}\left[(\partial_{m}\sigma)\sigma^{-1}\right]=\partial_{m}|\sigma|/|\%sigma|
%=\partial_m \log|\sigma|$. 
%This condition is expressed in terms of the matrix elements: 
%$(\partial_{m}\sigma_{11})\sigma_{22}-(\partial_{m}\sigma_{12})\sigma_{21}
%=-\left\{\sigma_{11}(\partial_{m}\sigma_{22})-\sigma_{12}(\partial_{m}\sigma_{21})\right\}$ %Then by applying this 
%which leads to  
which can be expressed in terms of the matrix elements: 
$(\partial_{m}\sigma_{11})\sigma_{22}-(\partial_{m}\sigma_{12})\sigma_{21}
=-\left\{\sigma_{11}(\partial_{m}\sigma_{22})-\sigma_{12}(\partial_{m}\sigma_{21})\right\}$ . Therefore, we have
%the following relation:
\begin{eqnarray}
\label{Tr(A_m A_n)_1}
\mbox{Tr}\left[(\partial_m \sigma)\sigma^{-1}(\partial_n \sigma)\sigma^{-1}\right]
=
\frac{-1}{|\sigma|}
\left(
\left|
\begin{array}{cc}
\partial_m \sigma_{11} & \partial_m \sigma_{12} \\
\partial_n \sigma_{21} & \partial_n \sigma_{22}
\end{array}
\right|
+
\left|
\begin{array}{cc}
\partial_n \sigma_{11} & \partial_n \sigma_{12} \\
\partial_m \sigma_{21} & \partial_m \sigma_{22}
\end{array}
\right|
\right).
\end{eqnarray}
In this paper, we always take the following parametrization for the soliton solution $\sigma$:
\begin{eqnarray}
\label{five}
\sigma=
\frac{-1}{\Delta}
\left(
\begin{array}{cc}
\Delta_{11} & \Delta_{12}  \\
\Delta_{21} & \Delta_{22}
\end{array}
\right), 
\end{eqnarray}
under the condition $\partial_{m}\left| \sigma \right|=0$.
Note that this reparametrization is not unique 
and there is a relation between the five variables: 
\begin{eqnarray}
\label{Relation of Delta and |J|}
\Delta_{11}\Delta_{22} - \Delta_{12}\Delta_{21}=\left| \sigma \right|\Delta^2.
%,~~\partial_{m}\left( \Delta_{11}\Delta_{22} - \Delta_{12}\Delta_{21} \right )=2\left| \sigma \right|\Delta\left(\partial_{m}\Delta\right). 
\end{eqnarray}
In this setting, the quadratic term \eqref{Tr(A_m A_n)_1} can be rewritten as
\begin{eqnarray}
&&\mbox{Tr}\left[(\partial_m \sigma)\sigma^{-1}(\partial_n \sigma)\sigma^{-1}\right]\nonumber \\
&\!\!\!\!=\!\!\!\!&
\label{Tr(A_m A_n)_2}
\frac{1}{|\sigma|\Delta^{2}}
\left\{
\left|
\begin{array}{cc}
\partial_m \Delta_{11} & \partial_m \Delta_{12} \\
\partial_n \Delta_{21} & \partial_n \Delta_{22}
\end{array}
\right|
+
\left|
\begin{array}{cc}
\partial_n \Delta_{11} & \partial_n \Delta_{12} \\
\partial_m \Delta_{21} & \partial_m \Delta_{22}
\end{array}
\right|
-
2|\sigma|(\partial_m \Delta)(\partial_n \Delta)
\right\}.
\end{eqnarray}
%where $A_{mn}+{\mbox{Perm}} (m, n):=A_{mn}+A_{nm}$. 
%Here we use the relation \eqref{Relation of Delta and |J|} to obtain the last equality. 
%in the last line. 
%By using the relation \eqref{Relation of Delta and |J|}, 
Similarly, the cubic term is: 
\begin{eqnarray}
&&\mbox{Tr}\left[(\partial_m \sigma)\sigma^{-1}(\partial_n \sigma)\sigma^{-1}(\partial_p \sigma)\sigma^{-1}\right]   \\
&&=
\frac{1}{2|\sigma|^2\Delta^{4}}
\left|
\begin{array}{cccc}
\Delta_{11} & \Delta_{12} & \Delta_{21} & \Delta_{22} \\
\partial_m \Delta_{11} & \partial_m \Delta_{12} & \partial_m \Delta_{21} & \partial_m \Delta_{22} \\
\partial_n \Delta_{11} & \partial_n \Delta_{12} & \partial_n \Delta_{21} & \partial_n \Delta_{22} \\
\partial_p \Delta_{11} & \partial_p \Delta_{12} & \partial_p \Delta_{21} & \partial_p \Delta_{22}
\end{array}
\right|   \label{Tr(A_m A_n A_p)_1}=
\frac{1}{2|\sigma|^2\Delta^{4}}
%\!\!\left\{
%\begin{array}{l}
(A_{mnp}+A_{npm}+A_{pmn}),
\smallskip \nonumber\\
%{\mbox{where}}~
&&A_{mnp}
:=
\left|
\!\!\begin{array}{cc}
\Delta_{11}\!\! & \Delta_{22}\!\! \\
\partial_m \Delta_{11}\!\! & \partial_m \Delta_{22}\!\!
\end{array}
\right|
\left|
\!\!\begin{array}{cc}
\partial_n \Delta_{12}\!\! & \partial_n \Delta_{21}\!\! \\
\partial_p \Delta_{12}\!\! & \partial_p \Delta_{21}\!\!
\end{array}
\right|
+
\left|
\!\!\begin{array}{cc}
\Delta_{12}\!\! & \Delta_{21}\!\! \\
\partial_m \Delta_{12}\!\! & \partial_m \Delta_{21}\!\!
\end{array}
\right|
\left|
\!\!\begin{array}{cc}
\partial_n \Delta_{11}\!\! & \partial_n \Delta_{22}\!\! \\
\partial_p \Delta_{11}\!\! & \partial_p \Delta_{22}\!\!
\end{array}
\right|.~~~
%(m, n, p) %\rightarrow (n, p, m), ~(p, m, n)
%\end{array}
%\!\!\right\}, 
\label{Tr(A_m A_n A_p)_2}~~~~~~
\end{eqnarray}
By the permutation property of determinants (Cf: \eqref{Tr(A_m A_n A_p)_1}), we have
\begin{eqnarray}
\label{Additive inverse of Tr(A_m A_n A_p)}
\mbox{Tr}\left[(\partial_m \sigma)\sigma^{-1}(\partial_n \sigma)\sigma^{-1}(\partial_p \sigma)\sigma^{-1}\right]
=
-\mbox{Tr}\left[(\partial_m \sigma)\sigma^{-1}(\partial_p \sigma)\sigma^{-1}(\partial_n \sigma)\sigma^{-1}\right].
\end{eqnarray}
Therefore under the condition $ \partial_{m}\left|\sigma \right|=0$, the Wess-Zumino term can be further simplified as
\begin{eqnarray}
S_{\scriptsize{\mbox{WZ}}}  \label{S_WZ}
&=& 
-\frac{i}{12\pi}
\int_{M_4} A \wedge 
\mbox{Tr}\left[
\left(\mathrm{d}\sigma\right)\sigma^{-1} \wedge \left(\mathrm{d}\sigma\right)\sigma^{-1} \wedge \left(\mathrm{d}\sigma\right)\sigma^{-1}
\right]   \\
&\!\!\!\!= \!\!\!\!&
\label{WZ term_complex space}
\frac{1}{8\pi}
\displaystyle{\int_{M_4}}\left\{
\begin{array}{l}
~~\mathrm{Tr}
\left( \theta_{w} \theta_{z}\theta_{\widetilde{z}}
\right)w
+\mathrm{Tr}
\left( \theta_{\widetilde{w}} \theta_{z}\theta_{\widetilde{z}}
\right)\widetilde{w}  \smallskip \\
-
\mathrm{Tr} 
\left( \theta_{z} \theta_{w}\theta_{\widetilde{w}}
\right)z
-
\mathrm{Tr} 
\left( \theta_{\widetilde{z}} \theta_{w}\theta_{\widetilde{w}}
\right)\widetilde{z}
\end{array}
\!\!\right\}
\mathrm{d}z \wedge \mathrm{d}\widetilde{z} \wedge \mathrm{d}w \wedge \mathrm{d}\widetilde{w}  \nonumber  \\
&\!\!\!\stackrel{\scriptsize{\mathbb{U}_1}}{=}\!\!\!\!& ~
\int_{\mathbb{U}_1}
{\cal{L}}_{\scriptsize{\mbox{WZ}}}
\mathrm{d}x^{1}\wedge\mathrm{d}x^{2}\wedge\mathrm{d}x^{3}\wedge\mathrm{d}x^{4},
\nonumber\\
{\cal{L}}_{\scriptsize{\mbox{WZ}}}
&\!\!\!\stackrel{\scriptsize{\mathbb{U}_1}}{=}\!\!\!\!&
-\frac{1}{8\pi}\left(
\mathrm{Tr}\left( \theta_1 \theta_2 \theta_4 \right)x^{1}
+\mathrm{Tr}\left( \theta_2 \theta_1 \theta_3 \right)x^{2}
+\mathrm{Tr}\left( \theta_3 \theta_2 \theta_4 \right)x^{3}
+\mathrm{Tr}\left( \theta_4 \theta_1 \theta_3 \right)x^{4}\right).~~~~~
\label{WZ term_real spaces}
\end{eqnarray}
%Here we introduce an informal notation $\stackrel{\scriptsize{\mbox{R}}}{=}$ to denote the cases when the coordinates are real which can be obtained by imposing the reality conditions on the complex coordinates. 

%\subsection{EOM}

\section{Darboux Transformation and Soliton Solutions}

In this section, we review the soliton solutions of 
the Yang equation %with $G=SL(2,\mathbb{C})$ 
which are constructed 
by applying the Darboux transformation \cite{GNO, GHHN}. 

%The Yang-Mills action densities of them are examined in detail and 
%They are actually solitons in the sense that 

\subsection{Darboux Transformation for the Yang Equation}

Let us assume that $G=GL(N,\mathbb{C})$ in this subsection.
The Yang equation (\ref{Yang}) can be rewritten as the following 
differential equation: 
\begin{eqnarray}
\label{yang}
\partial_{\widetilde{z}}((\partial_z \sigma) \sigma^{-1})
-\partial_{\widetilde{w}}((\partial_w \sigma) \sigma^{-1} )=0. 
\end{eqnarray} 
There exists a Lax representation of (\ref{yang}) 
given by the following linear system \cite{GNO}:
\begin{eqnarray}
L(f)&:=&
\sigma \partial_{w}(\sigma^{-1} f)
- (\partial_{\widetilde{z}}f)\zeta=0,\nonumber\\
M(f)&:=&
\sigma \partial_{z}(\sigma^{-1} f)
- (\partial_{\widetilde{w}}f)\zeta=0. 
\label{lin_yang}
\end{eqnarray}
The spectral parameter $\zeta$ here must be generalized to 
an $N\times N$ constant matrix otherwise the Darboux transformation would be a trivial transformation. This is a key point to define a nontrivial Darboux transformation as we will see later. 
%because $\zeta$ is a scalar matrix, the 

It is %easy to find  that
not hard to verify that 
%the linear system \eqref{lin_yang} satisfies} 
the compatibility condition $L(M(f))-M(L(f))=0$ implies the Yang equation \eqref{yang}. 
The existence of $N$-independent solutions of the linear system (\ref{lin_yang}) is an assumption here, however we will show later that it actually exists for the soliton solution cases. Then, $f$ can be rewritten as 
an $N\times N$ matrix which consists of the
$N$-independent solutions %$f_k ~(k=1\cdots,N)$
as column vectors of length $N$.
%The existence of $f$ is assumed here. 
%As we will see later, it actually exists for the soliton solutions. 

The Darboux transformation is defined as an auto-B\"acklund 
transformation of the linear system (\ref{lin_yang}). 
Firstly, we start with a solution $\sigma$ of the Yang equation, and a solution $f=f(\zeta)$ of the linear system \eqref{lin_yang}. 
Secondly, we
%Let $\sigma$ be a solution of the Yang equation, and 
%$f=f(\zeta)$ be a solution of the linear system (\ref{lin_yang}). 
%Let us 
prepare a special solution $\psi(\Lambda):=f(\Lambda)$ 
for a fixed spectral parameter matrix $\zeta=\Lambda$. 
Then the following Darboux transformation 
%is defined by the following transformation which leaves the linear system as it is \cite{GNO00}:
\begin{eqnarray}
\label{Darboux_phi}
f^\prime=
f \zeta - \psi \Lambda \psi^{-1} f,~~~
\label{Darboux_J}
\sigma^\prime= -\psi \Lambda \psi^{-1} \sigma,
\end{eqnarray}
keeps the linear system \eqref{lin_yang} invariant in form, 
that is, %they satisfy, 
\begin{eqnarray}
L^\prime(f^\prime)&:=&
\sigma^\prime \partial_{w}(\sigma^{\prime -1} f^\prime)
- (\partial_{\widetilde{z}}f^\prime)\zeta=0,\nonumber\\
M^\prime(f^\prime)&:=&
\sigma^\prime \partial_{z}(\sigma^{\prime -1} f^\prime)
- (\partial_{\widetilde{w}}f^\prime)\zeta=0.
\end{eqnarray}
As mentioned before, 
the transformation \eqref{Darboux_phi} becomes trivial if the spectral parameter is a scalar matrix where $\Lambda$ commutes with $\psi$. 
%where $\psi$ is an eigenfunction of the linear system
%\eqref{lin_yang}
%for the choice of eigenvalue $\zeta=\Lambda$. 
The Darboux transformation maps 
the input data $(\sigma, f(\zeta), \psi(\Lambda))$ to 
the output data $(\sigma^\prime, f^\prime(\zeta))$
and therefore we get a new solution 
$\sigma^\prime$ of the Yang equation successfully. In the same way, 
these output data can be reused as the next input data $(\sigma^\prime, f^\prime(\zeta), \psi^\prime(\Lambda^\prime))$ for the 
Darboux transformation. Here we define a special solution 
$\psi^\prime(\Lambda^\prime):= f^\prime(\Lambda^\prime)$
by choosing a suitable spectral parameter matrix $\zeta=\Lambda^\prime$. 
%In this way,  the Darboux transformation can be applied iteratively: 
Continuing this process, we get a series of input-output data :
$(\sigma, f, \psi)\mapsto (\sigma^\prime, f^\prime,\psi^\prime)\mapsto
\cdots$. Therefore by applying $n$ iterations of the Darboux transformation, we can get $n$ exact solutions $\sigma_{[j]}$ of the Yang equation and express them in terms of the quasideterminants in a compact form \cite{GNO, GHHN}.
For our purpose in this paper, it is sufficient to choose a trivial seed solution $\sigma_{[1]}=1$:
%we just mention the result when the choice of the trivial seed solution is $\sigma =1$ as follow :
%From a trivial seed solution $\sigma=1$, we can get an exact solution $\sigma$ 
%by $n$ iterations of the Darboux transformation. 
%The solutions can be expressed 
%in terms of the quasideterminants in a compact form \cite{GNO00, GHHN}:
%Here we use the terminology Wronskian type quasideterminants, or quasi-Wronskian for short.
%The specific form of $\sigma$ is as follows : 
\begin{eqnarray}
\label{Jn}
\sigma_{[n+1]}=
\left|
\begin{array}{cccc}
\psi_1&\cdots&\psi_n& 1\\
\psi_1\Lambda_1&\cdots &\psi_n\Lambda_n& 0\\
\vdots   && \vdots& \vdots\\
\psi_1\Lambda_1^{n-1}&\cdots&\psi_n\Lambda_n^{n-1}& 0\\
\psi_1\Lambda_1^{n}&\cdots& \psi_n\Lambda_n^{n}& \fbox{$0$}
\end{array}\right|, ~~~n \in \mathbb{N}
\end{eqnarray}
%which consists of the solutions 
%$\psi_j=\psi_j(\Lambda_j),~(j=1,2,\cdots,n)$ 
%of the initial linear system 
%with the trivial seed solution $\sigma=1$, 
where each $\psi_j=\psi_j(\Lambda_j)~(j=1,2,\cdots,n)$ is a solution ($N \times N$ matrix) of the initial linear system ($\sigma =1$): 
\begin{eqnarray}
\label{chasing}
\partial_w \psi_j=(\partial_{\widetilde{z}}\psi_j)\Lambda_j,~~~
\partial_z \psi_j=(\partial_{\widetilde{w}}\psi_j)\Lambda_j.
\end{eqnarray}
Hence the problem of solving the Yang equation reduces to solving the equation (\ref{chasing}). 
(The label $_{[n+1]}$ in the $n$-soliton solution 
is omitted in most part of this paper except for Appendix \ref{unitarity}.)

In the main part of this paper, we do not %discuss 
explain details of the quasideterminant, but provide the definition and properties of the quasideterminant in Appendix A. The detailed computations can be found in Appendix B and C.

%We remark that in $G=GL(N, \mathbb{C})$ case, all the $\psi_j$ and $\Lambda_j$ are $N \times N$ matrices. Therefore, $\sigma_n$ is exactly a $N \times N$ matrix if we expand it term by term as the Schur complement \eqref{Schur complement}.

\subsection{Soliton Solutions for $G=GL(2,\mathbb{C})$}

{}From now on, we focus only on the soliton solutions 
for $G=GL(2,\mathbb{C})$. An example of the multi-soliton solution is given by \cite{HaHu2}: 
\begin{eqnarray}
\label{n-soliton solution}
\label{CS_n}
\psi_j=\left(
\begin{array}{cc}
e^{L_j}
& 
e^{-\overline{L}_j}
\\ 
-e^{-L_j}
& 
e^{\overline{L}_j}
\end{array}\right)
,~~~
\Lambda_j=
\left(
\begin{array}{cc}
\lambda_j & 0 \\
0 & \mu_j
\end{array}
\right),
\end{eqnarray}
where the two kinds of spectral parameters 
$\lambda_j, \mu_j$ ($j=1,2,\cdots,n$) are complex constants with the following mutual relationship on each real space:
\begin{eqnarray}
(\lambda_j, \mu_j)&\!\!\!=\!\!\!&
\left\{
\begin{array}{l}
(\lambda_j, \overline{\lambda}_j) ~~ \mbox{on}~ \mathbb{U}_1, ~~
(\lambda_j, 1/\overline{\lambda}_j) ~~ \mbox{on}~ \mathbb{U}_2,~~ 
\smallskip \smallskip \\
(\lambda_j, -1 / \overline{\lambda}_j) ~~ \mbox{on} ~ \mathbb{E}.
%\smallskip \smallskip \\
%(\lambda_j, \mu_j), ~\mu_j \neq \lambda_j ~~\mbox{on} ~\mathbb{M}.
\end{array} \label{(lambda_j, mu_j)}
\right.
\end{eqnarray}
The powers $L_j$ of the exponential function are linear in the complex coordinates: 
$L_j:=\lambda_j\alpha_jz+\beta_j\widetilde{z}+\lambda_j\beta_jw+\alpha_j\widetilde{w}$, { where } { $\alpha_j, \beta_j \in \mathbb{C}$}. %where $\alpha_j,\beta_j$ are complex constants. 
The representations of $L_j$ in real coordinates are 
\begin{eqnarray}
\label{Lj}
L_j  
&\!\!\!\stackrel{\scriptsize{\mathbb{U}_1}}{=}\!\!\!&
\displaystyle{\frac{1}{\sqrt{2}}}
\left\{
(\lambda_j\alpha_j+\beta_j)x^{1} + (\lambda_j\beta_j-\alpha_j)x^{2} + (\lambda_j\alpha_i-\beta_j)x^{3} + (\lambda_j\beta_j+\alpha_j)x^{4}
\right\}, ~~~~~~~~\label{L_j_U1} \\
%&\!\!\!\stackrel{\scriptsize{\mathbb{U}_2}}{=}\!\!\!&
%\displaystyle{\frac{1}{\sqrt{2}}}
%\left\{(\lambda_j\alpha_j+\beta_j)x^{1} + i(\lambda_j\alpha_j-\beta_j)x^{2} + (\alpha_j+\lambda_j\beta_j)x^{3} + i(\alpha_j-\lambda_j\beta_j)x^{4}
%\right\},  \label{L_j_U2} \\
&\!\!\!\stackrel{\scriptsize{\mathbb{E}}}{=}\!\!\!&
\displaystyle{\frac{1}{\sqrt{2}}}
\left\{
(\lambda_j\alpha_j+\beta_j)x^{1} + i(\lambda_j\alpha_j-\beta_j)x^{2} + (\lambda_j\beta_j-\alpha_j)x^{3} + i(\lambda_j\beta_j+\alpha_j)x^{4}
\right\}. \label{L_j_E}
%\\\alpha_j, \beta_j, \lambda_j, \mu_j \in \mathbb{C}
\end{eqnarray}
We use the notation $\ell_{\mu}^{(j)}$ to simplify the coefficients of $L_j$ in the following sections, that is, $L_j:=\ell_\mu^{(j)}x^\mu$. 
%where $\alpha_j,\beta_j,\gamma_j, \delta_j$ are complex constants. 
%The coefficients in the real spaces are  denoted by $L_j=\ell_\mu^{(j)}x^\mu$. 

We remark that the determinant of the $n$-soliton solution $\sigma$ is constant \cite{HaHu2, Huang}:
%satisfies the requirement of $\partial_{m}\left| \sigma \right|=0$ since
\begin{eqnarray}
\label{Determinant of n-soliton solution}
\left|\sigma\right|=\prod_{j=1}^{n}\lambda_j\mu_j.
%~\lambda_j, \mu_j \in \mathbb{C}.
\end{eqnarray}
which satisfies the requirement $\partial_{\mu}|\sigma|=0$. 
Therefore, we can apply the formulas \eqref{Tr(A_m A_n)_2} and \eqref{Tr(A_m A_n A_p)_1} to the $n$-soliton solutions. Especially on the Ultrahyperbolic space $\mathbb{U}_1$, the $n$-soliton solution $\sigma$ satisfies $\sigma\sigma^\dagger=\sigma^\dagger\sigma=\vert\sigma\vert$ \cite{HaHu2, Huang} and hence after the scale transformation: $\sigma\mapsto \vert\sigma\vert^{1/2} \sigma$, $\sigma$ belongs to $SU(2)$. 
On the Euclidean space $\mathbb{E}$, $\sigma$ can take values in $U(2)$, which is proved in Appendix \ref{unitarity}.

%Especially, $\Lambda_j={\mbox{diag}}\left(e^{i\theta_j},e^{-i\theta_j}\right),~\theta_{j}\in \mathbb{R}\Rightarrow \sigma \in SU(2)$. 
%In fact, $\sigma \in \mbox{U}(2) \iff \displaystyle{|\sigma|=\prod_{j=1}^{n}|\lambda_j|^2 =1} \iff \sigma \in \mbox{SU}(2)$. 

%In fact,
%\begin{eqnarray}
%\sigma \in U(2)   \iff \sigma \in SU(2)  \iff 
%\begin{array}{cc}
%e^{i\theta}  &  0 \\
%0 & e^{-i\theta}
%\end{array}
%\right), ~\theta \in \mathbb{R}.
%\end{eqnarray} 

By the definition \eqref{2x2} of the quasideterminant, 
the $n$-soliton solution $\sigma$ (Cf: \eqref{Jn} and \eqref{n-soliton solution}) can be represented in the form of (\ref{five}):
%\begin{eqnarray}
%\sigma&=&
%\frac{-1}{\Delta}
%\left(
%\begin{array}{cc}
%\Delta_{11} & \Delta_{12}  \\
%\Delta_{21} & \Delta_{22}
%\end{array}
%\right), \\
%&&\Delta=\left|
%\begin{array}{ccc}
%\psi_1&\cdots&\psi_n\\
%\psi_1\Lambda_1&\cdots &\psi_n\Lambda_n\\
%\vdots   && \vdots\\
%\psi_1\Lambda_1^{n-1}&\cdots&\psi_n\Lambda_n^{n-1}
%\end{array}\right|,\\
%&&\Delta_{11}=
%{-}\left|
%\begin{array}{cccc}
%\psi_1&\cdots&\psi_n& {\bf e}_1\\
%\psi_1\Lambda_1&\cdots &\psi_n\Lambda_n& {\bf 0}\\
%\vdots   && \vdots& \vdots\\
%\psi_1\Lambda_1^{n-1}&\cdots&\psi_n\Lambda_n^{n-1}& {\bf 0}\\
%(\psi_1\Lambda_1^{n})_1&\cdots&(\psi_n\Lambda_n^{n})_1&0
%\end{array}\right|,~~~
%\Delta_{12}=
%{-}\left|
%\begin{array}{cccc}
%\psi_1&\cdots&\psi_n& {\bf e}_2\\
%\psi_1\Lambda_1&\cdots &\psi_n\Lambda_n& {\bf 0}\\
%\vdots   && \vdots& \vdots\\
%\psi_1\Lambda_1^{n-1}&\cdots&\psi_n\Lambda_n^{n-1}&{\bf 0}\\
%(\psi_1\Lambda_1^{n})_1&\cdots& (\psi_n\Lambda_n^{n})_1&0
%\end{array}\right|,\nonumber\\
%&&\Delta_{21}=
%{-}\left|
%\begin{array}{cccc}
%\psi_1&\cdots&\psi_n& {\bf e}_1\\
%\psi_1\Lambda_1&\cdots &\psi_n\Lambda_n& {\bf 0}\\
%\vdots   && \vdots& \vdots\\
%\psi_1\Lambda_1^{n-1}&\cdots&\psi_n\Lambda_n^{n-1}& {\bf 0}\\
%(\psi_1\Lambda_1^{n})_2&\cdots&(\psi_n\Lambda_n^{n})_2&0
%\end{array}\right|,~~~
%\Delta_{22}=
%{-}\left|
%\begin{array}{cccc}
%\psi_1&\cdots&\psi_n& {\bf e}_2\\
%\psi_1\Lambda_1&\cdots &\psi_n\Lambda_n& {\bf 0}\\
%\vdots   && \vdots& \vdots\\
%\psi_1\Lambda_1^{n-1}&\cdots&\psi_n\Lambda_n^{n-1}& {\bf 0}\\
%(\psi_1\Lambda_1^{n})_2&\cdots&(\psi_n\Lambda_n^{n})_2&0
%\end{array}\right|. \nonumber
%\end{eqnarray}
\begin{eqnarray}
&&\Delta=\left|
\begin{array}{ccc}
\psi_1&\cdots&\psi_n\\
\psi_1\Lambda_1&\cdots &\psi_n\Lambda_n\\
\vdots   && \vdots\\
\psi_1\Lambda_1^{n-1}&\cdots&\psi_n\Lambda_n^{n-1}
\end{array}\right|
 =\left|
\begin{array}{ccc}
(\psi_1)_1 & \cdots & (\psi_n)_1 \\
(\psi_1)_2 & \cdots & (\psi_n)_2 \\
& \Psi_{(n-1) \times n} &
\end{array}\right|,
\nonumber\\
&&\Delta_{11}=-\left|
\begin{array}{cccc}
\psi_1&\cdots&\psi_n& {\bf e}_1\\
\psi_1\Lambda_1&\cdots &\psi_n\Lambda_n& {\bf 0}\\
\vdots   && \vdots& \vdots\\
\psi_1\Lambda_1^{n-1}&\cdots&\psi_n\Lambda_n^{n-1}& {\bf 0}\\
(\psi_1\Lambda_1^{n})_1&\cdots&(\psi_n\Lambda_n^{n})_1&0
\end{array}\right|
=
\left|
\begin{array}{cccc}
(\psi_1\Lambda_1^{n})_1 & \cdots & (\psi_n\Lambda_n^{n})_1 \\
(\psi_1)_2 & \cdots & (\psi_n)_2 \\
 & \Psi_{(n-1) \times n} &
\end{array}\right|,~~~ 
\nonumber  \\
&&\Delta_{12}=-\left|
\begin{array}{cccc}
\psi_1&\cdots&\psi_n& {\bf e}_2\\
\psi_1\Lambda_1&\cdots &\psi_n\Lambda_n& {\bf 0}\\
\vdots   && \vdots& \vdots\\
\psi_1\Lambda_1^{n-1}&\cdots&\psi_n\Lambda_n^{n-1}&{\bf 0}\\
(\psi_1\Lambda_1^{n})_1&\cdots& (\psi_n\Lambda_n^{n})_1&0
\end{array}\right|
=\left|
	\begin{array}{cccc}
	(\psi_1)_1 & \cdots & (\psi_n)_1 \\
	(\psi_1\Lambda_1^{n})_1 & \cdots & (\psi_n\Lambda_n^{n})_1 \\
	& \Psi_{(n-1) \times n} &
	\end{array}\right|,~~~ 
\nonumber\\
&&\Delta_{21}=-\left|
\begin{array}{cccc}
\psi_1&\cdots&\psi_n& {\bf e}_1\\
\psi_1\Lambda_1&\cdots &\psi_n\Lambda_n& {\bf 0}\\
\vdots   && \vdots& \vdots\\
\psi_1\Lambda_1^{n-1}&\cdots&\psi_n\Lambda_n^{n-1}& {\bf 0}\\
(\psi_1\Lambda_1^{n})_2&\cdots&(\psi_n\Lambda_n^{n})_2&0
\end{array}\right|
=\left|
	\begin{array}{cccc}
	(\psi_1\Lambda_1^{n})_2 & \cdots & (\psi_n\Lambda_n^{n})_2 \\
	(\psi_1)_2 & \cdots & (\psi_n)_2 \\
    & \Psi_{(n-1) \times n} &
	\end{array}\right|,~~~ 
\nonumber \\
&&\Delta_{22}=-\left|
\begin{array}{cccc}
\psi_1&\cdots&\psi_n& {\bf e}_2\\
\psi_1\Lambda_1&\cdots &\psi_n\Lambda_n& {\bf 0}\\
\vdots   && \vdots& \vdots\\
\psi_1\Lambda_1^{n-1}&\cdots&\psi_n\Lambda_n^{n-1}& {\bf 0}\\
(\psi_1\Lambda_1^{n})_2&\cdots&(\psi_n\Lambda_n^{n})_2&0
\end{array}\right|
=\left|
	\begin{array}{cccc}
	(\psi_1)_1 & \cdots & (\psi_n)_1 \\
	(\psi_1\Lambda_1^{n})_2 & \cdots & (\psi_n\Lambda_n^{n})_2 \\
	& \Psi_{(n-1) \times n}&
	\end{array}\right|,
\label{data}
\end{eqnarray}
%\nonumber\\&&
where
%{Here we introduce a new symbol:} 
\begin{eqnarray*}
 \Psi_{(n-1) \times n}:=
\left(
\begin{array}{ccc}
%\psi_1&\cdots&\psi_n\\
\psi_1\Lambda_1&\cdots &\psi_n\Lambda_n\\
\vdots   && \vdots\\
\psi_1\Lambda_1^{n-1}&\cdots&\psi_n\Lambda_n^{n-1}
\end{array}\right),
\end{eqnarray*}
and ${\bf e}_1:=(1,0)^t$, ${\bf e}_2:=(0,1)^t$, ${\bf 0}:=(0,0)^t$, 
and $(A)_k$ is the $k$-th row of a square matrix $A$. 
The data $\Delta$ and $\Delta_{jk}$ 
are determinants of $2n\times 2n$ matrices.
%and of a $(2n+1)\times (2n+1)$ matrix, respectively. 

We remark that $\psi_j$ can be decomposed into, for instance
%there is the following identity:
%\begin{eqnarray}
%\label{XTheta}
%\psi_j
%=\left(
%\begin{array}{cc}
%e^{L_j}
%& 
%e^{-\overline{L}_j}
%\\ 
%-e^{-L_j}
%& 
%e^{\overline{L}_j}
%\end{array}\right)
%=\left(
%\begin{array}{cc}
%1
%& 
%0
%\\ 
%0
%& 
%e^{i\Theta_j}
%\end{array}\right)
%\left(
%\begin{array}{cc}
%e^{X_j}
%& 
%e^{-X_j}
%\\ 
%-1
%& 
%1
%\end{array}\right)
%\left(
%\begin{array}{cc}
%e^{-\overline{L}_j}
%& 
%0
%\\ 
%0
%&
%e^{L_j} 
%\end{array}\right), 
%\end{eqnarray}
\begin{eqnarray}
\label{XTheta}
\psi_j
=\left(
\begin{array}{cc}
e^{L_j} & e^{-\overline{L}_j}
\\ 
-e^{-L_j} & e^{\overline{L}_j}
\end{array}\right)
=\left(
\begin{array}{cc}
e^{X_j}  &   e^{i\Theta_j}\\ 
-e^{-i\Theta_j}  &  e^{X_j} 
\end{array}\right)
\left(
\begin{array}{cc}
e^{-\overline{L}_j} & 0  \\ 
0  &  e^{-L_j} 
\end{array}\right), 
\end{eqnarray}
where $X_j:=L_j+\overline{L}_j$, 
$i\Theta_j:=L_j-\overline{L}_j$. 
The second factor diag$(e^{-\overline{L}_j},e^{-L_j})$
can be eliminated in the $n$-soliton solutions (\ref{Jn})
due to the property of the quasideterminant (\ref{Rmulti}).
Hence the $n$-soliton solutions (\ref{Jn}) 
depend only on $X_j$ and $\Theta_j$. 
The expansion coefficients for the real coordinates 
are denoted by: $X_j= r_{\mu}^{(j)}x^{\mu}, 
i\Theta_j= s_{\mu}^{(j)}x^{\mu}$, that is, 
$r_{\mu}^{(j)}:= \ell_{\mu}^{(j)} + \overline{\ell}_{\mu}^{(j)}\in \mathbb{R}, 
s_{\mu}^{(j)}:= \ell_{\mu}^{(j)} - \overline{\ell}_{\mu}^{(j)}\in i\mathbb{R}$, 
where $L_j=\ell_\mu^{(j)}x^\mu$. It is obvious that 
\begin{eqnarray}
\label{Derivative of exponential functions_1}
\partial_{\mu}e^{\pm X_j} = \pm r_{\mu}^{(j)}e^{\pm X_j},~~
\partial_{\mu}e^{\pm i\Theta_j} = \pm s_{\mu}^{(j)}e^{\pm i\Theta_j}, ~~
\partial_{\mu}e^{\pm i\Theta_{jk}} = \pm (s_{\mu}^{(j)} -s_{\mu}^{(k)})e^{\pm i\Theta_{jk}},
\end{eqnarray}
{ where $\Theta_{jk} := \Theta_{j} - \Theta_{k}$.}
Note that the flip of space-time coordinates $x\!\rightarrow\! -x
\!\Leftrightarrow\! 
(x^1,\!x^2,\!x^3,\!x^4)\!\rightarrow\! (\!-x^1,\!-x^2,\!-x^3,\!-x^4)$
corresponds to the following flips of the new variables:
\begin{eqnarray}
x\rightarrow -x \Longleftrightarrow 
L_j \rightarrow -L_j  ~ \Longleftrightarrow ~ (X_j, \Theta_j) \rightarrow  (-X_j, -\Theta_j).
\end{eqnarray}
Under this flip, we find the following symmetry 
(Cf: \ref{Symmetry of Delta, Delta_ij}):
\begin{eqnarray}
\label{Symmetry of Delta, Delta_ij (2)}
%\Delta(-x)=\Delta(x), \Delta_{11}(-x)=\Delta_{22}(x), \Delta_{12})(-x)=\Delta_{21}(x), 
(\Delta, \Delta_{11}, \Delta_{12}, \Delta_{21}, \Delta_{22})\Big|_{(X_j, \Theta_j) \rightarrow (-X_j, -\Theta_j)}
&\!\!\!\!=\!\!\!\!&(\Delta, \Delta_{22}, \Delta_{21}, \Delta_{12}, \Delta_{11}),
\\
%\partial_{\mu}\Delta(-x)=-\partial_{\mu}\Delta(x), \partial_{\mu}\Delta_{11}(-x) =-\partial_{\mu}\Delta_{22}(x), \partial_{\mu}\Delta_{12})(-x)=-\partial_{\mu}\Delta_{21}(x), 
\partial_{\mu}(\Delta, \Delta_{11}, \Delta_{12}, \Delta_{21}, \Delta_{22})\Big|_{(X_j, \Theta_j) \rightarrow (-X_j, -\Theta_j)}
&\!\!\!\!=\!\!\!\!&-\partial_{\mu}(\Delta, \Delta_{22}, \Delta_{21}, \Delta_{12}, \Delta_{11}).
\end{eqnarray}

Here let us discuss the singularities of the solution $\sigma$. 
Under the decomposition (\ref{five}), possible singularities correspond to 
zeros of $\Delta$. For the one-soliton solution (Cf: \eqref{Data of J_2}), 
$\Delta=2 \cosh X_1$ and 
hence there is no singularity.\footnote{If the 2-1 component $-e^{-L_1}$ 
in the one-soliton solution $\psi$ is replaced with $+e^{-L_1}$, then 
$\Delta=2\sinh X_1$ and $\Delta$ has zero on $X_1=0$. %(See Section 6.) 
This corresponds to the following singular one-soliton solution of the KP equation: $u=2\partial_x^2 \log (e^X-e^{-X})\propto \mbox{csch}^2 X$ 
where $X$ is a linear combination of the space-time coordinates $t,x,y$. 
On the other hand, a non-singular one-soliton solution is given by $u(t,x,y)=2\partial_x^2 \log (e^X+e^{-X})\propto \mbox{sech}^2 X$.} 
For the two-soliton solution (Cf: \eqref{Data of J_3}), 
%On the Ultrahyperbolic spaces $\mathbb{U}_1$ and $\mathbb{U}_2$, 
we can evaluate the value of $\Delta$ on the Ultrahyperbolic space $\mathbb{U}_1$ as follows:
%use the triangular inequality (?) to show that
\begin{eqnarray}
\frac{1}{2} \Delta&=& a~\!\mbox{cosh}(X_1 + X_2) + b~\!\mbox{cosh}(X_1 - X_2) + c~\!\mbox{cos}\Theta_{12}    \nonumber  \\
&\!\!\!\!\!\! \geq \!\!\!\!\!\!&
%\left\{
%\begin{array}{l}
\left| \lambda_1-\lambda_2 \right|^2 + \left| \lambda_1-\overline{\lambda}_2 \right|^2 - \left| \left( \lambda_1-\overline{\lambda}_1 \right)\!\left( \lambda_2-\overline{\lambda}_2 \right) \right| % ~~~\mbox{on}~ \mathbb{U}_1   \smallskip \\
%\left| \lambda_1-\lambda_2 \right|^2 + \left| \lambda_1\overline{\lambda}_2 -1 \right|^2 - \left( \left| \lambda_1 \right|^2 - 1 \right)\!\left( \left| \lambda_2 \right|^2 -1  \right) ~\mbox{on}~ \mathbb{U}_2
%\end{array}
%\right.    
\nonumber\\
&\!\!\!\!\!\! = \!\!\!\!\!\!&
\left\{
\begin{array}{l}
2\left| \lambda_1-\overline{\lambda}_2 \right|^2 \textgreater 0  ~~~~\mbox{if}~  c \textgreater 0   \smallskip \\
2\left| \lambda_1-\lambda_2 \right|^2 \textgreater 0  ~~~~\mbox{if}~ c \textless 0
\end{array},
\right.% ~~\mbox{on}~ \mathbb{U}_1   \smallskip \\
%~~~~2\left| \lambda_1-\lambda_2 \right|^2 \textgreater 0  ~~~~~~~~~~~~~~~~~~\!\mbox{on}~ \mathbb{U}_2.
\end{eqnarray}
where $a,b,c$ are real constants defined in Table \ref{Table_1} of section 4.2. 
Therefore, the denominator is anywhere positive and 
$\sigma$ is proved to be non-singular on $\mathbb{U}_1$. 
%Additionally, %by \eqref{NL Sigma term_2-Soliton_form 1} or \eqref{NL Sigma term_2-Soliton_form 2} 
%we find that ${\cal{L}}_\sigma$ is real-valued %on $\mathbb{U}_1$ and $\mathbb{U}_2$ since the coefficients $a$, $b$, $c$, $d_{11}$, $d_{22}$ are real and $d_{12}$, $d_{21}$ are complex conjugates (Cf: Table \ref{Table_1}). 

On the other hand, on the Euclidean space $\mathbb{E}$, 
$\sigma$ has singularities because it has zero locus due to the fact that $\mbox{cosh}(X_1 \pm X_2) \geq 1$, $\left| \mbox{cos}\Theta_{12} \right| \leq 1$ and $a, b$ have opposite signs (See Table 1 in section 4.2). However this problem can be solved successfully by choosing suitable initial data $\psi$, which will be discussed in section 5. 

Finally we comment on an asymptotic behavior in the region that $r^2:=(x^1)^2+(x^2)^2+(x^3)^2+(x^4)^2$ is large enough in order to prove that the Wess-Zumino action density decays exponentially in the asymptotic region. We note that the $n$-soliton solution (\ref{Jn}) is a meromorphic function of $(\xi_1, \cdots, \xi_n, \eta_1, \cdots, \eta_n)$ where $\xi_K:=e^{X_K}, \eta_K:=e^{i\Theta_K}$. Let us discuss the absolute value of the Wess-Zumino action density. In fact, we will see in section 5 that %the Wess-Zumino action density for the $n$-soliton solution $\sigma$ is asymptotic to a zero, 
the action density tends to zero in the asymptotic region. 
%$\mathscr{R}_K$ and  $\mathscr{R}$. 
This implies that for any variable $\xi_K$, the polynomial degree of the denominator is greater than that of the numerator. ($\eta_K$ is not essential because of $\vert \eta_K \vert=1$.) Let us consider a specific asymptotic direction where the most dominant terms are $\xi_1^{i_1}\cdots \xi_n^{i_n}$ in the numerator and $\xi_1^{j_1}\cdots \xi_n^{j_n}$ in the denominator where $i_k \leq j_k$. Then the action density behaves ${\cal{O}}(\xi_1^{i_1-j_1}\cdots \xi_n^{i_n-j_n})$. At least for one $K$, $i_K<j_K$ and hence this implies that it decays exponentially. 

Let us take the two-soliton case for example. If we consider the asymptotic limit such that $X_1$ is finite and $\vert X_2\vert \gg 1$ (Cf: \eqref{Density of the WZ term_one-third part}), 
%For example, in the two-soliton case in the asymptotic region where $X_1$ is finite and $\vert X_2\vert \gg 1$, this is seen around (\ref{Density of the WZ term_one-third part}) where 
the most dominant factor is $e^{\pm X_2}$ and the denominator and the numerator have the same order of $\xi_1\equiv e^{\pm X_2}$. 
However, due to the identity (\ref{C}), the most dominant term in the numerator vanishes and hence the Wess-Zumino action density is ${\cal{O}}(\xi_1^{-k})$, where $k$ is some positive integer. Therefore we can conclude the Wess-Zumino action density decays exponentially. 

Therefore on $\mathbb{U}_1$, 
the Wess-Zumino action converges for the one- and two-soliton solutions 
because it has no singularity and decays exponentially. 
For the $n$-soliton solution, this is an open problem 
which is discussed in section 7.

\section{Evaluation of Action Density}

In this section, we compute the action density of the $SU(2)$ WZW$_4$ model for the one- and two-soliton solutions and find that the corresponding action densities are real-valued on each space. 
We also find that for the one-soliton solution, the NL$\sigma$M action density is localized on a three-dimensional hyperplane and the Wess-Zumino action density identically vanishes. %We mainly discuss in the case of $\mathbb{U}_1$, however 
For the two-soliton solution, we complete the calculation of the NL$\sigma$M term, and reach to a compact form. 
In particular, 
%we show that the NL$\sigma$M action density is real-valued on the Ultrahyperbolic space $\mathbb{U}_1$. 
the two peaks of the action density are localized on nonparallel two three-dimensional hyperplanes. 
%which can be interpreted to be localized on two intersecting three-dimensional hyperplanes in the asymptotic region. 
As for the Wess-Zumino term, we show that the action density is asymptotic to zero. 
Especially on the Ultrahyperbolic space $\mathbb{U}_1$, 
no singularity appears in the action densities for the two-soliton case as indicated in section 3. 
%{Hence the integration of the Wess-Zumino action density converges because it is exponentially decayed in the asymptotic region as shown in section 3.2. }

\subsection{One-Soliton Solutions}

In this subsection, we compute the action densities of the $SU(2)$ WZW$_4$ model for the one-soliton solutions.

%\subsubsection{NL$\sigma$M Action Density}

%From now on, we discuss the non-singular case of $a=b=c=d=1$ {(\color In fact, a, b, c, d were not mentioned in previous sections ??)} which correspond to our solution (\ref{Jn}).
%Let us compute the NL$\sigma$M action density. 

To calculate the NL$\sigma$M action density explicitly, we substitute the data of one-soliton \eqref{Data of J_2} and \eqref{Determinant of n-soliton solution} into (\ref{Tr(A_m A_n)_2}) for $m=n=\mu$ and then obtain the following result:
\begin{eqnarray}
\label{L_sigma}
{\cal{L}}_\sigma 
\!\!\!
&=&
\!\!\!
-\frac{1}{16\pi}
\mbox{Tr}\left[(\partial_{\mu}\sigma)\sigma^{-1}\left(\partial^{\mu}\sigma\right)\sigma^{-1} \right]
=
\frac{1}{8\pi}
d_{11}~\!\mbox{sech}^2 X_1,
%\!\!\!\!\!\!
%\mbox{where}~
%d_{11}:=\left\{\begin{array}{l}
%\left(\alpha_1\overline{\beta}_1 - \overline{\alpha}_1\beta_1\right)
%\left(\lambda_1-\overline{\lambda}_1\right)^3 / \left| \lambda_1 \right|^2~~~\mbox{on} ~ \mathbb{U}_1 
%~ \left( \alpha_1, \beta_1, \lambda_1 \notin \mathbb{R} \right) 
%\smallskip \\
%\left(\left| \alpha_1 \right|^2 - \left| \beta_1 \right|^2\right)
%\left(\left| \lambda_1 \right|^2 - 1 \right)^3 / \left| \lambda_1 \right|^2 ~~\mbox{on}~ \mathbb{U}_2 
%~\left( \left| \alpha_1\right|\neq\left| \beta_1\right|, ~\left| \lambda_1 \right| \neq 1 \right) 
%\smallskip \\
%\left(\left| \alpha_1 \right|^2 + \left| \beta_1 \right|^2 \right)
%\left(\left| \lambda_1 \right|^2 + 1 \right)^3 / \left| \lambda_1 \right|^2 ~~\mbox{on} ~ \mathbb{E}% ~~~~\!   
%\smallskip \\
%-\left(\lambda_1 - \mu_1 \right)^2\left| \lambda_1 - \mu_1 \right|^2\left| \beta_1 \right|^2 / \lambda_1\mu_1 ~~\! \mbox{on}~\mathbb{M} ~
%\left( \mu_1 \neq \lambda_1 \right)
%\end{array}.
%\right.
%\\
%&& ~~~~~{ \mbox{(Cf : $d_{ij}$ in table \ref{Table_1})}}  \nonumber 
\end{eqnarray}
where $d_{11}$ is determined by \eqref{(lambda_j, mu_j)} and \eqref{L_j_U1} $\sim$ \eqref{L_j_E}, for instance, $d_{11}^{~\!\mathbb{U}_1}=(\alpha_1\overline{\beta}_1 - \overline{\alpha}_1\beta_1)(\lambda_1-\overline{\lambda}_1)^3 / \left| \lambda_1 \right|^2$ and $d_{11}^{~\!\mathbb{E}}=(\left| \alpha_1 \right|^2 + \left| \beta_1 \right|^2 )(\left| \lambda_1 \right|^2 + 1 )^3 / \left| \lambda_1 \right|^2$ (Cf: $d_{jk}$ in Table \ref{Table_1}). Hence $d_{11}$ and ${\cal{L}}_\sigma$ are clearly real-valued. 
%where $d_{11}:=\left(\alpha_1\overline{\beta}_1 - \overline{\alpha}_1\beta_1\right)\left(\lambda_1-\overline{\lambda}_1\right)^3 / \left| \lambda_1 \right|^2$ which is calculated by \eqref{(lambda_j, mu_j)} and \eqref{L_j_U1} $\sim$ \eqref{L_j_E}. 
%In fact, $d_{11}$ takes real value on each real space. 
%On the other hand, 
Note that the action density vanishes identically in the case of $\alpha_1, \beta_1, \lambda_1 \in \mathbb{R}$ on $\mathbb{U}_1$ and hence the result is trivial. 
%and $\left|\alpha_1\right|=\left|\beta_1\right|$ 
%or $\left|\lambda_1\right|=1$ on $\mathbb{U}_2$.
For the nontrivial cases, the peak of the action density lies on the three-dimensional hyperplane described by the linear equation $X_1=0$ on each space. %This behavior is almost the same as the typical one-solitons such as the KP solitons and the ASDYM solitons. 

%In the case of $\mathbb{E}$, the action density has the same form as (\ref{L_sigma}) except for the coefficient: $d_{11}^{\mathbb{E}}=\left(\left| \alpha_1 \right|^2 + \left| \beta_1 \right|^2 \right)\left(\left| \lambda_1 \right|^2 + 1 \right)^3 / \left| \lambda_1 \right|^2$. 

%\subsubsection{Wess-Zumino Action Density}

The Wess-Zumino action density can be calculated by substituting the data of one-soliton \eqref{Data of J_2} and \eqref{Determinant of n-soliton solution} 
into (\ref{Tr(A_m A_n A_p)_2}) directly. 
Then we have 
\begin{eqnarray}
\left|
\begin{array}{cc}
\Delta_{11} & \Delta_{22} \\
\partial_{\mu}\Delta_{11} & \partial_{\mu}\Delta_{22}
\end{array}
\right|
&\!\!\!\! = \!\!\!\!&
\left|
\begin{array}{cc}
\lambda_1 e^{X_1} + \mu_1 e^{-X_1} &
\mu_1 e^{X_1} + \lambda_1 e^{-X_1}  \\
r_{\mu}^{(1)}(\lambda_1 e^{X_1} - \mu_1 e^{-X_1}) &
r_{\mu}^{(1)}(\mu_1 e^{X_1} - \lambda_1 e^{-X_1})  \\
\end{array}
\right|
\nonumber \\
&\!\!\!\!=\!\!\!\!&
-2r_{\mu}^{(1)}(\lambda_1^2 - \mu_1^2), ~~
\\
\left|
\begin{array}{cc}
\partial_{\nu} \Delta_{11} & \partial_{\nu} \Delta_{22} \\
\partial_{\rho} \Delta_{11} & \partial_{\rho} \Delta_{22}
\end{array}
\right|
&\!\!\!\!=\!\!\!\!&
\left|
\begin{array}{cc}
r_{\nu}^{(1)}
(\lambda_1 e^{X_1} - \mu_1 e^{-X_1}) &
r_{\nu}^{(1)}(\mu_1 e^{X_1} - \lambda_1 e^{-X_1})
\\
r_{\rho}^{(1)}(\lambda_1 e^{X_1} - \mu_1 e^{-X_1}) &
r_{\rho}^{(1)}(\mu_1 e^{X_1} - \lambda_1 e^{-X_1})
\end{array}
\right|=0, ~~~~
\\
\left|
\begin{array}{cc}
\Delta_{12} & \Delta_{21} \\
\partial_{\mu}\Delta_{12} & \partial_{\mu}\Delta_{21}
\end{array}
\right|
&\!\!\!\!=\!\!\!\!&
(\lambda_1 - \mu_1)^2
\left|
\begin{array}{cc}
e^{i\Theta_1}  &  e^{-i\Theta_1}  \\
s_{\mu}^{(1)}e^{i\Theta_1} &
-s_{\mu}^{(1)}e^{-i\Theta_1}
\end{array}
\right|  
=
-2s_{\mu}^{(1)}(\lambda_1 - \mu_1)^2,   
\\ 
\left|
\begin{array}{cc}
\partial_{\nu} \Delta_{12} & \partial_{\nu} \Delta_{21} \\
\partial_{\rho} \Delta_{12} & \partial_{\rho} \Delta_{21}
\end{array}
\right|
&\!\!\!\!=\!\!\!\!&
(\lambda_1 - \mu_1)^2
\left|
\begin{array}{cc}
s_{\nu}^{(1)} e^{i\Theta_1} & -s_{\nu}^{(1)} e^{-i\Theta_1} \\
s_{\rho}^{(1)} e^{i\Theta_1} & -s_{\rho}^{(1)} e^{-i\Theta_1}
\end{array}
\right|=0.  
\end{eqnarray}
These facts imply that 
\begin{eqnarray}
\label{Trivial action density_WZW4}
\mbox{Tr}\left[(\partial_{\mu}\sigma)\sigma^{-1}(\partial_{\nu}\sigma)\sigma^{-1}(\partial_{\rho}\sigma)\sigma^{-1}\right]=0, 
%~~\mbox{where}~~m, n, p = z, \widetilde{z}, w, \widetilde{w},
\end{eqnarray}
and therefore the Wess-Zumino action density is identical to zero for the one-soliton case. 
In fact, the identity \eqref{Trivial action density_WZW4} holds 
%for more {extensive} $\sigma$ 
even when $\sigma$ is not a classical solution of the WZW$_4$ model because the condition \eqref{chasing} is not used in our discussion. 
More explicitly, as long as the power $L_j$ of exponential function in $\psi_j$ is an arbitrary linear function of $x^{\mu}$, 
the Wess-Zumino action density vanishes identically. 

All the above results also hold in the case of $\mathbb{E}$. 
%The same results are true 
%any $\psi_j$ which is in the form of \eqref{CS_n} implies that
%In summary, the WZW$_4$ action density for the one-soliton solutions is localized on the three-dimensional hyperplane $X_1=0$. 

%\subsection{More General Solutions and Singularity}

\subsection{Two-Soliton Solutions}

In this subsection, we calculate the action densities of $SU(2)$ WZW$_4$ model explicitly for the two-soliton solutions. 

%\subsubsection{NL$\sigma$M Action Density}

By the result of Appendix \ref{Exact calculation of the NL Sigma Model Term (2-Soliton)} together with \eqref{(lambda_j, mu_j)} and \eqref{L_j_U1} $\sim$ \eqref{L_j_E}, we get the following compact form of the NL$\sigma$M action density for two-soliton solution :
\begin{eqnarray}
{\cal{L}}_\sigma&=&
%\!\!\!\!
-\frac{1}{16\pi}
\mbox{Tr}\left[\left(\partial_{\mu}\sigma\right)\sigma^{-1}\left(\partial^{\mu}\sigma\right)\sigma^{-1}\right] \nonumber  \\
&\!\!\!\! = \!\!\!\!&
\label{NL Sigma term_2-Soliton_form 1}
\frac{
	\left\{ 
	\begin{array}{l}
	~ab
	\left[
	d_{11}\cosh^2 X_2 + d_{22}\cosh^2 X_1
	\right]
	\medskip \\
	\!\!+ ac\left[
	d_{12}~\!\displaystyle{\cosh^2 \left(\frac{X_1 + X_2 - i\Theta_{12}}{2} \right)}
	+ d_{21}~\!\displaystyle{\cosh^2 \left(\frac{X_1 + X_2 + i\Theta_{12}}{2} \right)}
	\right]
	\medskip \\
	\!\!- bc\left[
	e_{12}~\!
	\displaystyle{\sinh^2 \left(\frac{X_1 - X_2 - i\Theta_{12}}{2}
		\right)}
	+\overline{e}_{12}~\!
	\displaystyle{\sinh^2 \left(\frac{X_1 - X_2 + i\Theta_{12}}{2} \right)}
	\right]
	\end{array}
	\!\!\!\right\}
}
{2\pi\displaystyle{
		\left[
		a\cosh(X_1 + X_2)
		+
		b\cosh(X_1 - X_2)
		+
		c\cos \Theta_{12}
		\right]^2}
	}~~~~~~
\nonumber\\
&\!\!\!\! = \!\!\!\!&
\label{NL Sigma term_2-Soliton_form 2}
\frac{\left\{
	\begin{array}{l}
	~~2ab\left[d_{11}~\!\mbox{cosh}^2X_2 + d_{22}~\!\mbox{cosh}^2 X_1 \right]
	\smallskip \\
	~\!+c\left[a(d_{12} + d_{21}) - b(e_{12} + \overline{e}_{12})\right]\mbox{cosh}X_1\mbox{cosh}X_2\mbox{cos}\Theta_{12} 
	\smallskip \\
	~\!+c\left[a(d_{12} + d_{21}) + b(e_{12} + \overline{e}_{12})\right](\mbox{sinh}X_1\mbox{sinh}X_2 + 1)\mbox{cos}\Theta_{12}
	\smallskip \\
	-ic\left[a(d_{12}- d_{21}) + b(e_{12} - \overline{e}_{12}) \right]\mbox{sinh}X_1\mbox{cosh}X_2\mbox{sin}\Theta_{12}  
	\smallskip \\
	-ic\left[a(d_{12}- d_{21}) - b(e_{12} - \overline{e}_{12}) \right]\mbox{cosh}X_1\mbox{sinh}X_2\mbox{sin}\Theta_{12}  
	\end{array}
	\right\}}
{4\pi
	\displaystyle{
		\left[
		a\cosh(X_1 + X_2)
		+
		b\cosh(X_1 - X_2)
		+
		c\cos \Theta_{12}
		\right]^2
	}
} 
\end{eqnarray} 
where $a, b, c, d_{jk}, e_{jk}$ are defined in the following Table \ref{Table_1} for each space. (The difference between $\mathbb{E}$ and $\mathbb{U}_1$ appears only in the coefficients like the one-soliton case.) 
Note that the coefficients in Table \ref{Table_1} also guarantee the NL$\sigma$M action density to be real-valued on $\mathbb{U}_1$ 
and $\mathbb{E}$. %-\left\{ \mbox{singularities}\right\}$.
%Apparently, ${\cal{L}}_\sigma$ for two-soliton solution is in a fractional form unlike the one-soliton case and therefore the problem of singularities remains. More precisely, we need to check whether the denominator of ${\cal{L}}_\sigma$ is nonzero  on each real space.
\begin{table}[h]
\!\!\!\!\caption{Summary of Coefficients}
\label{Table_1}
\medskip
\begin{center}
\begin{tabular}{|c|c|c|}
		%\hspace{-1cm}
		\hline
		Space & $\mathbb{U}_1$ & $\mathbb{E}$   \\ 
		(Metric) & $(+,+,-,-)$ & $(+,+,+,+)$   \\
		\hline
		\hline
		$a \in \mathbb{R^{+}}$ & $\left| \lambda_1 - \lambda_2 \right|^2 \textgreater 0$ & %$\left| \lambda_1 - \lambda_2 \right|^2\textgreater 0$ & 
$\left| \lambda_1 - \lambda_2 \right|^2\textgreater 0$ \\ 
		\hline 
		$b \in \mathbb{R}$ & $\left| \lambda_1 - \overline{\lambda}_2 \right|^2 \textgreater 0$ %& $\left| \lambda_1\overline{\lambda}_2 -1 \right|^2 \textgreater 0$ 
& $-\left| \lambda_1\overline{\lambda}_2 +1 \right|^2 \textless 0$ \\
		\hline 
		$c \in \mathbb{R}$ & $\left( \lambda_1 - \overline{\lambda}_1 \right)\!\!\left( \lambda_2 - \overline{\lambda}_2 \right)$ 
%& $-\left( \left| \lambda_1 \right|^2 - 1 \right)\!\!\left( \left| \lambda_2 \right|^2 - 1 \right)$ 
& $\left( \left| \lambda_1 \right|^2 + 1 \right)\!\!\left( \left| \lambda_2 \right|^2 + 1 \right)$ \\ 
		\hline
		$d_{jk}$  & $\underline{\left( \alpha_j\overline{\beta}_k - \beta_j\overline{\alpha}_k \right)
			\!\left( \lambda_j - \overline{\lambda}_k \right)^3}$ 
%& 		$\underline{\left( \alpha_j\overline{\alpha}_k - \beta_j\overline{\beta}_k \right)			\!\left( \lambda_j\overline{\lambda}_k - 1 \right)^3}$ 
&  $\underline{\left( \alpha_j\overline{\alpha}_k + \beta_j\overline{\beta}_k \right)
			\!\left( \lambda_j\overline{\lambda}_k + 1 \right)^3}$  \\ 
		$(=\overline{d}_{kj})$ & $\lambda_j\overline{\lambda}_k$ %& $\lambda_j\overline{\lambda}_k$ 
& $\lambda_j\overline{\lambda}_k$ \\ 
		\hline
		$e_{jk}$ &  $\underline{(\alpha_j\beta_k - \beta_j\alpha_k)( \lambda_j - \lambda_k)^3}$ %& $\underline{(\alpha_j\beta_k - \beta_j\alpha_k)( \lambda_j - \lambda_k)^3}$ 
& $\underline{(\alpha_j\beta_k - \beta_j\alpha_k)( \lambda_j - \lambda_k)^3}$ \\
		& $\lambda_j\lambda_k$ %& $\lambda_j\lambda_k$ 
& $\lambda_j\lambda_k$  \\
\hline 
%Singularity & non-singular & singular \\\hline
\end{tabular}
\end{center}
\end{table}
 
Next, let us explain why \eqref{NL Sigma term_2-Soliton_form 1} can be interpreted as two intersecting one-solitons in the asymptotic region.
Due to the solitonic property, individual one-solitons will regain all their features (wave shape, velocity, amplitude, etc.) outside the scattering region except for respective differences of an additional position shift. Theoretically, each one-soliton can be separated completely from the other one-soliton in the asymptotic region in which it dominates the asymptotic behavior mainly. 
%Now let us consider the the asymptotic behavior of the action density \eqref{NL Sigma term_2-Soliton_form 1} which can be interpreted as intersecting two one-solitons. 
%We show that on $\mathbb{U}_1$ and $\mathbb{U}_2$ describes the behavior of 2-soliton exactly. 
%In the soliton scatterings, it is interesting to consider the following type of asymptotic limit:
Therefore, we can consider the following type of asymptotic limit:
\begin{eqnarray}
\label{asymp_1}
\left\{
\begin{array}{l}
X_1 ~ \mbox{is finite}
\smallskip \\
|X_2|  \gg  |X_1|
\end{array}
\right.
\end{eqnarray}
in which the first one-soliton, localized on the hyperplane $X_1=0$, dominates the asymptotic behavior mainly. 
Such asymptotics will be discussed more systematically in section 5. 
In the asymptotic limit (\ref{asymp_1}), 
the action density 
\eqref{NL Sigma term_2-Soliton_form 2} is dominated by
\begin{eqnarray}
&&\!\!\!\! \mbox{Tr}\left[\left(\partial_{\mu}\sigma\right)\sigma^{-1}\left(\partial^{\mu}\sigma\right)\sigma^{-1} \right]_{|X_2| \gg |X_1|}
\nonumber \\
&\!\!\!\!=\!\!\!\!&
\frac{8abd_{11}\cosh^2{X_2} + \mathcal{O}(\cosh{X_2})}
{\left[
	a~\!\mbox{cosh}(X_1 + X_2) + b~\!\mbox{cosh}(X_1 - X_2) + \mathcal{O}(1)
	\right]^2
}
\nonumber\\
&\!\!\!\!=\!\!\!\!&
\frac{8abd_{11} + \mathcal{O}(\mathrm{sech}X_2)}
{
	\left[
	(a+b)~\!\mbox{cosh}X_1 + (a-b)~\!\mbox{sinh}X_1 ~\!\mbox{tanh}X_2 + \mathcal{O}(\mathrm{sech}X_2)
	\right]^2
}.  \label{NL Sigma term_2-Soliton_X_i_3} ~~~~~~~ 
\end{eqnarray}
Since $\mathrm{sinh}X_1$ and $\mathrm{cosh}X_1$ are finite and
%\begin{eqnarray}
%\left\{
%\begin{array}{l}
$\mbox{sech}X_2 \rightarrow 0$ and $\mbox{tanh}X_2 \rightarrow \pm 1$ as $X_2 \rightarrow \pm \infty$, 
%\end{array}
%\end{eqnarray}
we have
\begin{eqnarray}
\mbox{Tr}\left[\left(\partial_{\mu}\sigma\right)\sigma^{-1}\left(\partial^{\mu}\sigma\right)\sigma^{-1} \right]
&\!\!\!\! \stackrel{X_2 \rightarrow \pm \infty}{\longrightarrow} \!\!\!\!&
~~\frac{8abd_{11}}{\left[(a+b)~\!\mbox{cosh}X_1 \pm  (a-b)~\!\mbox{sinh}X_1\right]^2} %~~~\mbox{as $X_2 \longrightarrow \pm \infty$.}
\nonumber\\
&\!\!\!\! = \!\!\!\!&
\left\{
\begin{array}{l}
\displaystyle{\frac{8abd_{11}}{(ae^{X_1} + be^{-X_1})^2}}
~~~\mbox{as $X_2 \longrightarrow +\infty$}
\smallskip \\
\displaystyle{\frac{8abd_{11}}{(be^{X_1} + ae^{-X_1})^2}}
~~~\mbox{as $X_2 \longrightarrow -\infty$}
\end{array}.
\right.
\end{eqnarray} 
Now we conclude that
\begin{eqnarray}
\label{Asymptotic_2-Soliton}
%&&\!\!\!\!\mbox{Tr}\left[\left(\partial_{\mu}\sigma\right)\sigma^{-1}\left(\partial^{\mu}\sigma\right)\sigma^{-1}\right]   \nonumber \\
-8\pi {\cal{L}}_\sigma 
 \longrightarrow 
\left\{
\begin{array}{l}
(1)~X_1 ~\mbox{is finite}, ~X_2  \rightarrow  +\infty :  
~
d_{11}~\!{\mathrm{sech}^2\left( X_1 + \delta_1 \right)}
\medskip \\
(2)~X_1 ~\mbox{is finite}, ~X_2  \rightarrow  -\infty :  
~
d_{11}~\!{\mathrm{sech}^2\left( X_1 - \delta_1 \right)}
\medskip \\
(3)~X_2 ~\mbox{is finite}, ~X_1  \rightarrow  +\infty :  
~
d_{22}~\!{\mathrm{sech}^2\left( X_2 + \delta_2 \right)}
\medskip \\
(4)~X_2 ~\mbox{is finite}, ~X_1  \rightarrow  -\infty :  
~
d_{22}~\!{\mathrm{sech}^2\left( X_2 - \delta_2 \right)}
\end{array},
\right.
\end{eqnarray}
where the position shift factors (or the phase shift factors) are  
\begin{eqnarray}
\delta_1\equiv\delta_2:=\displaystyle{\frac{1}{2}~\!\log\left(\frac{a}{b}\right)} 
= \frac{1}{2}\log \left[ \frac{a(1, 1)}{a(1, -1)} \right] ~~\mbox{(Cf: \eqref{Coefficients of Delta, Delta_ij})}.
\end{eqnarray}
The cases $(3)$ and $(4)$ are obtained by the same argument and we just skip the detail here. 
By the above analysis, we find that the NL$\sigma$M action density \eqref{NL Sigma term_2-Soliton_form 1} has two peaks which are localized on nonparallel two three-dimensional hyperplanes described by the linear equation $X_1 \pm \delta_1=0$ and $X_2 \pm \delta_2=0$. More general discussion for $n$-soliton case is mentioned in Appendix \ref{Asymptotic Form of WZW_4 Action Density (n-Soliton)}. 

As for other asymptotic regions which differ from the cases $(1) \sim (4)$, no solitonic effect contributes to the action density, that is, the action density is asymptotic to zero in these regions. This will be proved in section 5.

%\subsubsection{Wess-Zumino Action Density}

Let us proceed to calculate the Wess-Zumino action density for the two-soliton solution. 
By substituting \eqref{Determinant of n-soliton solution}, \eqref{Data of J_3} and \eqref{Derivative of exponential functions_1} into \eqref{Tr(A_m A_n A_p)_2} for $(m, n, p) = (\mu, \nu, \rho)$, we have
\begin{eqnarray}
\label{Density of the WZ Term_decomposition}
\!\!\!\!\!\!\!\!&&\mbox{Tr}\left[(\partial_{\mu}\sigma)\sigma^{-1}(\partial_{\nu}\sigma)\sigma^{-1}(\partial_{\rho}\sigma)\sigma^{-1}\right]
\smallskip 
=
\displaystyle{\frac{1}{2}}
%\left\{
%\begin{array}{l}
(B_{\mu\nu\rho}+B_{\nu\rho\mu}+B_{\rho\mu\nu}),\\
\!\!\!\!\!\!\!\!&&
B_{\mu\nu\rho}:=
\displaystyle{\frac{1}{|\sigma|^2\Delta^4}}
\left(
\left|
\!\!\begin{array}{cc}
\Delta_{11}\!\! & \Delta_{22}\!\! \\
\partial_{\mu} \Delta_{11}\!\! & \partial_{\mu} \Delta_{22}\!\!
\end{array}
\right|
%\cdot
\left|
\!\!\begin{array}{cc}
\partial_{\nu} \Delta_{12}\!\! & \partial_{\nu} \Delta_{21}\!\! \\
\partial_{\rho} \Delta_{12}\!\! & \partial_{\rho} \Delta_{21}\!\!
\end{array}
\right|
+
\left|
\!\!\begin{array}{cc}
\Delta_{12}\!\! & \Delta_{21}\!\! \\
\partial_{\mu} \Delta_{12}\!\! & \partial_{\mu} \Delta_{21}\!\!
\end{array}
\right|
%\cdot
\left|
\!\!\begin{array}{cc}
\partial_{\nu} \Delta_{11}\!\! & \partial_{\nu} \Delta_{22}\!\! \\
\partial_{\rho} \Delta_{11}\!\! & \partial_{\rho} \Delta_{22}\!\!
\end{array}
\right|\right). 
\nonumber
%\end{array}
%\right\}\!\!+
%{\mbox{CycPerm}}(\mu, \nu, \rho)
%\!\!. ~~~~~~~~
\end{eqnarray}
Here each ingredient of $B_{\mu\nu\rho}$ can be calculated in the same way as the previous section. 
%The complete results { can be referred to} Appendix \ref{WZ2}. The denominator of the WZ term is proportional to $[A\cosh(X_1+X_2)+B\cosh(X_1-X_2)+C\cos\Theta_{12}]^4$ where $A,B$ and $C$ are constants. The numerator of the WZ term is a polynomial of (hyperbolic) cosine and sine functions as in the NL$\sigma$M term. 
For example, the result of the first determinant factor in (\ref{Density of the WZ Term_decomposition})
is 
%calculated as 
\begin{eqnarray}
%\label{WZ term_part1}
&&\displaystyle{\frac{1}{|\sigma|\Delta^2}}
\left|
\!\!\begin{array}{cc}
\Delta_{11}\!\! & \Delta_{22}\!\! \\
\partial_{\mu} \Delta_{11}\!\! & \partial_{\mu} \Delta_{22}\!\!
\end{array}
\right|
\nonumber \\
&\!\!\!\!=\!\!\!\!&
\frac{
-\left\{
\begin{array}{l}
 ~~~2r_{\mu}^{(1)}
ab\mathcal{D}_{11}~\!\mathrm{cosh}(2X_2)
+2r_{\mu}^{(2)}
ab\mathcal{D}_{22}~\!\mathrm{cosh}(2X_1)
\smallskip \\
+\left[ (r_{\mu}^{(1)} + r_{\mu}^{(2)}) + (s_{\mu}^{(1)} - s_{\mu}^{(2)}) \right]
ac\mathcal{D}_{12}~\!\mathrm{cosh}(X_1 + X_2 -i\Theta_{12})
\smallskip \\
+\left[ (r_{\mu}^{(1)} + r_{\mu}^{(2)}) - (s_{\mu}^{(1)} - s_{\mu}^{(2)}) \right]
ac\mathcal{D}_{21}~\!\mathrm{cosh}(X_1 + X_2 +i\Theta_{12})
\smallskip \\
+\left[ (r_{\mu}^{(1)} - r_{\mu}^{(2)}) + (s_{\mu}^{(1)} - s_{\mu}^{(2)}) \right]
bc\mathcal{E}_{12}~\!\mathrm{cosh}(X_1 - X_2 -i\Theta_{12})
\smallskip \\
-\left[ (r_{\mu}^{(1)} - r_{\mu}^{(2)}) - (s_{\mu}^{(1)} - s_{\mu}^{(2)}) \right]
bc\widetilde{\mathcal{E}}_{12}~\!\mathrm{cosh}(X_1 - X_2 +i\Theta_{12}) 
\end{array}
\right\} -F
}
{
	2\left[
	\begin{array}{l}
	a~\!\mbox{cosh}(X_1 + X_2)
	+
	b~\!\mbox{cosh}(X_1 - X_2)
	+
	c~\!\mbox{cos}\Theta_{12}
	\end{array}
	\right]^2   
}
\end{eqnarray}
The definition of the coefficients and the result of the remaining determinant factors can be found in Appendix \ref{WZ2}. 
Furthermore, we can also show that the Wess-Zumino action density is real-valued on 
$\mathbb{U}_1$ and $\mathbb{E}$. (Cf: Appendix \ref{WZ2}).
%and that there is no singularity on $\mathbb{U}_1$ because $\Delta$ has no zero.  

%Furthermore, we can clarify the asymptotic behavior of the Wess-Zumino action density \eqref{Density of the WZ Term_decomposition} as in the previous section. Let us consider the asymptotic limit such that $|X_2|  \gg  |X_1|$ and $X_1$ is finite. 
%This is also given in Appendix \ref{WZ2}.
%By the same evaluation as in the previous section, 
By the same technique used in the previous section, we
consider the asymptotic limit such that $|X_2|  \gg  |X_1|$ for finite $X_1$, and find that 
%we have the asymptotic form of $B_{\mu\nu\rho}$ in (\ref{Density of the WZ Term_decomposition}):
\begin{eqnarray}
\label{Density of the WZ term_one-third part} 
%&&\!\!\!\!\!\!
%\displaystyle{\frac{1}{2|\sigma|^2\Delta^4}}
%\left\{\begin{array}{l}\left|\!\!\begin{array}{cc}
%\Delta_{11}\!\! & \Delta_{22}\!\! \\
%\partial_{\mu} \Delta_{11}\!\! & \partial_{\mu} \Delta_{22}\!\!
%\end{array}\right|\left|
%\!\!\begin{array}{cc}
%\partial_{\nu} \Delta_{12}\!\! & \partial_{\nu} \Delta_{21}\!\! \\
%\partial_{\rho} \Delta_{12}\!\! & \partial_{\rho} \Delta_{21}\!\!
%\end{array}\right|+\left|\!\!\begin{array}{cc}
%\Delta_{12}\!\! & \Delta_{21}\!\! \\
%\partial_{\mu} \Delta_{12}\!\! & \partial_{\mu} \Delta_{21}\!\!
%\end{array}\right|\left|\!\!\begin{array}{cc}
%\partial_{\nu} \Delta_{11}\!\! & \partial_{\nu} \Delta_{22}\!\! \\
%\partial_{\rho} \Delta_{11}\!\! & \partial_{\rho} \Delta_{22}\!\!
%\end{array}\right|  \smallskip \\\end{array}\!\!\right\}
B_{\mu\nu\rho}
\stackrel{|X_2| \gg |X_1|}{\simeq}
\frac{-4a^2b^2 C_{\mu\nu\rho}\mathcal{D}_{11}\mathcal{d}_{11}\mathrm{tanh}X_2 + \mathcal{O}(\mathrm{sech}X_2)}{\left[(a+b)~\!\mbox{cosh}X_1 + (a-b)~\!\mbox{sinh}X_1\mbox{tanh}X_2 + \mathcal{O}(\mathrm{sech}X_2)\right]^4},
\end{eqnarray}
where $C_{\mu\nu\rho}:=\left(r_{\mu}^{(1)}s_{\nu}^{(1)} + s_{\mu}^{(1)}r_{\nu}^{(1)}\right)r_{\rho}^{(2)}
-\left(r_{\mu}^{(1)}s_{\rho}^{(1)} + s_{\mu}^{(1)}r_{\rho}^{(1)}\right)r_{\nu}^{(2)}$. This is asymptotic to
\begin{eqnarray}
\label{WZ2_asymp}
B_{\mu\nu\rho}\stackrel{X_2 \rightarrow \pm \infty}{\longrightarrow}
\mp 4C_{\mu\nu\rho}\mathcal{D}_{11}\mathcal{d}_{11} \mathrm{sech}^4(X_1 \pm \delta_1),
\end{eqnarray}
where the phase shift factor is 
$\displaystyle{\delta_1:=(1/2)\mathrm{log}(a/b)}$. 
%{ We note that $B_{\mu\nu\rho}$ behaves like an odd function with respect to $X_1$}. This might suggests the integration of the Wess-Zumino action density would be zero.  

In fact, the coefficient $C_{\mu\nu\rho}$ 
in (\ref{WZ2_asymp}) satisfies the following relation:
\begin{eqnarray}
\label{C}
C_{\mu\nu\rho}+C_{\nu\rho\mu}+C_{\rho\mu\nu} =0.
\end{eqnarray} 
Therefore the cubic term 
(\ref{Density of the WZ Term_decomposition}) identically vanishes in the asymptotic region, and the Wess-Zumino action density is asymptotic to zero for the two-soliton case:
\begin{eqnarray}
\label{WZ0}
\mbox{Tr}\left[(\partial_{\mu}\sigma)\sigma^{-1}(\partial_{\nu}\sigma)\sigma^{-1}(\partial_{\rho}\sigma)\sigma^{-1}\right]
\longrightarrow 0.
\end{eqnarray}
More general discussion for $n$-soliton case can be found in Appendix \ref{Asymptotic Form of WZW_4 Action Density (n-Soliton)}. 

The Wess-Zumino action density for the two-soliton is a smooth function and non-singular and hence bounded. 
Moreover, it decays to zero exponentially as mentioned in section 3.2. Therefore, we conjecture that the Wess-Zumino action $S_{\scriptsize{\mbox{WZ}}}$ would be zero exactly.
%in $\mathbb{U}_1$. 
%As is proved in section 3.2, the Wess-Zumino action $S_{\scriptsize{\mbox{WZ}}}$ converges for the two-soliton solution. We conjecture that it would be zero.

\section{Asymptotic Analysis of $n$-Soliton Solutions}

Due to the problem of singularity of two-soliton solution on the Euclidean space $\mathbb{E}$, in this section, we consider a modified $n$-soliton solution and discuss the corresponding asymptotic behaviors in a systematic way. The modified $n$-soliton solution is
\begin{eqnarray}
\label{n-soliton solutions_enlarged}
\sigma=
\left|
\begin{array}{ccccc}
\psi_1 & \psi_2 & \cdots & \psi_n & 1 \\
\psi_1 \Lambda_1 & \psi_2 \Lambda_2 & \cdots & \psi_n \Lambda_{n} & 0 \\
\psi_1 \Lambda_1^2 & \psi_2 \Lambda_2^2 & \cdots & \psi_n \Lambda_{n}^2 & 0 \\
\vdots & \vdots & \ddots & \vdots & \vdots \\
\psi_1 \Lambda_1^n & \psi_2 \Lambda_2^n & \cdots & \psi_n \Lambda_{n}^n & \fbox{$0$}
\end{array}
\right|,~
	\begin{array}{l}
	\psi_j=
	\left(
	\begin{array}{cc}
	e^{L_j} & e^{-\overline{L}_j} \\
	-\epsilon e^{-L_j} & e^{\overline{L}_j}
	\end{array}
	\right),~L_j=\ell_{\mu}^{(j)}x^{\mu}
	\smallskip \\
	\Lambda_j
	=
	\left(
	\begin{array}{cc}
	\lambda_j^{(+)} & 0 \\
	0 & \lambda_j^{(-)}
	\end{array}
	\right)
	\end{array},
\end{eqnarray}
where the spectral parameters $(\lambda_j,\mu_j)$ are 
rewritten by $(\lambda_j^{(+)},\lambda_j^{(-)})$ for later convenience.
The slight difference between \eqref{n-soliton solutions_enlarged} and \eqref{n-soliton solution} is an additional constant factor $\epsilon$ taking values in $\{\pm 1\}$. The case of $\epsilon=+1$ coincides with \eqref{n-soliton solution}. 
%{ existence of the constant $\epsilon$ which is $+1$ or $-1$. The latter case $\epsilon=-1$ is our previous choice.}  
%coefficients $a_j$, $b_j$, $c_j$ and $d_j$.  
We will show that the non-singular $n$-solitons can be constructed completely for all $n \in \mathbb{N}$ by suitable choices of the constant $\epsilon$ 
with respect to the Ultrahyperbolic space $\mathbb{U}_{1}$ and the Euclidean space $\mathbb{E}$.

First of all, we define two types of the asymptotic region for the $n$-soliton solutions. Let us consider the asymptotic region of the four-dimensional space 
where $r^2=(x^1)+(x^2)+(x^3)+(x^4)$ is large enough for the spacial point $x=(x^1,x^2,x^3,x^4)$. The asymptotic region is divided into $2^n$ regions 
by the $n$ hyperplanes $X_j=0~(j=1,2\cdots,n)$
%, that is, according to 
depending on $X_j>0$ or $X_j<0$. In order to label these regions, it is convenient to introduce a new notation $\varepsilon_j \in \left\{ \pm 1, 0\right\}$.
Then the $2^n$ asymptotic regions can be denoted by 
${\mathscr{R}}(\varepsilon_1,\cdots,\varepsilon_n)$ 
%(or ${\mathscr{R}_K}(\varepsilon_1,\cdots,\varepsilon_n)$) 
%where 
in which $\varepsilon_j=+ 1$, %$\varepsilon_j=0$ 
and $\varepsilon_j=- 1$ correspond to the following cases $(+)$ and $(-)$ respectively:\footnote{Here we suppose that $\vert X_j\vert$ are large enough in the asymptotic region and hence $X_j>0$ implies $X_j\gg +1$.}
\begin{eqnarray}
\label{two_asymp}
(+)&&X_j\gg +1 \Leftrightarrow 
{\mbox{Re}} L_j\gg +1 \Leftrightarrow 
\left|e^{L_j}\right| \gg 1 \Leftrightarrow 
\left|e^{-L_j}\right| \ll 1,\nonumber \\
(-)&&X_j\ll -1 \Leftrightarrow 
{\mbox{Re}} L_j\ll -1 \Leftrightarrow 
\left|e^{L_j}\right| \ll 1 \Leftrightarrow 
\left|e^{-L_j}\right| \gg 1. 
\end{eqnarray}
%{ Then we can define one type of asymptotic region for the $n$-soliton solution as follows:}
Then we can unify the asymptotic regions as
\begin{eqnarray}
{\mathscr{R}}:=\bigcup_{\varepsilon_j=\pm1}
{\mathscr{R}}(\varepsilon_1,\cdots,\varepsilon_n).
\end{eqnarray}
We will see that the Wess-Zumino action density vanishes in ${\mathscr{R}}$. 
%This corresponds to asymptotoc region in a general direction. 

On the other hand, there is the other type of the asymptotic region along the hyperplane $X_j$ %in the tangent directions of the $n$-solitons 
which corresponds to the case of $\varepsilon_j=0$. 
To make the asymptotic region be four-dimensional, let us define the asymptotic region along $X_K$ as a tubular neighborhood of 
${\mathscr{R}}(\varepsilon_1,\cdots,\varepsilon_K=0,\cdots,\varepsilon_n)$
which is denoted by ${\mathscr{R}_K}(\varepsilon_1,\cdots,\varepsilon_n)$. 
In this region, the value of $X_K$ is considered to be finite. 
We will see that the NL$\sigma$M and the Wess-Zumino 
action densities coincide with a one-soliton configuration 
in ${\mathscr{R}}_K$.

The two type of asymptotic regions can be expressed in terms of the following sets:
\begin{eqnarray*}
{\mathscr{R}}\!\!\!&:=&\!\!\!\left\{x%k_1e^1+k_2e^2+k_3e^3+k_4e^4
\in M_4
\left|
\begin{array}{l}
%e^1,e^2,e^3,e^4 {\mbox{ are linearly independent vectors}}. 
%\\
x_1^2+x_2^2+x_3^2+x_4^2 {\mbox{ is large enough.}} 
\\
X_j{\mbox{ are all positive or negative. }} (j=1,\cdots,n)
\end{array}
\right.
\right\}
\\ 
{\mathscr{R}}_K\!\!\!&:=&\!\!\!\left\{x=k_1e^1+k_2e^2+k_3e^3+a\in M_4
\left|
\begin{array}{l}
e^1,e^2,e^3 {\mbox{ are linearly independent vectors}}\\
{\mbox{ tangent to the hyperplane: }} X_K=0 \\
a {\mbox{ is a finite vector.}} \\
k_1^2+k_2^2+k_3^2 {\mbox{ is large enough.}} 
\\
X_j (j\neq K){\mbox{ are all positive or negative.}}
\end{array}
\right.
\right\}
\end{eqnarray*}
A simple example of the asymptotic regions are shown in the figure below.
($\mathscr{R}_3$ is shown by pink.)
%\begin{wrapfigure}[1]{r}{14cm}
\begin{figure}[h]
%\vspace{-15cm}
\vspace{-10mm}
\hspace{2.8cm}
\includegraphics[width=11cm]{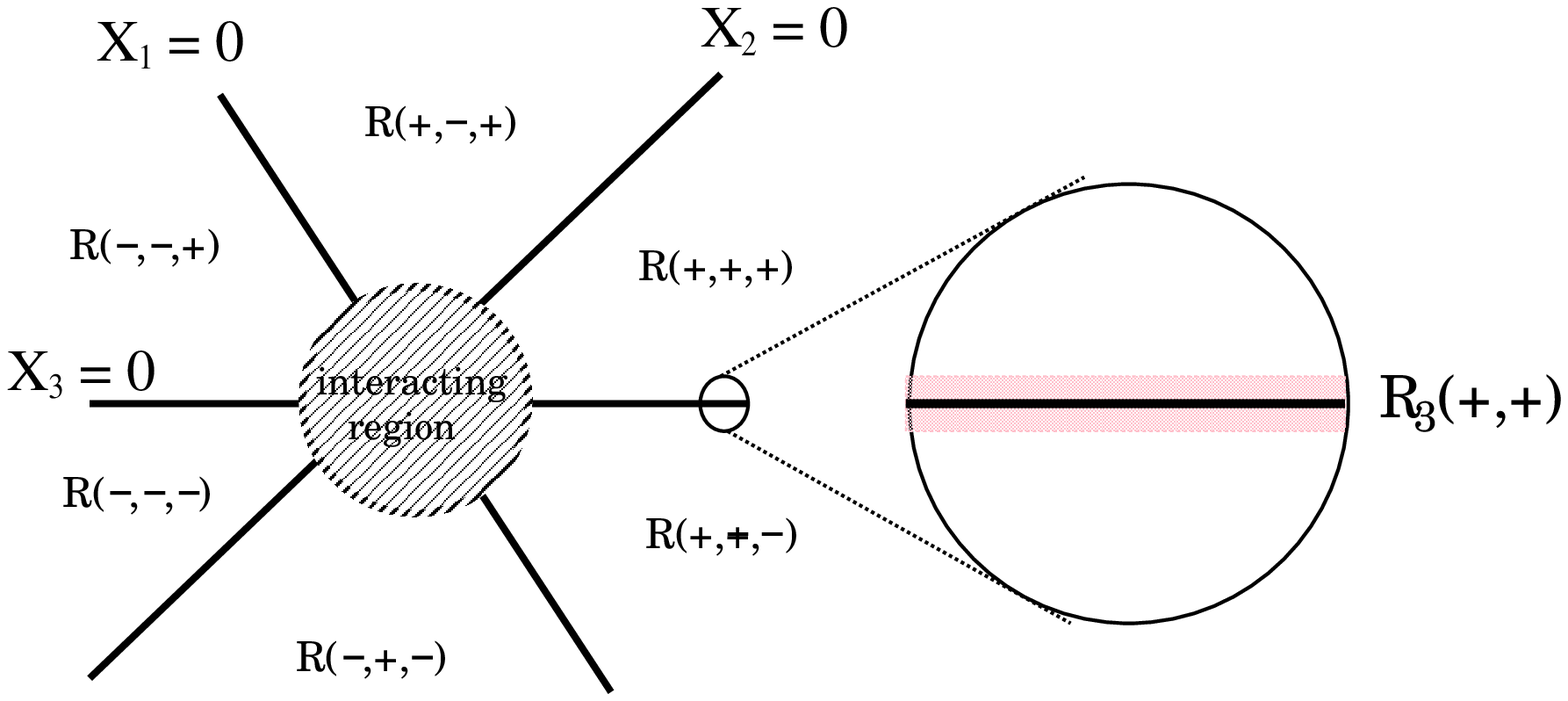}
% \caption{Asymptotic Regions ($R_3$ is denoted by the pinky region.)}
\label{asymp}
\end{figure}
%\end{wrapfigure}

\vspace{17cm}
%In the asymptotic region ${\mathscr{R}}$, 
For the asymptotic regions of type ${\mathscr{R}}$, the behavior of \eqref{n-soliton solutions_enlarged} is dominated by all $X_j$ for $|X_j|$ is large enough.
%the value of $\vert X_j \vert$ are all large enough. 
By \eqref{two_asymp}, we find that $\sigma$ is asymptotic to a constant matrix for each asymptotic region of the type $\mathscr{R}$: 
%the asymptotic form of the $n$-soliton solution is asymptotic to constant:
\begin{eqnarray*}
\sigma 
\stackrel{{\mathscr{R}}}{\simeq}
\left|
\begin{array}{cccc}
C_1^{(\pm)}
&
\cdots
&
C_n^{(\pm)}
& 
1
\\
C_1^{(\pm)} \Lambda_1
&
\cdots
& 
C_n^{(\pm)}\Lambda_n
&0
\\
\vdots
&
& 
\vdots
&
\vdots
\\
C_1^{(\pm)}\Lambda_1^n
& 
\cdots
&
C_n^{(\pm)}\Lambda_n^n
& \fbox{0}
\end{array}
\right|,
~~~
\mbox{where }
~
C_i^{(+)}:=\left(
\begin{array}{cc}
1&0 \\
0&1 
\end{array}
\right),
~~
C_i^{(-)}:=
\left(
\begin{array}{cc}
0&1 \\
-\epsilon & 0 
\end{array}
\right).
\end{eqnarray*}
The suffix $(\pm)$ in $C_j$ corresponds to the signature of $\varepsilon_j$. Therefore, the action densities 
${\cal{L}}_{\sigma}$ and ${\cal{L}}_{\scriptsize{\mbox{WZ}}}$ 
identically vanish in the type ${\mathscr{R}}$ asymptotic regions.
%Hence action densities of this identically vanish. 

On the other hand, 
since $X_K$ is kept to be finite for the type ${\mathscr{R}}_K$ asymptotic regions, we have \cite{HaHu2}
%in the asymptotic region ${\mathscr{R}}_K$, 
%$X_K$ is kept to be finite 
%the value of $\vert X_j \vert$ are large enough
%and therefore we can find the asymptotic behavior of 
%the $n$-soliton solution \cite{HaHu2} as follows:
\begin{eqnarray*}
\sigma \stackrel{{\mathscr{R}}_K}{\simeq}
\left|
\begin{array}{cccccc}
C_1^{(\pm)}
&
\cdots
&
\psi_K
& 
\cdots
&
C_n^{(\pm)}
& 
1
\\
C_1^{(\pm)} \Lambda_1
&
\cdots
&
\psi_K\Lambda_K
&
\cdots
& 
C_n^{(\pm)}\Lambda_n
&0
\\
\vdots
&
&
\vdots
&
& 
\vdots
&
\vdots
\\
C_1^{(\pm)}\Lambda_1^n
& 
\cdots
&
\psi_K\Lambda_K^n
& 
\cdots
&
C_n^{(\pm)}\Lambda_n^n
& \fbox{0}
\end{array}
\right|. 
\end{eqnarray*}
This actually leads to the following one-soliton type 
solution \cite{HaHu2}:\footnote{We note that the two operations of taking limit and of taking derivation do not commute in general, however, in our case, do commute. This is proved in Appendix \ref{commute}} 
\begin{eqnarray}
\label{one-soliton_asymp}
\sigma\stackrel{{\mathscr{R}}_K}{\simeq}
\left|  
\begin{array}{cc}
\check{\psi}_K & 1 \\
\check{\psi}_K\Lambda_{K} & \fbox{0}
\end{array}
\right|D^{(k)},~~ \mbox{where}~~
\left\{
\begin{array}{l}
\check{\psi}_K
:=
\left(
\begin{array}{cc}
a_{K}e^{L_{K}} & b_{K}e^{-\overline{L}_{K}} \\
-c_{K}e^{-L_{K}} & d_{K}e^{\overline{L}_{K}}
\end{array}
\right)
\smallskip \\
%\Lambda_{K}:=\left(
%\begin{array}{cc}
%\lambda_K  &  0 \\
%0 & \mu_K
%\end{array}\right)
D_K : \mbox{a constant matrix},
\end{array}
\right.
\end{eqnarray}
and the coefficients $a_K, b_K, c_K, d_K$ 
can be expressed in terms of the spectral parameters as:
\begin{eqnarray}
%\left\{
%\begin{array}{l}
&&\displaystyle{a_{K}=
\!\!\!\prod_{j=1, j \neq K}^{n}
\left( \lambda_K^{(+)} - \lambda_j^{(\pm)} \right)},~~
\displaystyle{b_{K}=
\!\!\!\prod_{j=1, j \neq K}^{n}
\left( \lambda_K^{(-)} - \lambda_j^{(\pm)} \right),
}
\nonumber\\
\label{abcd}
&&\displaystyle{c_{K}=
\!\!\!\prod_{j=1, j \neq K}^{n}
\left( \lambda_K^{(+)} - \lambda_j^{(\mp)} \right)}\epsilon,~~
\displaystyle{d_{K}=
\!\!\!\prod_{j=1, j \neq K}^{n}
\left( \lambda_K^{(-)} - \lambda_j^{(\mp)} \right).
}
%\end{array},
%~\left( \lambda_i^{(+)}, \lambda_i^{(-)} \right):=\left( \lambda_i, \mu_i \right).~~~
%\right.
\end{eqnarray}
%This actually satisfies Eq.(\ref{chasing}) and 
%gives a one-soliton solution of the Yang equation. 
%The difference with the one-soliton solution in Section 4 is seen in the existence of $a_K,b_K,c_K,d_K$. 
In fact, we will see later that the coefficients $a_K, b_K, c_K, d_K$
%This difference actually 
determine the position shift (known as the phase shift) of the one-soliton solution \eqref{one-soliton_asymp} in each asymptotic region of type $\mathscr{R}_K$. 
%in the asymptotic region, which is known as the phase shift. 
Furthermore, these coefficients also determine whether the singularities of $n$-solitons exist in the action density of the WZW$_4$ model. 
%Furthermore, in $\mathbb{E}$, the NL$\sigma$M action density can be singular due to these coefficients. In order to discuss the singularities in detail, let us proceed the calculation of the NL$\sigma$M action density in terms of  $a_K,b_K,c_K,d_K$. 
%The five data are easily obtained as: 
%\begin{eqnarray}\label{}
%\begin{array}{l}
%\Delta = (a_Kd_K)e^{X_K} + (b_Kc_K)e^{-X_K}, ~~~~~~~~
%\partial_{\mu}\Delta = \left( \ell_{\mu} + \overline{\ell}_{\mu} \right)\left( (a_Kd_K)e^{X} - (b_Kc_K)e^{-X}\right)\smallskip \\
%\Delta_{11} = \lambda(ad)e^{X} + \mu(b_Kc_K)e^{-X}, ~~\partial_{\mu}\Delta_{11} = \left( \ell_{\mu} + \overline{\ell}_{\mu} \right)\left(\lambda(ad)e^{X} - \mu(b_Kc_K)e^{-X} \right) \smallskip \\
%\Delta_{22} = \mu(a_Kd_K)e^{X} + \lambda(b_Kc_K)e^{-X}, ~~\partial_{\mu}\Delta_{22} = \left( \ell_{\mu} + \overline{\ell}_{\mu} \right)\left( \mu(ad)e^{X} - \lambda(b_Kc_K)e^{-X} \right)\smallskip \\
%\Delta_{12} = -(\lambda - \mu)(ab)e^{i\Theta},~~~~~~~~~~~~\partial_{\mu}\Delta_{12}= -\left( \ell_{\mu} - \overline{\ell}_{\mu} \right)(\lambda - \mu)(a_Kb_K)e^{i\Theta}\smallskip \\
%\Delta_{21} = -( \lambda - \mu )(cd)e^{-i\Theta},~~~~~~~~~\partial_{\mu}\Delta_{21} = \left( \ell_{\mu} - \overline{\ell}_{\mu} \right)(c_Kd_K)e^{-i\Theta}.
%\end{array}~~
%\end{eqnarray}
%By expressing the ratio $a_Kd_K/b_Kc_K$ in the polar form: $a_Kd_K/b_Kc_K=r_K e^{i\varphi_K}$ where $r_K:=\vert a_Kd_K/b_Kc_K \vert$, 
%the quadratic term is computed in the same way:
First of all, we can calculate the asymptotic form of the following $\mu$-th component of the quadratic term by using \eqref{one-soliton_asymp}. 
(The summation is not taken over $\mu$.) The result is
\begin{eqnarray}
\label{quadratic}
\mbox{Tr}\left[ (\partial_{\mu}\sigma)\sigma^{-1}\right]^{2}
&:=&\mbox{Tr}\left[(\partial_{\mu}\sigma)\sigma^{-1}(\partial_{\mu}\sigma)\sigma^{-1}\right]
%~~~(\mbox{{not sum}})
\nonumber \\
&\!\!\!\! \stackrel{{\mathscr{R}}_K}{\simeq}\!\!\!\!&
\frac{8\left| \ell_{\mu}^{(K)} \right|^2(\lambda_K^{(+)} - \lambda_K^{(-)})^2}{\lambda_K\mu_K}\cdot
\frac{a_K b_K c_K d_K}{\left(a_K d_Ke^{X_K} + b_K c_Ke^{-X_K} \right)^2}
\nonumber \\
&\!\!\!\!=\!\!\!\!&
\frac{8\left| \ell_{\mu}^{(K)} \right|^2(\lambda_K^{(+)} - \lambda_K^{(-)})^2}{\lambda_K^{(+)}\lambda_K^{(-)}}\cdot
\frac{1}{\left(\displaystyle{\frac{a_K d_K}{b_K c_K}}e^{2X_K} +  \displaystyle{\frac{b_K c_K}{a_K d_K}}e^{-2X_K} + 2 \right)}
\nonumber \\
&\!\!\!\!=\!\!\!\!&
\frac{8\left| \ell_{\mu}^{(K)} \right|^2(\lambda_K^{(+)} - \lambda_K^{(-)})^2}{\lambda_K^{(+)}\lambda_K^{(-)}}\cdot
\frac{1}{e^{i\varphi_K}\left(\displaystyle e^{X_K+\delta_K} +e^{-i\varphi_K} 
\displaystyle e^{-(X_K+\delta_K)}\right)^2},
\nonumber
\end{eqnarray}
where $\delta_K:=(1/2)\log r_K$,
$r_K:=\vert a_K d_K/ b_K c_K \vert$ and the ratio $a_K d_K / b_K c_K:=r_K e^{i\varphi_K}$. 
In particular, $a_K d_K / b_K c_K \in \mathbb{R}^{+}$ if $\varphi_K=0$ and 
$a_K d_K / b_K c_K \in \mathbb{R}^{-}$ if $\varphi_K=\pi$. This fact implies 
%$a_Kd_K/b_Kc_K$ belongs to positive real number $\mathbb{R}^{+}$ ($\varphi_K=0$) or to negative real number $\mathbb{R}^{-}$ ($\varphi_K=\pi$), the quadratic term becomes: 
\begin{eqnarray}
\mbox{Tr}\left[ (\partial_{\mu}\sigma)\sigma^{-1} \right]^{2}
\stackrel{{\mathscr{R}}_K}{\simeq}
\left\{
\begin{array}{l}
~\displaystyle{\frac{2\left|\ell_{\mu}^{(K)}\right|^2\!(\lambda_K^{(+)} - \lambda_K^{(-)} )^2}{\lambda_K^{(+)}\lambda_K^{(-)}}\mbox{ sech}^2(X_K +\delta_K )}~~\mbox{if}~~
a_K d_K / b_K c_k \in \mathbb{R^{+}}
%{\epsilon=-1}
\smallskip \\
\!\!\displaystyle{\frac{-2\left|\ell_{\mu}^{(K)}\right|^2\!(\lambda_K^{(+)} - \lambda_K^{(-)})^2}{\lambda_K^{(+)}\lambda_K^{(-)}}\mbox{ csch}^2(X_K +\delta_K )}~~
\mbox{if}~~
%{\epsilon=+1}
a_K d_K / b_K c_k \in \mathbb{R^{-}}
\end{array}
\right.
\label{The condition of 1-Solitons}
\end{eqnarray}
where $\mbox{csch} x:=1/\sinh x$. 
%Therefore, in the case of 
Apparently, for $a_K d_K / b_K c_k <0 $, the singularities exist on the entire three-dimensional hyperplane $X_K+\delta_K=0$.
%There is a similar discussion in the KP solitons 

%{ On the other hand, either the position of peak (in $a_K d_K / b_K c_K > 0$ case) or the position of singularity (in $a_K d_K / b_K c_K<0$ case) are shfted by $\delta_K$ in the asymptotic regions of type ${\mathscr{R}}_K$. }
%We can see that the position{s of the peaks are in any case}  shifted by $\delta_K$ in the asymptotic region ${\mathscr{R}}_K$. 
%Now let us check whether the NL$\sigma$M action density has singularity and evaluate the phase shift factor in the three kinds of the real spaces.
Now let us find out the condition such that the NL$\sigma$M action density is non-singular. For the Ultrahyperbolic space $\mathbb{U}_1$, 
the reality condition is $\lambda_j^{(-)}=\overline{\lambda}_j^{(+)}$. 
%and $\displaystyle{\lambda_j^{(-)}=1/\overline{\lambda}_j^{(+)}}$ respectively. 
By \eqref{abcd}, we have
%and the ratio is found to be
\begin{eqnarray}
\label{abcdU}
\frac{a_K d_K}{b_K c_K}=
%\stackrel{\mathbb{U}_1}{=}
%\left\{ 
%\begin{array}{l}
%{(-\epsilon)}
\frac{1}{\epsilon}
\prod_{j=1,j\neq K}^{n}
\left| \frac{\lambda_K - \lambda_j}{\lambda_K - \overline{\lambda}_j} 
\right|^{2\varepsilon_j}
~~~ \mbox{on}~~\mathbb{U}_1
%\smallskip \\
%{(-\epsilon)}
%{\color{green} \displaystyle{\frac{1}{\epsilon}}}
%\displaystyle{\prod_{j=1, j\neq K}^{n}
%\left| \frac{\lambda_K - \lambda_j}{\lambda_K \overline{\lambda}_j - 1} \right|%^{2\varepsilon_j}
%~~~ \mbox{on}~~\mathbb{U}_2}
%\end{array}.
%\right.
\end{eqnarray}
%where the suffix $(\pm)_j2$ corresponds to $+2$ if $\varepsilon_j=+1$ and $-2$  if $\varepsilon_j=-1$. 
%They are positive and hence there is no singularity. 
Comparing it with \eqref{The condition of 1-Solitons}, 
we can conclude that in the case of $\epsilon=+1$, the NL$\sigma$M action density of the $n$-soliton is definitely asymptotic to a non-singular one-soliton for the Ultrahyperbolic signature. 
This fact implies that for all $n \in \mathbb{N}$, the $n$-soliton solution \eqref{n-soliton solution} gives a class of non-singular NL$\sigma$M action densities for the Ultrahyperbolic signature.
%since each $\psi_j ~(a_j=b_j=c_j=d_j=1)$  fits the requirement of $a_j d_j / b_j c_j >0$.

Similarly, the reality condition of the Euclidean space $\mathbb{E}$ : $\displaystyle\lambda_j^{(-)}=-1/\overline{\lambda}_j^{(+)}$ implies
\begin{eqnarray}
\label{abcdE}
\frac{a_K d_K}{b_K c_K}=
%\stackrel{\mathbb{E}}{=}
%{\epsilon(-1)^{n}}
\displaystyle{\frac{(-1)^{n-1}}{\epsilon}}
\displaystyle{\prod_{j=1,j\neq K}^{n}
	\left| \frac{\lambda_K - \lambda_j}{\lambda_K \overline{\lambda}_j + 1} \right|^{2\varepsilon_j}} 
 ~~~\mbox{on} ~~\mathbb{E}.  
%\\&\!\!\!\!>\!\!\!\!& 
%>0 ~~\mbox{if}~~
%\left\{\begin{array}{l}
%\mbox{(1) $n$ is odd and $\epsilon={\color{green} +1}$} ~
%\smallskip \\
%\mbox{(2) $n$ is even and $\epsilon=\color{green} -1$} ~
%\end{array}.
%\right.
\end{eqnarray}
The ratio (\ref{abcdE}) is positive in the following two cases:
(1) $n$ is odd and $\epsilon=+1$ or (2) $n$ is even and $\epsilon=-1$. 
Then the NL$\sigma$M action densities are non-singular. 
On the other hand, the ratio (\ref{abcdE}) is negative 
in the following two cases: (3) $n$ is even and $\epsilon=+1${\footnote{The case (3) for $n=2$ corresponds to the singular two-soliton solution on $\mathbb{E}$ (Cf: Table \ref{Table_1}). } or (4) $n$ is odd and $\epsilon=-1$.\footnote{The case (4) for $n=1$ corresponds to the singular solution in the footnote 5.} 
Then the NL$\sigma$M action densities are singular. 
%In fact, this missing part of non-singular solutions can be complemented by considering the condition (2) directly.
%If we replace the coefficients $(a_j, b_j, c_j, d_j)=(1, 1, 1, 1)$ of $\psi_j$ in \eqref{n-soliton solutions_enlarged} by $(a_j, b_j, c_j, d_j)=(1, 1, -1, 1)$,
%We can also get a class of non-singular NL$\sigma$M action densities for even $n$ immediately. 
%Actually, there is no intrinsically difference between solutions generated by the conditions (1) and (2) in the asymptotic regions of type $\mathscr{R}_K$ (Cf: \eqref{The condition of 1-Solitons}). 
It is quite interesting that in the Euclidean signature, singular and non-singular solutions are generated alternately by the Darboux transformations with respect to initial solutions $\psi_j$ for $\epsilon=\pm 1$. 
%This would give a new insight into the study of integrable systems.

%Theoretically, 
In summary, for all $n \in \mathbb{N}$, 
non-singular NL$\sigma$M action densities of the $n$-soliton can be constructed 
by taking $\epsilon= +1$ for all $n \in \mathbb{N}$ (Cf: (\ref{abcdU})) on the Ultrahyperbolic space $\mathbb{U}_{1}$, and 
by taking $\epsilon= +1$ for all odd $n$ and $\epsilon=-1$ for all even $n$ (Cf: (\ref{abcdE}) and the cases (1) and (2)) on the Euclidean space $\mathbb{E}$. 
They would share the same asymptotic form in $\mathscr{R}_K$ on each real space:
%Without loss of generality, we just show the simplest example depending on the choices of $\psi_j$ (Cf: \eqref{n-soliton solutions_enlarged}) in which the constant are
%More precisely,
\begin{eqnarray}
\label{action density_asym}
{\cal{L}}_\sigma =
-\frac{1}{16\pi}\mathrm{Tr}
\left[
(\partial_{\mu}\sigma)\sigma^{-1}  
(\partial^{\mu}\sigma)\sigma^{-1} \right]
\stackrel{{\mathscr{R}}_K}{\simeq}
-\frac{1}{8\pi} d_{KK}~\!\mbox{sech}^2\left( X_K + \delta_K \right),
\end{eqnarray}
where $d_{KK}$ is defined in Table \ref{Table_1} (Cf : $\mathbb{U}_1$ and $\mathbb{E}$) and the phase shift factor is
\begin{eqnarray}
\label{phase shifts}
\delta_K= \displaystyle{ \frac{1}{2}\log \left(\frac{a_K d_K}{b_K c_K} \right)}
=\left\{
\begin{array}{l}
\displaystyle{\sum_{j=1, j \neq K}^{n}\varepsilon_j\log\left| \frac{\lambda_K - \lambda_j}{\lambda_K - \overline{\lambda}_j} \right|}  ~~~~ \mbox{on}~~\mathbb{U}_1 
\smallskip \\
%\displaystyle{\sum_{j=1, j \neq K}^{n}\varepsilon_j\log\left| \frac{\lambda_K - \lambda_j}{\lambda_K \overline{\lambda}_j - 1} \right|} ~~~ \mbox{on} ~~\mathbb{U}_2
%\smallskip \\
\displaystyle{\sum_{j=1, j \neq K}^{n}\varepsilon_j\log\left| \frac{\lambda_K - \lambda_j}{\lambda_K \overline{\lambda}_j + 1} \right|}
~~~ \mbox{on} ~~\mathbb{E}
\end{array}.
\right.
\end{eqnarray}
Since the result of \eqref{action density_asym} is valid for arbitrary $K$ in $\left\{ 1, 2, \dots, n\right\}$,
we can regard the behavior of non-singular $n$-soliton as a ``non-linear superposition'' of $n$ non-singular and mutually nonparallel one-solitons on each real space in which each one-soliton in the asymptotic region $\mathscr{R}_K$ maintains its form invariant but is shifted by $\delta_K$, called the phase shift factor which results from a non-linear effect.

%we find that for all $n \in \mathbb{Z}^{+}$, the $n$-soliton behaves as  { ``non-linear superposition of''} non-singular $n$ one-solitons in $\mathscr{R}_K$  up to a phase shift factor $\delta_K$. 

%More precisely, the asymptotic form in $\mathscr{R}_K$ of the NL$\sigma$M action density of the $n$-soliton solution is 

%Hence the $n$-soliton behaves as { ``non-linear superposition of''} singular $n$ one-solitons if $n$ is even and as that of non-singular $n$ one-soliton if $n$ is even, in $\mathscr{R}_K$ due to the fact of \eqref{The condition of 1-Solitons}.
%{On the other hand, starting with another initial solutions $\psi_j$ as not $a_j=b_j=c_j=d_j=1$ but $a_j=b_j=-c_j=d_j=1$, then we can obtain non-singular solutions if $n$ is even. This is easily seen in (\ref{c_K}) by the replacement of $c_K$ with $-c_K$. Therefore, we can get non-singular $n$-solitons solution \eqref{n-soliton solution} for all $n$ on $\mathbb{E}$ as well. It is interesting that singular and smooth solutions are generated alternately by the Darboux transformations.  This would be new insight in integrable systems.
%The phase shift factors are all the same as in $\mathbb{U}$. }

In conclusion, in the asymptotic region, the $n$-soliton solution possesses $n$ isolated and localized lumps of the NL$\sigma$M action density, and we can interpret it as $n$ intersecting soliton walls. The phase shift factors are also obtained explicitly. The scattering process of the $n$-soliton solution is quite similar to that of the KP solitons \cite{OhWa, Kodama, Kodama2}. 
On the other hand, the Wess-Zumino action density identically vanishes in the asymptotic region because in the asymptotic region $\mathscr{R}$, the action density identically vanishes, and in the asymptotic region $\mathscr{R}_K$, the $n$-soliton solution is reduced to the one-soliton (\ref{one-soliton_asymp}) whose Wess-Zumino action density is identically zero as proved in section 4.1.

%Finally we comment on the singularities of the $n$-soliton solution $\sigma$. The singularities correspond to zero-points of the datum $\Delta$. 

\section{Reduction to $(1+2)$-Dimensions}

So far, we discuss the action density of the WZW$_4$ model for $n$-soliton solutions and find that it is localized on nonparallel $n$ codimension-one hyperplanes in four dimensions. 
However, to understand better the physical meaning of our soliton solutions, it would be a good idea to calculate the energy density of the soliton solutions 
and compare with the action density. 
%definitions and properties of solitons are usually discussed not by action densities but by energy densities. 
For this purpose, we assume the translation invariance in the $x^2$ direction. 

The WZW$_4$ model Lagrangian is reduced to the following one (Cf: \eqref{S_sigma} and \eqref{S_WZ}):
\begin{eqnarray}
\label{monopole}
\mathcal{L}_{\scriptsize{\mbox{tot}}}=                   
{-\frac{1}{16\pi}\left(
\mbox{Tr}(\theta_t)^2 - \mbox{Tr}(\theta_x)^2 - \mbox{Tr}(\theta_y)^2
+{ 2} \mbox{Tr}\left( 
\theta_t \theta_x \theta_y
\right)y\right),}
\end{eqnarray} 
where we reset $(t,x,y):=(x_1, x_3, x_4)$ and 
$\theta_{\mu}:= (\partial_{\mu}\sigma)\sigma^{-1}
%=\left[~\!(\theta_{\mu})_{ij}~\!\right]_{2 \times 2}, 
~(\mu=t, x, y)$. %take values in the tangent space of $G=\mbox{SU}(2)$.  
The equation of motion is the Ward chiral model \cite{Ward_JMP} 
or the space-time monopole equation \cite{DTU,GOS} 
in $(1+2)$ dimensions in the Yang form. 
The $n$-soliton solution of (\ref{monopole}) is obtained by imposing the condition $\alpha_j = \lambda_j\beta_j$ (Cf: \eqref{L_j_U1}) on the powers $L_j$ of the $n$-soliton solution \eqref{n-soliton solution}. 
Then, the powers of $n$-soliton solution is actually reduced to
$
{ L_j}
=(\beta_j/\sqrt{2})
\left[~\!
(\lambda_j^2 +1)t + (\lambda_j^2-1)x + 2\lambda_jy
~\! \right]$ 
in the (1+2)-dimensional space-time. 
%In this section, we make a reduction to $(1+2)$-dimensional real spaces where time evolution and energy density are well-defined. 
%We compute the energy densities of the one-soliton and two-soliton solutions to confirm that they are localized on the same hyperplanes as the action densities are. This would imply that analysis of the action density have the same significance as the energy density. The locus where the action density is localized suggests existence of a physical object. 
%Let us consider the real space $\mathbb{U}_1$ and 
%Without loss of generality, we 
%In this case, we use the reduced $n$-soliton solution to distinguish it from \eqref{n-soliton solution}. For convenience, and introduce the notations $X_j := L_j + \overline{L}_j$, $i\Theta_j := L_j - \overline{L}_j$ and $\Theta_{jk} := \Theta_j - \Theta_k$ like we used in the previous sections. 
%By \eqref{Density of the NL sigma model term} and \eqref{WZ term_real spaces}, 

Let us consider three angular coordinates $\phi_i(x)~ (i=1,2,3)$ which parametrize $SU(2)\thickapprox S^3$ where $\sigma(x)$ belongs to. Then the Hamiltonian density can be obtained by the Legendre transformation of the Lagrangian:
\begin{eqnarray*}
\mathcal{H}_{\scriptsize{\mbox{tot}}}&=&
\sum_{i=1}^{3}
\frac{\partial {\mathcal{L}_{\scriptsize{\mbox{tot}}}}}{\partial (\partial_t\phi_i)}\partial_t\phi_i
-\mathcal{L}_{\scriptsize{\mbox{tot}}}
=-\frac{1}{16\pi}\left(
\mbox{Tr}(\theta_t)^2 + \mbox{Tr}(\theta_x)^2 + \mbox{Tr}(\theta_y)^2\right),
\end{eqnarray*}
This Hamiltonian physically makes sense because it is positive definite due to the fact that $\theta_\mu$ is an anti-hermitian matrix. This is a conserved energy density by definition. Note that the contribution of the Wess-Zumino term to $\mathcal{H}_{\scriptsize{\mbox{tot}}}$ vanishes identically: ${\cal{H}_{\scriptsize{\mbox{WZ}}}}=0$, or equivalently, ${\mathcal{H}}_{\scriptsize{\mbox{tot}}} = {\mathcal{H}}_{\sigma}$. %(the NL$\sigma$M Hamiltonian density). 

Let us calculate the energy density of the reduced soliton solution from the Hamiltonian density ${\mathcal{H}}_{\scriptsize{\mbox{tot}}}$. For the one-soliton solution, the Hamiltonian $\mathcal{H}_{\scriptsize{\mbox{tot}}}$ is:
\begin{eqnarray}
\mathcal{H}_{\scriptsize{\mbox{tot}}}
&\!\!\!\!=\!\!\!\!&
{-\frac{1}{8\pi}}
d_{11}\mbox{sech}^2{ X_1},~~~
d_{11} := \frac{(|\lambda_1|^2 + 1)^2(\lambda_1 - \overline{\lambda}_1)^2}{{|\lambda_1|^2}}|\beta_1|^2.
\end{eqnarray}
This is in the same form as the reduced NL$\sigma$M action density $\mathcal{L}_{\sigma}$ up to an overall coefficients (Cf: \eqref{L_sigma}). Therefore, the peaks of $\mathcal{H}_{\scriptsize{\mbox{tot}}}$ and $\mathcal{L}_{\sigma}$ are localized on the same two-dimensional hyperplane $X=0$ in the $(1+2)$-dimensional space-time. In this sense, $\mathcal{L}_{\sigma}$ also can be interpreted as an analogue of the energy density in physical reality.

For the two-soliton solution, the Hamiltonian density for the two-soliton solution is calculated by using the result of Appendix \eqref{Exact calculation of the NL Sigma Model Term (2-Soliton)}, we have
\begin{eqnarray}
\label{Hamiltonian density_2-Soliton}
\mathcal{H}_{\scriptsize{tot}}
%=-\mbox{Tr}\left(\dot{\sigma}\sigma^{-1} \right)^{2} \nonumber \\
={-}
\frac{\!\left\{
	\begin{array}{l}
	~ab
	\left[
	d_{11}\cosh^2 { X_2} + d_{22}\cosh^2 { X_1}
	\right]
	\medskip \\
	\!\!+ ac\left[
	d_{12}~\!\displaystyle\cosh^2 {{\!\left(\frac{X_1 + X_2 - i\Theta_{12}}{2} \right)}}
	\!+ d_{21}~\!\displaystyle{\cosh^2 {{\!\left(\frac{X_1 + X_2 + i\Theta_{12}}{2} \right)}}}
	\right]
	\medskip \\
	\!\!- bc\left[
	e_{12}~\!
	\displaystyle{\sinh^2 {{\!\left(\frac{X_1 - X_2 - i\Theta_{12}}{2} \right)}}}
	\!+\overline{e}_{12}~\!
	\displaystyle{\sinh^2 {{\!\left(\frac{X_1 - X_2 + i\Theta_{12}}{2} \right)}}}
	\right]
	\end{array}
	\!\!\!\right\}
}
{{2\pi}
	\displaystyle{
		\left[
		a\cosh { (X_1 + X_2)}
		+
		b\cosh { (X_1 - X_2)}
		+
		c\cos { \Theta_{12}}
		\right]^2
	}
} 
, ~
\end{eqnarray}
where $a, b, c$ are the same coefficients defined in Table \ref{Table_1} and
\begin{eqnarray}
\displaystyle{d_{jk}
	:=\frac{(\lambda_j \overline{\lambda}_k + 1)^2(\lambda_j-\overline{\lambda}_k)^2} {\lambda_j\overline{\lambda}_k}
}\beta_j \overline{\beta}_k,~~ 
\displaystyle{e_{jk}
	:=\frac{(\lambda_j \lambda_k + 1)^2(\lambda_j-\lambda_k)^2} {\lambda_j\lambda_k
	}\beta_j \beta_k}.
\end{eqnarray}

%\noindent
As for the NL$\sigma$M term, the Hamiltonian density $\mathcal{H}_{\scriptsize{\mbox{tot}}}$ for the two-soliton is also in the same form as the reduced NL$\sigma$M action density $\mathcal{L}_{\sigma}$ up to the differences of the coefficients $d_{ij}$ and $e_{ij}$ (Cf: \eqref{NL Sigma term_2-Soliton_form 1}). Therefore, the two peaks of $\mathcal{H}_{\scriptsize{\mbox{tot}}}$ are localized on the same hyperplanes $X_1\pm \delta_1=0$ and $X_2\pm \delta_2=0$ as those of $\mathcal{L}_{\sigma}$ (Cf: (\ref{Asymptotic_2-Soliton})). The phase shift factors are also perfectly the same. There is no singularity as well. This result implies that there is no essential difference between $\mathcal{H}_{\scriptsize{\mbox{tot}}}$ and $\mathcal{L}_{\sigma}$ for describing the solitonic properties. 

The peaks of the energy density of the two-soliton solutions 
is shown in Figure below.

%\begin{wrapfigure}[1]{r}{7cm}
\begin{figure}[h]
%\begin{center}
%\vspace{-15.4cm}
%\vspace{-0.7cm}
\hspace{5cm}
\includegraphics[width=6.8cm]{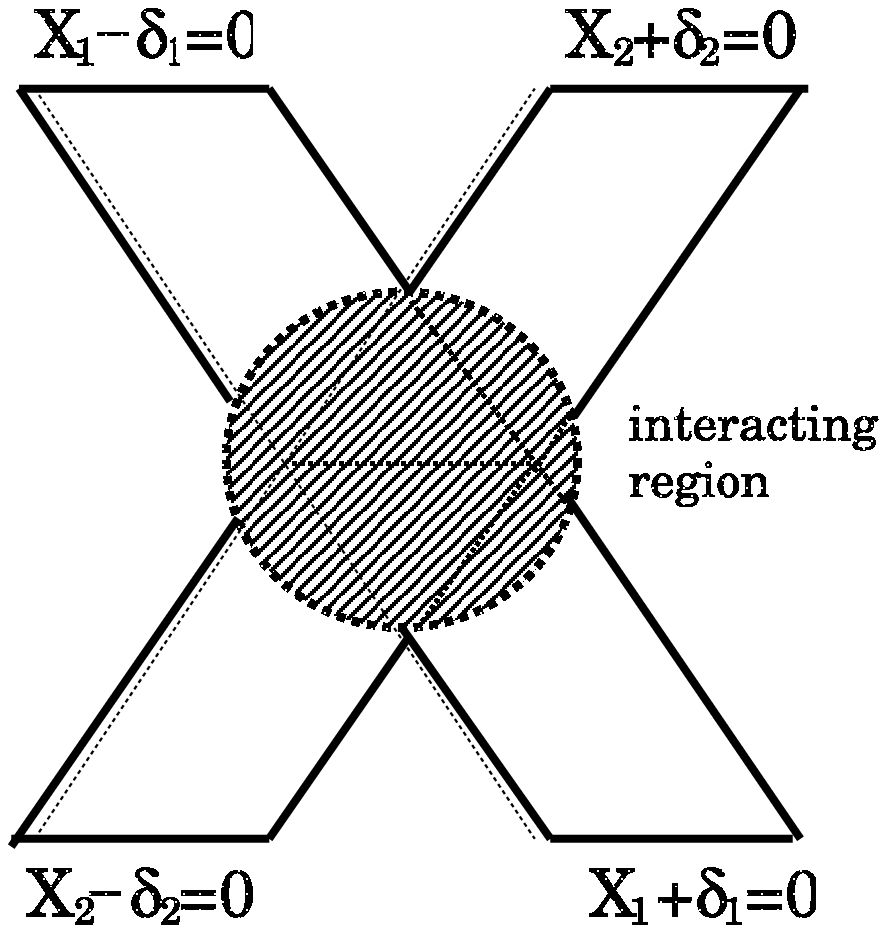} 
%\end{center}
% \caption{Asymptotic Regions ($R_3$ is denoted by the pinky region.)}
%\label{2-soliton}
\end{figure}
%\end{wrapfigure}

%Therefore we could say that for the  NL$\sigma$M term, 
%the action density would describe physical realty and 
%solitonic properties as well as the Hamiltonian density. 
%for the two-soliton solution. 
%the Hamiltonian density has the same distribution as 

%\vspace{5cm}
On the other hand, as for the Wess-Zumino term, there is a mismatch that the Hamiltonian density $\mathcal{H}_{\scriptsize{\mbox{WZ}}}$ is identical to zero, while we cannot confirm the action density $\mathcal{L}_{\scriptsize{\mbox{WZ}}}$ is. The physical meaning of this mismatch should be clarified in the future work.

%for the Wess-Zumino term, 
%there is difference between the Wess-Zumino action density and 
%the WZ Hamiltonian density in the interacting region. 
%The meaning of this mismatch should be clarified in the future work.  

%This might implies that the non-zero contribution of the Wess-Zumino action density 
%would be unphysical bubble in the sense that it has no energy. 

%\newpage

\section{Conclusion and Discussion}

In this paper, we calculated the action density of the WZW$_4$ model for the classical soliton solutions. We found that for the one-soliton solutions, 
the NL$\sigma$M action density is localized on a three-dimensional hyper-plane
and the Wess-Zumino action density identically vanishes. This suggests the existence of a three-brane in the open N=2 string theory. For the two-soliton solutions, the NL$\sigma$M action density has a beautiful compact form which represents an intersecting two one-solitons. The Wess-Zumino action density does not vanish in the interaction region but does vanish in the asymptotic region. For the $n$-soliton solutions, we clarified asymptotic behavior and found that the NL$\sigma$M action density describes ``nonlinear superposition'' of intersecting $n$ one-solitons and the Wess-Zumino action density asymptotically vanishes. The nonlinear interaction gives rise to phase shifts which were evaluated explicitly. We also calculated the Hamiltonian (energy) density of the one and two-soliton solutions of the reduced WZW model in $(1+2)$-dimensions. We found that the energy density of the Wess-Zumino term identically vanishes, and the energy density of the NL$\sigma$M term has the same profile as the action density for our soliton solutions. The peaks of the energy densities perfectly coincide with those of the action density including the phase shift factor. 

We also discussed whether singularities exist for the $n$-soliton solutions. For the one and two-soliton solutions, we proved that there is no singularity. For the $n$-soliton solutions ($n\geq 3$), it is unsolved, however, we can argue as follows. The existence of singularities in the solution is equivalent to the existence of zeros in the data $\Delta$ which is a polynomial of $e^{X_j}$ and $e^{i\Theta_j}$. %The data $\Delta$ could have zeros when $e^{X_j}$ are constants and/or $e^{i\Theta_j}$ are constants. 
Because $X_j$ and $\Theta_j$ are linear functions of the real coordinates, possible singularities would lie on the intersection of $X_j=C_j$ and/or $\Theta_j=D_j$ where $C_j$ and $D_j$ are constants. These possibilities are mostly forbidden because there is no singularity of the action density of the $n$-soliton solutions in the asymptotic region where the intersection still exists. The intersection of just four hyperplanes defined by $X_j=C_j$ or $\Theta_j=D_j~(j=1,2,3,4)$ gives rise to an isolated singularity. This possibility might arise when the parameters in the solutions are appropriately tuned, which should be clarified in the future.   

The next step is to clarify roles and properties of the soliton solutions in the open N=2 string theory. At least we can see that they are not D-branes because the number of the solitons is not related to the rank of gauge group. It is worth studying the topological charge and mass of the solitons, and explicit calculation of infinite conserved densities \cite{CGSW,Dolan,IvLe,IKU} for the $n$-soliton solutions. 
It is also interesting to construct resonance solutions of the solitons which represent the three-brane reconnections, or in other words, annihilation and creation of the three-branes. Then a classification of the soliton solutions could be possible like the positive Grassmannian description of the KP solitons by Kodama and Williams \cite{KoWi}. The moduli space of the $n$-soliton solutions could be described in a geometrical framework. Extension of the model to noncommutative spaces would allow the presence of background $B$-fields in the open N=2 string theory \cite{SeWi, LPS, HaTo, Hamanaka_NPB}. The isolated singularities mentioned above might be resolved and new physical objects appear on the noncommutative spaces such as noncommutative $U(1)$ instantons \cite{NeSc}. Sen's conjecture on the tachyon condensation (for a review see \cite{Ohmori}) could be confirmed by the solution generating technique \cite{HKL} in the context of the open N=2 string theory. 

Furthermore, the WZW$_4$ model can be realized in the context of the twistor string theory \cite{Witten}. Recently Bittleston and Skinner show that a meromorphic Chern-Simons theory on the twistor space in six dimensions has a double fibration structure which gives rise to the WZW$_4$ model by solving along fibers in one direction and the four-dimensional Chern-Simons theory by symmetry reduction in another direction \cite{BiSk}. These models are connected to each other and have a close relationship to integrable systems \cite{CoYa}. The KP equation has not yet obtained as a symmetry reduction of the anti-self-dual Yang-Mills equation so far, however, this six-dimensional Chern-Simons theory might give a ``unified theory'' of integrable systems including both the Sato theory \cite{Sato} of the KP equation and the twistor descriptions of classical integrable systems. This might give a stringy viewpoint to various aspects of integrability and duality. The relation to mirror symmetry is also exciting \cite{NeVa}.

% \vspace{-4mm}
\subsection*{Acknowledgments}
% \vspace{-2mm}

MH thanks string group members at Nagoya university for useful comments at the String Journal Club on July 21, 2022. MH is also grateful to the YITP at Kyoto University, where he had fruitful discussions at the conference on Strings and Fields 2022 (YITP-W-22-09) on August 19, 2022. 
The work of HK and SCH is supported in part by Grant-in-Aid for Scientific Research (\#18K03274). The work of SCH is supported by 
%the scholarship of Japan-Taiwan Exchange Association and 
the Iwanami Fujukai Foundation.

%\medskip

\begin{appendix}
 
\section{Brief Review of Quasideterminants}

In this subsection, 
we excerpt some necessary pre-knowledge of quasideterminant mentioned in section 2 of the previous paper \cite{HaHu2}. It is a brief review of the work of Gelfand and Retakh \cite{GeRe} (See also e.g. \cite{GGRW, Huang}).
%we give a brief introduction to  
%the quasideterminant defined first by Gelfand and Retakh
%\cite{GeRe} (See also e.g. \cite{GGRW, Huang}).

The quasideterminant is defined for an $n \times n$ matrix X where 
matrix elements belong to a noncommutative ring. 
The quasideterminant is a noncommutative generalization of 
the matrix determinant in this sense, 
however,  rather has a direct relation to the inverse matrix of $X$. 

Let $X=(x_{ij})$ be an $n\times n$ invertible matrix 
over a noncommutative ring and 
$Y=(y_{ij})$ be the inverse matrix of $X$: $X Y=Y X =1$. 
The existence of $Y$ is assumed. 
Then the $(i,j)$-th quasideterminant of $X$ is defined 
as the inverse of an element of $Y=X^{-1}$:
\begin{eqnarray}
\vert X \vert_{ij}:=y_{ji}^{-1}.
\end{eqnarray}
This has a convenient expression as follows:
\begin{eqnarray}
\label{Qdet}
\vert X\vert_{ij}=
\left|
\begin{array}{ccccc}
x_{11}&\cdots &x_{1j} & \cdots& x_{1n}\\
\vdots & & \vdots & & \vdots\\
x_{i1}&\cdots & {\fbox{$x_{ij}$}}& \cdots& x_{in}\\
\vdots & & \vdots & & \vdots\\
x_{n1}& \cdots & x_{nj}&\cdots & x_{nn}
\end{array}\right|.
\end{eqnarray}
%To expand \eqref{Qdet},
When the matrix elements belong to a commutative ring, e.g. $\mathbb{C}$, 
the quasideterminant can be represented as a ratio of ordinary determinants 
by virtue of the Laplace formula on inverse matrices:
\begin{eqnarray}
\label{laplace}
\left|
X
\right|_{ij}
=y_{ji}^{-1}
=(-1)^{i+j}\frac{\det X}{\det X^{ij}},
\end{eqnarray}
where $X^{ij}$ is a matrix obtained from $X$ 
by deleting $i$-th row and $j$-th column. 

In order to find another representation of the quasideterminant, 
let us consider the inverse matrix formula 
for the $2\times 2$ block matrix divided as follows:
\begin{eqnarray*}
	X^{-1}=\left(
	\begin{array}{cc}
		A&B \\C&d
	\end{array}
	\right)^{-1}
	=\left(\begin{array}{cc}
		A^{-1}+A^{-1} B s^{-1} C  A^{-1}
		&-A^{-1} B s^{-1}\\
		-s^{-1} C A^{-1}
		&s^{-1}
	\end{array}\right),
\end{eqnarray*}
where $A$ is a square matrix, $d$ is a single element and 
$s:=d-C A^{-1} B$ is called the Schur complement. 
The quantity $s^{-1}$ is just the $(n,n)$-element of $X^{-1}$
and hence the quasideterminant $\vert X \vert_{nn}$ is $s$. 
If we decompose $X$ into a $2\times 2$ block matrix where 
$x_{ij}$ corresponds to the single element $d$, 
the $(i,j)$-th quasideterminant can be expressed 
in the form of  the Schur complement:
\begin{eqnarray}
\label{Schur complement}
\vert X \vert_{ij} 
= x_{ij} -  \sum_{k (\neq i), l (\neq j)}
x_{ik}  ( {X}^{ij})^{-1}_{kl}x_{l j},
%&=&x_{ij}-\sum_{i^\prime (\neq i), j^\prime (\neq j)}
%x_{ii^\prime}  (({X}^{ij})^{-1})_{i^\prime j^\prime} 
%x_{j^\prime	j}
=x_{ij}-\sum_{k(\neq i), l (\neq j)}
x_{ik}  \vert {X}^{ij}\vert_{lk}^{-1}
x_{l j},
\end{eqnarray}
%where $X^{ij}$ is the submatrix obtained from $X$ by deleting $i$-th row and $j$-th column. 
By using this, explicit representations of the quasideterminants can be obtained iteratively.

We note that the quasideterminant is well-defined in the case that each matrix element $x_{ij}$ in \eqref{Qdet} take values in GL$(N,\mathbb{C})$. (Then, $X$ is an $nN \times nN$ matrix.) The following example of the $N=2$ case can be expressed finally by the ratios of determinants due to (\ref{Schur complement}) and (\ref{laplace}):  
\begin{eqnarray}
\label{2x2}
\begin{vmatrix}
M\!\!\!\!\!&\begin{array}{cc}C_1&C_2\end{array}\\
\begin{array}{c}
R_1\!\!\!\!\!\\R_2\!\!\!\!\!
\end{array}
&\fbox{$
	\begin{array}{cc}
	a&b\\c&d
	\end{array}$}
\end{vmatrix}
&=&
\left(\begin{array}{cc}a&b\\c&d\end{array}\right)
-
\left(
\begin{array}{c}
R_1\\R_2
\end{array}
\right)
M^{-1}
\left(
\begin{array}{cc}C_1&C_2\end{array}
\right)
\nonumber\\
&=&
\left(
\begin{array}{cc}
\begin{vmatrix}
M&C_1\\
R_1&\fbox{$a$}
\end{vmatrix}&
\begin{vmatrix}
M&C_2\\
R_1&\fbox{$b$}
\end{vmatrix}
\smallskip \\
\begin{vmatrix}
M&C_1\\
R_2&\fbox{$c$}\\
\end{vmatrix}&
\begin{vmatrix}
M&C_2\\
R_2&\fbox{$d$}\\
\end{vmatrix}
\end{array}
\right)
=
\dfrac{1}{\vert M\vert}
\left(
\begin{array}{cc}
\begin{vmatrix}
M&C_1\\
R_1&a
\end{vmatrix}&
\begin{vmatrix}
M&C_2\\
R_1&b
\end{vmatrix}
\smallskip \\
\begin{vmatrix}
M&C_1\\
R_2&c\\
\end{vmatrix}&
\begin{vmatrix}
M&C_2\\
R_2&d\\
\end{vmatrix}
\end{array}
\right). ~~~~~
\end{eqnarray}
The final form corresponds to 
{ 
the parametrization \eqref{five} of $\sigma$ for 
the soliton solution (\ref{Jn}):}
\begin{eqnarray}
\label{A6}
{ 
\Delta \!=\! \vert M \vert,}~
{ 
-\Delta_{11} \!=\!
\begin{vmatrix}
M&C_1\\
R_1&a
\end{vmatrix},~
-\Delta_{12} \!=\!
\begin{vmatrix}
M&C_2\\
R_1&b
\end{vmatrix},~
-\Delta_{21} \!=\!
\begin{vmatrix}
M&C_1\\
R_2&c\\
\end{vmatrix},~
-\Delta_{22} \!=\!
\begin{vmatrix}
M&C_2\\
R_2&d\\
\end{vmatrix}},~~~
\end{eqnarray}
{ 
which leads to the soliton data (\ref{data}).}

Here we summarize some properties and identities of the quasideterminant, 
which are relevant to discussions in this paper. 

\vspace{2mm}
\noindent
%\begin{proposition}
%\end{proposition}
{\bf Proposition A.1 \cite{GGRW, GeRe, Huang}}  \\
Let $A=(a_{ij})$ be a square matrix of order $n$ in {(1), 
while in (2) and (3),} appropriate partitions are made 
so that all matrices in quasideterminants are square. 
\begin{enumerate}
%	\item [{(}1{)}]  Permutation of Rows and Columns	
%	The quasideterminant $\vert A\vert_{ij}$
%	does not depend on permutations of rows and columns
%	in the matrix $A$. 	
	\item  [{(}1{)}] The common multiplication of rows and columns
	
	For any invertible elements $\Lambda_{j}~(j=1,\cdots,n)$, 
	we have
	\begin{eqnarray}	
	\label{Rmulti}
	\left|
	\begin{array}{ccccc}
	a_{1,1}\Lambda_{1} & \cdots & a_{1,j}\Lambda_{j} & \cdots & a_{1,n}\Lambda_{n} \\
	\vdots &   & \vdots &  & \vdots  \\
	a_{i,1}\Lambda_{1} & \cdots & \fbox{$a_{i,j}\Lambda_{j}$} & \cdots & a_{i,n}\Lambda_{n} \\
	\vdots &   & \vdots &  & \vdots  \\
	a_{n,1}\Lambda_{1} & \cdots & a_{n,j}\Lambda_{j} & \cdots & a_{n,n}\Lambda_{n}
	\end{array}\right|
	&=&
	\left|
	\begin{array}{ccccc}
	a_{1,1} & \cdots & a_{1,j} & \cdots & a_{1,n} \\
	\vdots &   & \vdots &  & \vdots  \\
	a_{i,1} & \cdots & \fbox{$a_{i,j}$} & \cdots & a_{i,n} \\
	\vdots &   & \vdots &   & \vdots  \\
	a_{n,1} & \cdots & a_{n,j} & \cdots & a_{n,n}
	\end{array}\right|\Lambda_{j},~~~~~~~\\
	\label{Lmulti}
	\left|
	\begin{array}{ccccc}
	\Lambda_{1}a_{1,1} & \cdots & \Lambda_{1}a_{1,j}& \cdots & \Lambda_{1}a_{1,n} \\
	\vdots &   & \vdots &  & \vdots  \\
	\Lambda_{i}a_{i,1} & \cdots & \fbox{$\Lambda_{i}a_{i,j}$} & \cdots & \Lambda_{i} a_{i,n} \\
	\vdots &   & \vdots &  & \vdots  \\
	\Lambda_{n}a_{n,1} & \cdots & \Lambda_{n}a_{n,j} & \cdots & \Lambda_{n}a_{n,n}
	\end{array}\right|
	&=&
	\Lambda_{i} \left|
	\begin{array}{ccccc}
	a_{1,1} & \cdots & a_{1,j} & \cdots & a_{1,n} \\
	\vdots &   & \vdots &  & \vdots  \\
	a_{i,1} & \cdots & \fbox{$a_{i,j}$} & \cdots & a_{i,n} \\
	\vdots &   & \vdots &   & \vdots  \\
	a_{n,1} & \cdots & a_{n,j} & \cdots & a_{n,n}
	\end{array}\right|.
	\end{eqnarray}
%	We note that on the left hand side of \eqref{Rmulti}, the common elements $\Lambda_{j}$ must appear on the right side of the same column. 
%On the other hand, if the common elements appear on the left side of the same row, one can get the rule for common multiplication of rows.
	
%	\item [{(}3{)}] The addition of rows and columns	
%	Let the matrix $N=(n_{ij})$ be obtained from
%	the matrix $A$ by replacing the $k$-th column of $A$
%	with the sum of the $k$-th column and $l$-th column, that is,
%	$n_{ik}=a_{ik}+a_{{il}}$ and $n_{ij}=a_{ij}$
%	for $k\neq j$. Then $\vert A\vert_{ij}=\vert N \vert_{ij}$, 
%	for $j\neq k$. 
%	(The addition of rows is similar.) 
	
	\item  [{(}2{)}] Noncommutative Jacobi identity \cite{GiNi07} 
	%(An useful and simplified version of the noncommutative Sylvester's theorem\cite{GeRe}):
	\begin{equation}
	\label{jacobi}
	\begin{vmatrix}
	a&R&b\\
	P&M&Q\\
	c&S&\fbox{$d$}
	\end{vmatrix}=
	\begin{vmatrix}
	M&Q\\
	S&\fbox{$d$}
	\end{vmatrix}-
	\begin{vmatrix}
	P&M\\
	\fbox{$c$}&S
	\end{vmatrix}    
	\begin{vmatrix}
	\fbox{$a$}&R\\
	P&M
	\end{vmatrix}^{-1}
	\begin{vmatrix}
	R&\fbox{$b$}\\
	M&Q
	\end{vmatrix}.
	\end{equation}
	
	\item  [{(}3{)}] 
	Homological relations \cite{GeRe,GiNi07}
	\begin{eqnarray}
		\label{homological}
	\!\!    \begin{vmatrix}
	a&\!R\!&b\\
	P&\!M\!&Q\\
	\fbox{$c$}&\!S\!&d
	\end{vmatrix}
	\!=\!  \begin{vmatrix}
	a&\!R\!&b\\
	P&\!M\!&Q\\
	c&\!S\!&\fbox{$d$}
	\end{vmatrix}\!
	\begin{vmatrix}
	a&\!R\!&b\\
	P&\!M\!&Q\\
	\fbox{0}&\!0\!&1
	\end{vmatrix},~~
	\begin{vmatrix}
	a&\!R\!&\fbox{$b$}\\
	P&\!M\!&Q\\
	c&\!S\!&d
	\end{vmatrix}
	\!=\!      \begin{vmatrix}
	a&\!R\!&\fbox{0}\\
	P&\!M\!&0\\
	c&\!S\!&1
	\end{vmatrix}\!
	\begin{vmatrix}
	a&\!R\!&b\\
	P&\!M\!&Q\\
	c&\!S\!&\fbox{$d$}
	\end{vmatrix}.~~~~~
	\end{eqnarray}
%	 \mbox{If we use the homological relation again on the right hand side, we can obtain the}\\ \mbox{following inverse relation immediately:}
%		\begin{eqnarray} \label{inverse relation}
%		\begin{vmatrix}
%		a&\!R\!&b\\
%		P&\!M\!&Q\\
%		\fbox{0}&\!0\!&1
%		\end{vmatrix}^{-1}=
%		\begin{vmatrix}
%		a&\!R\!&b\\
%		P&\!M\!&Q\\
%		1&\!0\!&\fbox{0}
%		\end{vmatrix},~~
%		\begin{vmatrix}
%		a&\!R\!&\fbox{0}\\
%		P&\!M\!&0\\
%		c&\!S\!&1
%		\end{vmatrix}^{-1}\!=
%		\begin{vmatrix}
%		a&\!R\!&1\\
%		P&\!M\!&0\\
%		c&\!S\!&\fbox{0}
%		\end{vmatrix}\!
%		\end{eqnarray}
\end{enumerate}

%\newpage

\section{Proof of Statement in Footnote 8}
\label{commute}
{\bf Proposition B.1}
\smallskip \\
Let $\sigma$ be the $n$-soliton solution defined by
\begin{eqnarray}
\label{n-soliton solution_appendeix B}
\sigma=
\left|
\begin{array}{ccccc}
\psi_1 & \psi_2 & \cdots & \psi_n & 1 \\
\psi_1 \Lambda_1 & \psi_2 \Lambda_2 & \cdots & \psi_n \Lambda_{n} & 0 \\
\psi_1 \Lambda_1^2 & \psi_2 \Lambda_2^2 & \cdots & \psi_n \Lambda_{n}^2 & 0 \\
\vdots & \vdots & \ddots & \vdots & \vdots \\
\psi_1 \Lambda_1^n & \psi_2 \Lambda_2^n & \cdots & \psi_n \Lambda_{n}^n & \fbox{$0$}
\end{array}
\right|,~
\begin{array}{l}
\psi_j=
\left(
\begin{array}{cc}
e^{L_j} & e^{-\overline{L}_j} \\
-e^{-L_j} & e^{\overline{L}_j}
\end{array}
\right),~L_j=\ell_{\mu}^{(j)}x^{\mu} 
\smallskip \\
\Lambda_j
:=
\left(
\begin{array}{cc}
\lambda_j & 0 \\
0 & \mu_j
\end{array}
\right)
\end{array},~~
\end{eqnarray}
and $\mathscr{R}_K$ be the asymptotic region defined by the asymptotic limit
\begin{eqnarray}
\label{Asymptotic limit_L_1}
\left\{
\begin{array}{l}
\mathrm{Re}L_K ~\mbox{is fixed}  
\smallskip \\
\mathrm{Re}L_{j, j \neq K} \longrightarrow \pm \infty
\end{array}.
\right.
\end{eqnarray}
Then the operation of the partial derivative $\partial_{\mu}$ commutes with the operation of the asymptotic limit \eqref{Asymptotic limit_L_1} for $\sigma$. 
\smallskip \\
(Proof)
\smallskip \\
Without loss of generality, we consider $K=1$ case 
%The other cases are equivalent to $K=1$ due to the permutation rule of the quasideterminant.
{due to the fact that the quasideterminant { $\vert \sigma \vert_{ij}$} 
does not depend on permutations of rows and columns 
in the matrix { $\sigma$} \cite{GeRe}.}

For $j \neq 1$, $\psi_j \Lambda_{j}^{m}$ can be decomposed into
\begin{eqnarray}
\label{Decomposition of psi_i Lambda_i}
\psi_j \Lambda_j^{m}
=
\left\{
\begin{array}{l}
\left(
\begin{array}{cc}
1 & e^{-2\overline{L}_j} \\
-e^{-2L_j} & 1
\end{array}
\right)
\Lambda_j^{m}
\left(
\begin{array}{cc}
e^{L_j} & 0 \\
0 & e^{\overline{L}_j} 
\end{array}
\right)
=:f_j^{(+)} \Lambda_j^{m} E_j^{(+)}
\smallskip \\
\left(
\begin{array}{cc}
-e^{2L_j} & 1 \\
1 & e^{2\overline{L}_j}
\end{array}
\right)
\Lambda_j^{m}
\left(
\begin{array}{cc}
-e^{-L_j} & 0 \\
0 & e^{-\overline{L}_j} 
\end{array}
\right)
=:f_j^{(-)} \Lambda_j^{m} E_j^{(-)}
\end{array}
\right.,
\end{eqnarray}
and
\begin{eqnarray}
\label{f_i tends to C_i}
f_j^{(\pm)} \longrightarrow C_j^{(\pm)} ~~\mbox{as}~~ \mathrm{Re}L_j \longrightarrow \pm \infty,~~
\mbox{where}~
\left\{
\begin{array}{l}
C_j^{(+)}
:=
\left(
\begin{array}{cc}
1 & 0 \\
0 & 1
\end{array}
\right)
\smallskip \\
C_j^{(-)}
:=
\left(
\begin{array}{cc}
0 & 1 \\
1 & 0
\end{array}
\right)
\end{array}.
\right.
\end{eqnarray}
By \eqref{Decomposition of psi_i Lambda_i} { and \eqref{Rmulti}},
the right common factors $E_j^{(\pm)}$ of each column of $\sigma$ can be omitted completely, and hence
\begin{eqnarray}
\sigma=
\left|
\begin{array}{ccccc}
	\psi_1 & f_2^{(\pm)} & \cdots & f_n^{(\pm)} & 1 \\
	\psi_1 \Lambda_1 & f_2^{(\pm)} \Lambda_2 & \cdots & f_n^{(\pm)} \Lambda_{n} & 0 \\
	\psi_1 \Lambda_1^2 & f_2^{(\pm)} \Lambda_2^2 & \cdots & f_n^{(\pm)} \Lambda_{n}^2 & 0 \\
	\vdots & \vdots & \ddots & \vdots & \vdots \\
	\psi_1 \Lambda_1^n & f_2^{(\pm)} \Lambda_2^n & \cdots & f_n^{(\pm)} \Lambda_{n}^n & \fbox{$0$}
\end{array}
\right|
\end{eqnarray}
which is asymptotic to 
\begin{eqnarray}
\widetilde{\sigma}:=
\left|
\begin{array}{ccccc}
\psi_1 & C_2^{(\pm)} & \cdots & C_n^{(\pm)} & 1 \\
\psi_1 \Lambda_1 & C_2^{(\pm)} \Lambda_2 & \cdots & C_n^{(\pm)} \Lambda_{n} & 0 \\
\psi_1 \Lambda_1^2 & C_2^{(\pm)} \Lambda_2^2 & \cdots & C_n^{(\pm)} \Lambda_{n}^2 & 0 \\
\vdots & \vdots & \ddots & \vdots & \vdots \\
\psi_1 \Lambda_1^n & C_2^{(\pm)} \Lambda_2^n & \cdots & C_n^{(\pm)} \Lambda_{n}^n & \fbox{$0$}
\end{array}
\right|.
\end{eqnarray}
By the fact that
\begin{eqnarray}
\label{Reltation of C_iLambda_i}
C_j^{(\pm)} \Lambda_j^m
=
\Lambda_j^{(\pm)m} C_j^{(\pm)},~~
\left\{
\begin{array}{l}
\Lambda_j^{(+)} 
:=
\left(
\begin{array}{cc}
\lambda_j & 0 \\
0 & \mu_j
\end{array}
\right)
\smallskip \\
\Lambda_j^{(-)} 
:=
\left(
\begin{array}{cc}
\mu_j & 0 \\
0 & \lambda_j
\end{array}
\right)
\end{array},
\right. 
\end{eqnarray}
{ and \eqref{Rmulti}, }
the right common factors $C_j^{(\pm)}$ of each column of $\widetilde{\sigma}$ can be omitted completely, and hence
\begin{eqnarray}
\widetilde{\sigma}
=
\left|
\begin{array}{ccccc}
\psi_1 & 1 & \cdots & 1 & 1 \\
\psi_1 \Lambda_1 & \Lambda_2^{(\pm)} & \cdots & \Lambda_{n}^{(\pm)} & 0 \\
\psi_1 \Lambda_1^2 & \Lambda_2^{(\pm)2} & \cdots & \Lambda_{n}^{(\pm)2} & 0 \\
\vdots & \vdots & \ddots & \vdots & \vdots \\
\psi_1 \Lambda_1^n & \Lambda_2^{(\pm)n} & \cdots & \Lambda_{n}^{(\pm)n} & \fbox{$0$}
\end{array}
\right|,
\end{eqnarray}
which is called the asymptotic form of the $n$-soliton solution $\sigma$.
By the derivative formula \cite{GiNi07} of the quasideterminant
\begin{eqnarray}
\partial_{\mu}\left|
\begin{array}{cc}
A & B \\
C & \fbox{$d$}
\end{array}
\right|
&\!\!\!\!=\!\!\!\!&
\left|
\begin{array}{cc}
A & \partial_{\mu}B \\
C & \fbox{$\partial_{\mu}d$}
\end{array}
\right|
+\sum\limits_{j=1}^{n}
\left|
\begin{array}{cc}
A & \partial_{\mu}A_{j}\\
C & \fbox{$\partial_{\mu}C_{j}$}
\end{array}
\right| 
\left|
\begin{array}{cc}
A & B \\
E^{j} & \fbox{$0$}
\end{array}
\right|,  
\\
&&(\mbox{$A_{j}$ : $j$-th column of $A$, $E^{j}$ : $j$-th row of identity matrix $I$}.)   \nonumber 
\end{eqnarray}
we have
\begin{eqnarray}
&& \!\!\!\!\!\!\!\!\!\!\!\!\!\!\!\! \partial_{\mu}\widetilde{\sigma}
=
\begin{array}{l}
\left|
\begin{array}{ccccc}
\psi_1 & 1 & \cdots & 1 & \partial_{\mu}\psi_1 \\
\psi_1 \Lambda_1 & \Lambda_2^{(\pm)} & \cdots & \Lambda_{n}^{(\pm)} & (\partial_{\mu}\psi_1)\Lambda_1 \\
\psi_1 \Lambda_1^2 & \Lambda_2^{(\pm)2} & \cdots & \Lambda_{n}^{(\pm)2} &  (\partial_{\mu}\psi_1)\Lambda_1^2 \\
\vdots & \vdots & \ddots & \vdots & \vdots \\
\psi_1 \Lambda_1^n & \Lambda_2^{(\pm)n} & \cdots & \Lambda_{n}^{(\pm)n} & \fbox{$(\partial_{\mu}\psi_1)\Lambda_1^{n}$}
\end{array}
\right|
\left|
\begin{array}{ccccc}
\psi_1 & 1 & \cdots & 1 & 1 \\
\psi_1 \Lambda_1 & \Lambda_2^{(\pm)} & \cdots & \Lambda_{n}^{(\pm)} & 0 \\
\vdots & \vdots & \ddots & \vdots & \vdots \\
\psi_1 \Lambda_1^{n-1} & \Lambda_2^{(\pm)n-1} & \cdots & \Lambda_{n}^{(\pm)n-1} & 0 \\
1 & 0 & \cdots & 0 & \fbox{$0$}
\end{array}
\right|
\bigskip \\
+
\displaystyle{
\sum_{j=2}^{n}
\underbrace{
\left|
\begin{array}{ccccc}
\psi_1 & 1 & \cdots & 1 & 0 \\
\psi_1 \Lambda_1 & \Lambda_2^{(\pm)} & \cdots & \Lambda_{n}^{(\pm)} & 0 \\
\psi_1 \Lambda_1^2 & \Lambda_2^{(\pm)2} & \cdots & \Lambda_{n}^{(\pm)2} & 0 \\
\vdots & \vdots & \ddots & \vdots & \vdots \\
\psi_1 \Lambda_1^n & \Lambda_2^{(\pm)n} & \cdots & \Lambda_{n}^{(\pm)n} & \fbox{$0$}
\end{array}
\right|
}
\left|
\begin{array}{ccccc}
\psi_1 & \cdots & 1 &\cdots& 1 \\
\psi_1 \Lambda_1 & \cdots & \Lambda_j^{(\pm)} & \cdots & 0 \\
\vdots & \ddots & \vdots & \ddots & \vdots \\
\psi_1 \Lambda_1^{n-1} & \cdots & \Lambda_j^{(\pm)n-1} & \cdots & 0 \\
0 & \cdots & 1 & \cdots & \fbox{$0$}
\end{array}
\right|
}
\end{array}
\\
&&~~~~~~~~~~~~~~~~~~~~~~~~~~~~~~~~=0 ~~~(\mbox{$\partial_{\mu}\Lambda_{j}^{(\pm)m}=0$ on the last column}.)   
\nonumber 
\end{eqnarray}
Now we can conclude that 
\begin{eqnarray}
&&\partial_{\mu}\widetilde{\sigma}    \nonumber \\
&\!\!\!\!=\!\!\!\!&
\left|
\begin{array}{ccccc}
\psi_1 & 1 & \cdots & 1 & \!\partial_{\mu}\psi_1 \\
\psi_1 \Lambda_1 & \Lambda_2^{(\pm)} & \cdots & \Lambda_{n}^{(\pm)} & \!(\partial_{\mu}\psi_1)\Lambda_1 \\
\psi_1 \Lambda_1^2 & \Lambda_2^{(\pm)2} & \cdots & \Lambda_{n}^{(\pm)2} &  \!(\partial_{\mu}\psi_1)\Lambda_1^2 \\
\vdots & \vdots & \ddots & \vdots & \!\vdots \\
\psi_1 \Lambda_1^n & \Lambda_2^{(\pm)n} & \cdots & \Lambda_{n}^{(\pm)n} & \!\fbox{$(\partial_{\mu}\psi_1)\Lambda_1^{n}$}
\end{array}
\!\right|
\left|
\begin{array}{ccccc}
\psi_1 & 1 & \!\cdots & \!1 & \!1 \\
\psi_1 \Lambda_1 & \Lambda_2^{(\pm)} & \!\cdots & \!\Lambda_{n}^{(\pm)} & \!0 \\
\vdots & \vdots & \!\ddots & \!\vdots & \!\vdots \\
\psi_1 \Lambda_1^{n-1} & \Lambda_2^{(\pm)n-1} & \!\cdots & \!\Lambda_{n}^{(\pm)n-1} & \!0 \\
1 & 0 & \!\cdots & \!0 & \!\fbox{$0$}
\end{array}
\!\right|. ~~~~~~~~
\end{eqnarray}
On the other hand, by the derivative formula of the quasideterminant on $\sigma$ we have
\begin{eqnarray}
&&\partial_{\mu}\sigma   \nonumber \\
&\!\!\!\!=\!\!\!\!&
\sum_{j=1}^{n}
\left|
\begin{array}{ccccc}
\psi_1 & \psi_2 & \cdots & \psi_n & \partial_{\mu}\psi_j \\
\psi_1 \Lambda_1 & \psi_2\Lambda_2 & \cdots & \psi_n\Lambda_{n} & (\partial_{\mu}\psi_j)\Lambda_j \\
\psi_1 \Lambda_1^2 & \psi_2\Lambda_2^2 & \cdots & \psi_n\Lambda_{n}^2 &  (\partial_{\mu}\psi_j)\Lambda_j^2 \\
\vdots & \vdots & \ddots & \vdots & \vdots \\
\psi_1 \Lambda_1^n & \psi_2\Lambda_2^{n} & \cdots & \psi_n\Lambda_{n}^{n} & \fbox{$(\partial_{\mu}\psi_j)\Lambda_j^{n}$}
\end{array}
\!\right|
\left|
\begin{array}{ccccc}
\psi_1 & \cdots & \psi_j &\cdots& \!1 \\
\psi_1 \Lambda_1 & \cdots & \psi_j\Lambda_j & \cdots & \!0 \\
\vdots & \ddots & \vdots & \ddots & \!\vdots \\
\psi_1 \Lambda_1^{n-1} & \cdots & \psi_j\Lambda_j^{n-1} & \cdots & \!0 \\
0 & \cdots & 1 & \cdots & \!\fbox{$0$}
\end{array}
\!\right|. ~~~~~~~~ 
\end{eqnarray}
By a similar argument as \eqref{Decomposition of psi_i Lambda_i}, \eqref{f_i tends to C_i}, we have
\begin{eqnarray}
\label{Decomposition of partial derivative of psi_i Lambda_i}
(\partial_{\mu}\psi_j) \Lambda_j^{m}
=
\left\{
\begin{array}{l}
\left(
\begin{array}{cc}
1 & -e^{-2\overline{L}_j} \\
e^{-2L_j} & 1
\end{array}
\right)
\Lambda_j^{m}
\left(
\begin{array}{cc}
\ell_{\mu}^{(j)}e^{L_j} & 0 \\
0 & \overline{\ell}_{\mu}^{(j)}e^{\overline{L}_j} 
\end{array}
\right)
=:\widetilde{f}_j^{~\!(+)} \Lambda_j^{m} \widetilde{E}_j^{(+)}
\smallskip \\
\left(
\begin{array}{cc}
e^{2L_j} & 1 \\
1 & -e^{2\overline{L}_j}
\end{array}
\right)
\Lambda_j^{m}
\left(
\begin{array}{cc}
\ell_{\mu}^{(j)}e^{-L_j} & 0 \\
0 & -\overline{\ell}_{\mu}^{(j)}e^{-\overline{L}_j} 
\end{array}
\right)
=:\widetilde{f}_j^{~\!(-)} \Lambda_j^{m} \widetilde{E}_j^{(-)}
\end{array}
\right.,
\end{eqnarray}
and
\begin{eqnarray}
\label{tilde_f_i tends to C_i}
\widetilde{f}_j^{~\!(\pm)} \longrightarrow C_j^{(\pm)} ~~\mbox{as}~~ \mathrm{Re}L_j \longrightarrow \pm \infty,~~
\mbox{where}~
\left\{
\begin{array}{l}
C_j^{(+)}
:=
\left(
\begin{array}{cc}
1 & 0 \\
0 & 1
\end{array}
\right)
\smallskip \\
C_j^{(-)}
:=
\left(
\begin{array}{cc}
0 & 1 \\
1 & 0
\end{array}
\right)
\end{array}.
\right.
\end{eqnarray}
Now by \eqref{Decomposition of psi_i Lambda_i}, \eqref{Decomposition of partial derivative of psi_i Lambda_i} { and \eqref{Rmulti},}
we can omit the right common factor $E_j^{(\pm)}$  from the $j$-th column $(j=2 \sim n)$, and take the right common factor $\widetilde{E}_j^{(\pm)}$ of the last column out of the quasideterminant. Then we obtain
\begin{eqnarray}
&&\partial_{\mu}\sigma =   \nonumber \\
&&\begin{array}{l}
\left|
\begin{array}{ccccc}
\psi_1 & f_2^{(\pm)} & \cdots & f_n^{(\pm)} & \partial_{\mu}\psi_1 \\
\psi_1 \Lambda_1 & f_2^{(\pm)}\Lambda_2 & \cdots & f_n^{(\pm)}\Lambda_{n} & (\partial_{\mu}\psi_1)\Lambda_1 \\
\psi_1 \Lambda_1^2 & f_2^{(\pm)}\Lambda_2^2 & \cdots & f_n^{(\pm)}\Lambda_{n}^2 &  (\partial_{\mu}\psi_1)\Lambda_1^2 \\
\vdots & \vdots & \ddots & \vdots & \vdots \\
\psi_1 \Lambda_1^n & f_2^{(\pm)}\Lambda_2^{n} & \cdots & f_n^{(\pm)}\Lambda_{n}^{n} & \fbox{$(\partial_{\mu}\psi_1)\Lambda_1^{n}$}
\end{array}
\right|
\left|
\begin{array}{ccccc}
\psi_1 & \cdots & f_j^{(\pm)} &\cdots& 1 \\
\psi_1 \Lambda_1 & \cdots & f_j^{(\pm)}\Lambda_j & \cdots & 0 \\
\vdots & \ddots & \vdots & \ddots & \vdots \\
\psi_1 \Lambda_1^{n-1} & \cdots & f_j^{(\pm)}\Lambda_j^{n-1} & \cdots & 0 \\
1 & \cdots & 0 & \cdots & \fbox{$0$}
\end{array}
\right|
\bigskip \\
+ \displaystyle{
\sum_{j=2}^{n}
\left|
\begin{array}{ccccc}
\psi_1 & f_2^{(\pm)} & \cdots & f_n^{(\pm)} & \widetilde{f}_j^{~\!(\pm)} \\
\psi_1 \Lambda_1 & f_2^{(\pm)}\Lambda_2 & \cdots & f_n^{(\pm)}\Lambda_{n} & \widetilde{f}_j^{~\!(\pm)}\Lambda_j \\
\psi_1 \Lambda_1^2 & f_2^{(\pm)}\Lambda_2^2 & \cdots & f_n^{(\pm)}\Lambda_{n}^2 &  \widetilde{f}_j^{~\!(\pm)}\Lambda_j^2 \\
\vdots & \vdots & \ddots & \vdots & \vdots \\
\psi_1 \Lambda_1^n & f_2^{(\pm)}\Lambda_2^{n} & \cdots & f_n^{(\pm)}\Lambda_{n}^{n} & \fbox{$\widetilde{f}_j^{~\!(\pm)}\Lambda_j^{n}$}
\end{array}
\right|}
\widetilde{E}_j^{(\pm)} \times
\smallskip \\
~~~~~~~\left|
\begin{array}{ccc:c:cc:c}
\psi_1 & f_2^{(\pm)} & \cdots & \psi_j &\cdots& f_n^{(\pm)} & 1 \\
\hdashline
\psi_1 \Lambda_1  & f_2^{(\pm)}\Lambda_2 & \cdots & \psi_j\Lambda_j & \cdots & f_n^{(\pm)}\Lambda_{n} & 0 \\
\vdots & \vdots & \ddots & \vdots & \ddots & \vdots & \vdots \\
\psi_1 \Lambda_1^{n-1} & f_2^{(\pm)}\Lambda_2^{n-1} & \cdots & \psi_j\Lambda_j^{n-1} &  \cdots & f_n^{(\pm)}\Lambda_{n}^{n-1} & 0  \\
\hdashline 
0 & 0 & \cdots & 1 & \cdots & 0 & \fbox{$0$}
\end{array}
\right|
\end{array} 
\end{eqnarray}
By the Jacobi identity { \eqref{jacobi}}, we have
\begin{eqnarray}
&&\!\!\!\!\widetilde{E}_j^{(\pm)}
	\left|
	\begin{array}{ccc:c:cc:c}
	\psi_1 & f_2^{(\pm)} & \cdots & \psi_j &\cdots& f_n^{(\pm)} & 1 \\
	\hdashline
	\psi_1 \Lambda_1  & f_2^{(\pm)}\Lambda_2 & \cdots & \psi_j\Lambda_j & \cdots & f_n^{(\pm)}\Lambda_{n} & 0 \\
	\vdots & \vdots & \ddots & \vdots & \ddots & \vdots & \vdots \\
	\psi_1 \Lambda_1^{n-1} & f_2^{(\pm)}\Lambda_2^{n-1} & \cdots & \psi_j\Lambda_j^{n-1} &  \cdots & f_n^{(\pm)}\Lambda_{n}^{n-1} & 0 \\
	\hdashline 
	0 & 0 & \cdots & 1 & \cdots & 0 & \fbox{$0$}
	\end{array}
	\right|
	\\
&\!\!\!\!=\!\!\!\!&
-\widetilde{E}_j^{(\pm)}
\left|
\begin{array}{ccccccc}
\psi_1 &  f_2^{(\pm)} & \cdots & \fbox{$\psi_j$} & \cdots & f_n^{(\pm)} \\
\psi_1\Lambda_1 &  f_2^{(\pm)}\Lambda_2 & \cdots & \psi_j\Lambda_j & \cdots & f_n^{(\pm)}\Lambda_n \\
\vdots & \vdots & \ddots & \vdots & \ddots & \vdots \\
\psi_1\Lambda_1^{n-1} &  f_2^{(\pm)}\Lambda_2^{n-1} & \cdots & \psi_j\Lambda_j^{n-1} & \cdots & f_n^{(\pm)}\Lambda_n^{n-1} \\
\end{array}
\right|^{-1}
\\
&\!\!\!\!=\!\!\!\!&
-\widetilde{E}_j^{(\pm)}(E_j^{(\pm)})^{-1}
\left|
\begin{array}{ccccccc}
\psi_1 &  f_2^{(\pm)} & \cdots & \fbox{$f_j^{(\pm)}$} & \cdots & f_n^{(\pm)} \smallskip \\
\psi_1\Lambda_1 &  f_2^{(\pm)}\Lambda_2 & \cdots & f_j^{(\pm)}\Lambda_i & \cdots & f_n^{(\pm)}\Lambda_n \\
\vdots & \vdots & \ddots & \vdots & \ddots & \vdots \\
\psi_1\Lambda_1^{n-1} &  f_2^{(\pm)}\Lambda_2^{n-1} & \cdots & f_j^{(\pm)}\Lambda_j^{n-1} & \cdots & f_n^{(\pm)}\Lambda_n^{n-1} \\
\end{array}
\right|^{-1}
\\
&& (\mbox{By \eqref{Decomposition of psi_i Lambda_i} { and \eqref{Rmulti}}, we can take the right common factor $E_j^{(\pm)}$ out of the}
\nonumber \\
&& \mbox{quasideterminant}.)  \nonumber 
\end{eqnarray}
By \eqref{Decomposition of psi_i Lambda_i} and \eqref{Decomposition of partial derivative of psi_i Lambda_i}, 
\begin{eqnarray}
\widetilde{E}_j^{(\pm)}(E_j^{(\pm)})^{-1}
=
\left\{
\begin{array}{l}
\left(
\begin{array}{cc}
\!\ell_{\mu}^{(j)}e^{L_j}  & 0 \\
0 & \overline{\ell}_{\mu}^{(j)}e^{\overline{L}_j}
\end{array}
\!\!\right)
\left(
\begin{array}{cc}
\!e^{-L_j}  & 0 \\
0 & e^{-\overline{L}_j}
\end{array}
\right)
\smallskip \\
\left(
\begin{array}{cc}
\!\ell_{\mu}^{(j)}e^{-L_j}  & 0 \\
0 & \!\!\!\!\!\!-\overline{\ell}_{\mu}^{(j)}e^{-\overline{L}_j}
\end{array}
\!\!\right)
\left(
\begin{array}{cc}
\!-e^{L_j}  & 0 \\
0 & \!\!e^{\overline{L}_j}
\end{array}
\!\!\right)

\end{array}
\right.
=
\pm \left(
\begin{array}{cc}
\ell_{\mu}^{(j)} & 0 \\
0 & \overline{\ell}_{\mu}^{(j)}
\end{array}
\!\!\right)
=: \widetilde{\Lambda}_j^{(\pm)}  ~~~~
\end{eqnarray}
which are constant matrices.
Therefore, we can conclude that 
\begin{eqnarray}
&&
\!\!\!\!\!\!\!\!\!\!\!\!\partial_{\mu}\sigma =  
%\nonumber \\
%&&
%\begin{array}{l}
\left|
\begin{array}{ccccc}
\psi_1 & f_2^{(\pm)} & \cdots & f_n^{(\pm)} & \partial_{\mu}\psi_1 \\
\psi_1 \Lambda_1 & f_2^{(\pm)}\Lambda_2 & \cdots & f_n^{(\pm)}\Lambda_{n} & (\partial_{\mu}\psi_1)\Lambda_1 \\
%\psi_1 \Lambda_1^2 & f_2^{(\pm)}\Lambda_2^2 & \cdots & f_n^{(\pm)}\Lambda_{n}^2 &  (\partial_{\mu}\psi_1)\Lambda_1^2 \\
\vdots & \vdots & \ddots & \vdots & \vdots \\
\psi_1 \Lambda_1^n & f_2^{(\pm)}\Lambda_2^{n} & \cdots & f_n^{(\pm)}\Lambda_{n}^{n} & \fbox{$(\partial_{\mu}\psi_1)\Lambda_1^{n}$}
\end{array}
\right|
\left|
\begin{array}{ccccc}
\psi_1 & \cdots & f_j^{(\pm)} &\cdots& 1 \\
\psi_1 \Lambda_1 & \cdots & f_j^{(\pm)}\Lambda_j & \cdots & 0 \\
\vdots & \ddots & \vdots & \ddots & \vdots \\
\psi_1 \Lambda_1^{n-1} & \cdots & f_j^{(\pm)}\Lambda_j^{n-1} & \cdots & 0 \\
1 & \cdots & 0 & \cdots & \fbox{$0$}
\end{array}
\right| 
%\end{array}
%\end{eqnarray}
\nonumber \\
%\begin{eqnarray}
%\begin{array}{l}
&&\!\!\!\!\!\!\!\!\!\!\!\! - \displaystyle{
	\sum_{j=2}^{n}
	\!\left|
	\begin{array}{ccccc}
	\!\!\psi_1 & \!\!\!f_2^{(\pm)} & \!\!\!\cdots & \!\!\!\!\!f_n^{(\pm)} & \!\!\!\widetilde{f}_j^{~\!(\pm)} \\
	\!\!\psi_1 \Lambda_1 & \!\!\!f_2^{(\pm)}\Lambda_2 & \!\!\!\cdots & \!\!\!\!\!f_n^{(\pm)}\Lambda_{n} & \!\!\!\widetilde{f}_j^{~\!(\pm)}\Lambda_j \\
	%\psi_1 \Lambda_1^2 & f_2^{(\pm)}\Lambda_2^2 & \cdots & f_n^{(\pm)}\Lambda_{n}^2 &  \widetilde{f}_j^{~\!(\pm)}\Lambda_j^2 \\
	\!\!\vdots & \!\!\!\vdots & \!\!\!\ddots & \!\!\!\!\!\vdots & \!\!\!\vdots \\
	\!\!\psi_1 \Lambda_1^n & \!\!\!f_2^{(\pm)}\Lambda_2^{n} & \!\!\!\cdots & \!\!\!\!\!f_n^{(\pm)}\Lambda_{n}^{n} & \!\!\!\fbox{$\widetilde{f}_j^{~\!(\pm)}\Lambda_j^{n}$}
	\end{array}
	\!\!\right|}
\widetilde{\Lambda}_j^{(\pm)} 
%\times
%\nonumber \\
%&&~~~~~~~
\!\left|
\begin{array}{ccccccc}
\!\!\psi_1 &  \!\!\!\!f_2^{(\pm)} & \!\!\!\!\cdots & \!\!\!\!\fbox{$f_j^{(\pm)}$} & \!\!\!\!\cdots & \!\!\!\!\!f_n^{(\pm)}  \\
\!\!\psi_1\Lambda_1 &  \!\!\!\!f_2^{(\pm)}\Lambda_2 & \!\!\!\!\cdots & \!\!\!\!f_j^{(\pm)}\Lambda_i & \!\!\!\!\!\cdots & \!\!\!\!\!f_n^{(\pm)}\Lambda_n \\
\!\!\vdots & \!\!\!\!\vdots & \!\!\!\!\ddots & \!\!\!\!\vdots & \!\!\!\!\ddots & \!\!\!\!\!\vdots \\
\!\!\psi_1\Lambda_1^{n-1} &  \!\!\!\!\!f_2^{(\pm)}\Lambda_2^{n-1} & \!\!\!\!\cdots & \!\!\!\!f_j^{(\pm)}\Lambda_j^{n-1} & \!\!\!\!\cdots & \!\!\!\!\!f_n^{(\pm)}\Lambda_n^{n-1} \\
\end{array}
\!\!\right|^{-1}
\nonumber 
%\end{array}  
\end{eqnarray}
which is asymptotic (Cf: \eqref{f_i tends to C_i}, \eqref{tilde_f_i tends to C_i}) to  
\begin{eqnarray}
\!\!\!\!\!\!&&\widetilde{\partial_{\mu}\sigma}   \nonumber \\
\!\!\!\!\!\!&&\begin{array}{l}
=
\left|
\begin{array}{ccccc}
\psi_1 & C_2^{(\pm)} & \cdots & C_n^{(\pm)} & \partial_{\mu}\psi_1 \\
\psi_1 \Lambda_1 & C_2^{(\pm)}\Lambda_2 & \cdots & C_n^{(\pm)}\Lambda_{n} & (\partial_{\mu}\psi_1)\Lambda_1 \\
\psi_1 \Lambda_1^2 & C_2^{(\pm)}\Lambda_2^2 & \cdots & C_n^{(\pm)}\Lambda_{n}^2 &  (\partial_{\mu}\psi_1)\Lambda_1^2 \\
\vdots & \vdots & \ddots & \vdots & \vdots \\
\psi_1 \Lambda_1^n & C_2^{(\pm)}\Lambda_2^{n} & \cdots & C_n^{(\pm)}\Lambda_{n}^{n} & \fbox{$(\partial_{\mu}\psi_1)\Lambda_1^{n}$}
\end{array}
\right|
\left|
\begin{array}{ccccc}
\psi_1 & \cdots & \!\!\!C_j^{(\pm)} & \!\!\cdots& \!\!\!1 \\
\psi_1 \Lambda_1 & \cdots & \!\!\!C_j^{(\pm)}\Lambda_j & \!\!\cdots & \!\!\!0 \\
\vdots & \ddots & \vdots & \!\!\!\ddots & \!\!\!\vdots \\
\psi_1 \Lambda_1^{n-1} & \cdots & \!\!\!C_j^{(\pm)}\Lambda_j^{n-1} & \!\!\!\cdots & \!\!\!0 \\
1 & \cdots & 0 & \cdots & \!\!\!\fbox{$0$}
\end{array}
\!\right|   
\bigskip \\
- \displaystyle{
	\sum_{j=2}^{n}
	\underbrace{
	\left|
	\begin{array}{ccccc}
	\!\psi_1 & \!C_2^{(\pm)} & \!\!\cdots & \!\!\!C_n^{(\pm)} & \!C_j^{(\pm)} \\
	\!\psi_1 \Lambda_1 & \!C_2^{(\pm)}\Lambda_2 & \!\!\cdots & \!\!\!C_n^{(\pm)}\Lambda_{n} & \!C_j^{(\pm)}\Lambda_j \\
	\!\psi_1 \Lambda_1^2 & \!C_2^{(\pm)}\Lambda_2^2 & \!\!\cdots & \!\!\!C_n^{(\pm)}\Lambda_{n}^2 &  \!C_j^{(\pm)}\Lambda_j^2 \\
	\!\vdots & \!\vdots & \!\!\ddots & \!\!\!\vdots & \!\vdots \\
	\!\psi_1 \Lambda_1^n & \!C_2^{(\pm)}\Lambda_2^{n} & \!\!\cdots & \!\!\!C_n^{(\pm)}\Lambda_{n}^{n} & \!\fbox{$C_j^{(\pm)}\Lambda_j^{n}$}
	\end{array}
	\!\right|
}
	\!\widetilde{\Lambda}_j^{(\pm)}
	\left|
	\begin{array}{ccccc}
	\!\psi_1 & \!\!\cdots & \!\!\!\fbox{$C_j^{(\pm)}$} &\!\!\cdots& \!\!\!1 \\
	\!\psi_1 \Lambda_1 & \!\!\cdots & \!\!\!C_j^{(\pm)}\Lambda_j & \!\!\cdots & \!\!\!0 \\
	\!\vdots & \!\!\ddots & \!\!\vdots & \!\!\!\ddots & \!\!\!\vdots \\
	\!\psi_1 \Lambda_1^{n-1} & \!\!\cdots & \!\!\!C_j^{(\pm)}\Lambda_j^{n-1} & \!\!\cdots & \!\!\!0
	\end{array}
	\!\right|^{-1}
}
\end{array} 
  \\
&&~~~~~~~~~~~~~~ =0 ~~ (\mbox{The $j$-th column is identical to the last column})   
\nonumber 
\end{eqnarray}
By \eqref{Reltation of C_iLambda_i} { and \eqref{Rmulti}}, the right common factors $C_j^{(\pm)}$ of each column can be omitted completely, and hence
\begin{eqnarray}
&&\!\!\!\!\!\!\widetilde{\partial_{\mu}\sigma} 
\nonumber \\
&\!\!\!\!=\!\!\!\!&
\left|
\begin{array}{ccccc}
\psi_1 & 1 & \cdots & 1 & \partial_{\mu}\psi_1 \\
\psi_1 \Lambda_1 & \Lambda_2^{(\pm)} & \cdots & \Lambda_{n}^{(\pm)} & (\partial_{\mu}\psi_1)\Lambda_1 \\
\psi_1 \Lambda_1^2 & \Lambda_2^{(\pm)2} & \cdots & \Lambda_{n}^{(\pm)2} &  (\partial_{\mu}\psi_1)\Lambda_1^2 \\
\vdots & \vdots & \ddots & \vdots & \vdots \\
\psi_1 \Lambda_1^n & \Lambda_2^{(\pm)n} & \cdots & \Lambda_{n}^{(\pm)n} & \fbox{$(\partial_{\mu}\psi_1)\Lambda_1^{n}$}
\end{array}
\!\right|
\left|
\begin{array}{ccccc}
\psi_1 & 1 & \!\cdots & \!1 & \!1 \\
\psi_1 \Lambda_1 & \Lambda_2^{(\pm)} & \!\cdots & \!\Lambda_{n}^{(\pm)} & \!0 \\
\vdots & \vdots & \!\ddots & \!\vdots & \!\vdots \\
\psi_1 \Lambda_1^{n-1} & \Lambda_2^{(\pm)n-1} & \!\cdots & \!\Lambda_{n}^{(\pm)n-1} & \!0 \\
1 & 0 & \!\cdots & \!0 & \!\fbox{$0$}
\end{array}
\!\right| ~~~~~~~~
\\
&\!\!\!\!=\!\!\!\!&
\partial_{\mu}\widetilde{\sigma}. 
\end{eqnarray}
We can make the same proof in the asymptotic region $\mathscr{R}$ as well.

\section{Proof of Unitarity for $n$-Soliton Solutions on $\mathbb{E}$}
\label{unitarity}
{\bf Proposition C.1}
\smallskip \\
Let $\sigma_{[n+1]}$ be defined in \eqref{n-soliton solution_appendeix B} with reality condition \eqref{(lambda_j, mu_j)} on $\mathbb{E}$. Then
$\sigma_{[n+1]} \in U(2)$ on the Euclidean space if $|\lambda_j|=1$.
\medskip\\
\noindent
(Proof)
\smallskip \\
For $n=1$, we have
\begin{eqnarray}
\psi_{1}^{\dagger}\psi_1=\psi_1\psi_{1}^{\dagger}=(e^{L_1+\overline{L}_1}+e^{-(L_1 + \overline{L}_1)})I,~
\Lambda_1^{\dagger}\Lambda_1=\Lambda_1\Lambda_1^{\dagger}=|\lambda_1|^2I = I
\end{eqnarray}
which implies
\begin{eqnarray}
\sigma_{[2]}^{\dagger}\sigma_{[2]}=(-\psi_1\Lambda_1\psi_1^{-1})^{\dagger}(-\psi_1\Lambda_1\psi_1^{-1})
=
(\psi_1^{\dagger}\psi_1)(\psi_1\psi_1^{\dagger})^{-1}(\Lambda_1^{\dagger}\Lambda_1)=I
=\sigma_{[2]}\sigma_{[2]}^{\dagger},
\end{eqnarray}
that is, the one-soliton solution $\sigma_{[2]} \in U(2)$.
Assume that the $n$-soliton solution $\sigma_{[n+1]} \in U(2)$ for $1 \leq n \leq k-1$. For $n=k$ and by the Darboux transformation \cite{GHHN}, we have
\begin{eqnarray}
\label{Darboux transf}
\sigma_{[k+1]}= - \psi_{[k]} \Lambda_{[k]}\psi_{[k]}^{-1}\sigma_{[k]},
\end{eqnarray}   
where
\begin{eqnarray}
\psi_{[k]}
:=
\left|
\begin{array}{ccccc}
\psi_1 & \psi_2 & \cdots & \psi_{k-1} & \psi_{k}  \\
\psi_1\Lambda_1 & \psi_2\Lambda_2 & \cdots & \psi_{k-1}\Lambda_{k-1} & \psi_{k}\Lambda_k  \\
\vdots  & \vdots &  & \vdots & \vdots \\
\psi_1\Lambda_1^{k-2} & \psi_2\Lambda_2^{k-2} & \cdots & \psi_{k-1}\Lambda_{k-1}^{k-2} & \psi_{k}\Lambda_k^{k-2} \\
\psi_1\Lambda_1^{k-1} & \psi_2\Lambda_2^{k-1} & \cdots & \psi_{k-1}\Lambda_{k-1}^{k-1} & \fbox{$\psi_{k}\Lambda_k^{k-1}$}
\end{array}
\right|.
\end{eqnarray}
By the Jacobi identity { \eqref{jacobi}}, 
\begin{eqnarray}
\psi_{[k]}
&\!\!\!\!=\!\!\!\!&
\left|
\begin{array}{ccc}
\!\psi_2\Lambda_2 &  \!\!\cdots & \!\psi_{k}\Lambda_{k} \\
\!\vdots & & \vdots \\
\!\psi_2\Lambda_2^{k-1} & \!\!\cdots & \!\fbox{$\psi_{k}\Lambda_{k}^{k-1}$}
\end{array}
\right|
\nonumber \\
&&\!\!\!\!-
\left|
\begin{array}{ccc}
\!\psi_1\Lambda_1 &  \!\!\cdots & \!\!\psi_{k-1}\Lambda_{k-1} \\
\!\vdots & & \!\!\vdots \\
\!\fbox{$\psi_1\Lambda_1^{k-1}$}  &  \!\!\cdots & \!\!\psi_{k-1}\Lambda_{k-1}^{k-1}
\end{array}
\right|
\left|
\begin{array}{ccc}
\!\fbox{$\psi_1$} &  \!\!\cdots & \!\!\psi_{k-1} \\
\!\vdots & & \!\!\vdots \\
\!\psi_1\Lambda_1^{k-2}  &  \!\!\cdots & \!\!\psi_{k-1}\Lambda_{k-1}^{k-2}
\end{array}
\right|^{-1}
\left|
\begin{array}{ccc}
\!\psi_2 &  \!\!\cdots &  \!\!\fbox{$\psi_k$} \\
\!\vdots & & \vdots \\
\!\psi_2\Lambda_2^{k-2}  &  \!\!\cdots & \!\!\psi_k\Lambda_k^{k-2}
\end{array}
\right|
\nonumber \\
&\!\!\!\!=\!\!\!\!&
\underbrace{
\!\!\left[
\begin{array}{l}
~\left|
\begin{array}{ccc}
\psi_2\Lambda_2 &  \cdots & \!\psi_{k}\Lambda_{k} \\
\vdots & & \!\vdots \\
\psi_2\Lambda_2^{k-1} & \cdots & \!\fbox{$\psi_{k}\Lambda_{k}^{k-1}$}
\end{array}
\right|
\left|
\begin{array}{ccc}
\!\psi_2 &  \cdots &  \!\!\fbox{$\psi_k$} \\
\!\vdots & & \!\!\vdots \\
\!\psi_2\Lambda_2^{k-2}  &  \cdots &  \!\!\psi_k\Lambda_k^{k-2}
\end{array}
\right|^{-1}  
\\
\!\!-
\left|
\begin{array}{ccc}
\!\!\psi_1\Lambda_1 &  \!\!\cdots & \!\!\!\psi_{k-1}\Lambda_{k-1} \\
\!\!\vdots & & \!\!\!\vdots \\
\!\!\fbox{$\psi_1\Lambda_1^{k-1}$}  &  \!\!\cdots & \!\!\!\psi_{k-1}\Lambda_{k-1}^{k-1}
\end{array}
\right|
\left|
\begin{array}{ccc}
\!\!\fbox{$\psi_1$} &  \!\!\cdots & \!\!\! \psi_{k-1} \\
\!\!\vdots & & \vdots \\
\!\!\psi_1\Lambda_1^{k-2}  &  \!\!\!\cdots & \!\!\!\psi_{k-1}\Lambda_{k-1}^{k-2}
\end{array}
\!\right|^{-1}
\end{array}
\!\!\right]
}
\!\left|
\begin{array}{ccc}
\!\!\psi_2 &  \!\!\cdots &  \!\!\! \fbox{$\psi_k$} \\
\!\!\vdots & & \!\!\!\vdots \\
\!\!\psi_2\Lambda_2^{k-2}  &  \!\!\cdots &  \!\!\!\psi_k\Lambda_k^{k-2}
\end{array}
\!\right|
\nonumber \\ 
&& ~~~~~~~~~~~~~~~~~~~~\mbox{By the Jacobi identity { \eqref{jacobi}}}
\nonumber \\
&\!\!\!\!=\!\!\!\!&
\!\!\left[
\left|
\begin{array}{cccc}
\!\!\psi_2 & \!\!\!\cdots & \!\!\psi_k & \!\!\!1 \\
\!\!\psi_2\Lambda_2 & \!\!\!\cdots & \!\!\psi_k\Lambda_k & \!\!\!0 \\
\!\!\vdots & & \!\!\!\vdots & \!\!\!\vdots \\
\!\!\psi_2\Lambda_2^{k-1} & \!\!\!\cdots & \!\!\psi_k\Lambda_k^{k-1} & \!\!\!\fbox{$0$}
\end{array}
\right|
-
\left|
\begin{array}{cccc}
\!\!\psi_1 & \!\!\!\cdots & \!\!\psi_{k-1} & \!\!\!1 \\
\!\!\psi_1\Lambda_1 & \!\!\!\cdots & \!\!\psi_{k-1}\Lambda_{k-1} & \!\!\!0 \\
\!\!\vdots & & \vdots & \vdots \\
\!\!\psi_1\Lambda_1^{k-1} & \!\!\!\cdots & \!\!\psi_k\Lambda_{k-1}^{k-1} & \!\!\!\fbox{$0$}
\end{array}
\right|
\right]
\!\left|
\begin{array}{ccc}
\!\!\psi_2 &  \!\!\cdots &  \!\!\! \fbox{$\psi_k$} \\
\!\!\vdots & & \!\!\!\vdots \\
\!\!\psi_2\Lambda_2^{k-2}  &  \!\!\cdots &  \!\!\! \psi_k\Lambda_k^{k-2}
\end{array}
\!\right| 
\nonumber \\
&\!\!\!\!=:\!\!\!\!&
(\widetilde{\sigma}_{[k]} - \sigma_{[k]})\widetilde{\psi}_{[k-1]}, 
\label{Recurrece_psi}
\end{eqnarray}
where
$\widetilde{\sigma}_{[k]} :=\sigma_{[k]} \Big|_{(\psi_1, \Lambda_1) \rightarrow (\psi_k, \Lambda_k)} \in U(2)$.
On the other hand, 
\begin{eqnarray}
&&\!\!\!\! \widetilde{\psi}_{[k-1]}\Lambda_{k}\widetilde{\psi}_{[k-1]}^{-1}
\nonumber \\
&\!\!\!\! = \!\!\!\!&
-
\underbrace{
\!\left|
\begin{array}{ccc}
\!\!\psi_2 &  \!\!\cdots &  \!\!\! \fbox{$\psi_k$} \\
\!\!\vdots & & \!\!\!\vdots \\
\!\!\psi_2\Lambda_2^{k-2}  &  \!\!\cdots &  \!\!\! \psi_k\Lambda_k^{k-2}
\end{array}
\!\right|
} 
\Lambda_k
\!\left|
\begin{array}{ccc}
\!\!\psi_2 &  \!\!\cdots &  \!\!\! \fbox{$\psi_k$} \\
\!\!\vdots & & \!\!\!\vdots \\
\!\!\psi_2\Lambda_2^{k-2}  &  \!\!\cdots &  \!\!\! \psi_k\Lambda_k^{k-2}
\end{array}
\!\right|^{-1} 
\nonumber \\
&& ~~~~~~~~~~~~~~~~||~\mbox{By the homological relation { \eqref{homological}}}
\nonumber \\
&&~~~
\!\left|
\begin{array}{cccc}
\!\!\psi_2 &  \!\!\cdots &  \!\!\! \psi_{k-1} & \!\!\!\fbox{$0$} \\
\!\!\vdots & & \!\!\!\vdots  & \!\!\!\vdots \\
\!\!\psi_2\Lambda_2^{k-2}  &  \!\!\cdots &  \!\!\! \psi_{k-1}\Lambda_{k-1}^{k-2} & \!\!\! 1
\end{array}
\!\right|
\!\left|
\begin{array}{ccc}
\!\!\psi_2 &  \!\!\cdots &  \!\!\! \psi_k \\
\!\!\vdots & & \!\!\!\vdots \\
\!\!\psi_2\Lambda_2^{k-2}  &  \!\!\cdots &  \!\!\! \fbox{$\psi_k\Lambda_k^{k-2}$}
\end{array}
\!\right|
\nonumber \\
&\!\!\!\!=\!\!\!\!&
\!\left|
\begin{array}{cccc}
\!\!\psi_2 &  \!\!\cdots &  \!\!\! \psi_{k-1} & \!\!\!1 \\
\!\!\vdots & & \!\!\!\vdots  & \!\!\!\vdots \\
\!\!\psi_2\Lambda_2^{k-2}  &  \!\!\cdots &  \!\!\! \psi_{k-1}\Lambda_{k-1}^{k-2} & \!\!\! \fbox{$0$}
\end{array}
\!\right|^{-1}
\!\!\left\{
\underbrace{
-\!\left|
\begin{array}{ccc}
\!\!\psi_2\Lambda_2 &  \!\!\cdots &  \!\!\! \psi_k\Lambda_k \\
\!\!\vdots & & \!\!\!\vdots \\
\!\!\psi_2\Lambda_2^{k-1}  &  \!\!\cdots &  \!\!\! \fbox{$\psi_k\Lambda_k^{k-1}$}
\end{array}
\!\right|
\!\left|
\begin{array}{ccc}
\!\!\psi_2 &  \!\!\cdots &  \!\!\! \fbox{$\psi_k$} \\
\!\!\vdots & & \!\!\!\vdots \\
\!\!\psi_2\Lambda_2^{k-2}  &  \!\!\cdots &  \!\!\! \psi_k\Lambda_k^{k-2}
\end{array}
\!\right|^{-1} 
}
\right\}
\nonumber \\
&& ~~~~~~~~~~~~~~~~~~~~~~~~~~~~~~~~~~~~~~~~~~~~~~~~~~~~~~~~~~~~~~~~~|| ~\mbox{By the Jacobi identity { \eqref{jacobi}}}
\nonumber \\
&&~~~~~~~~~~~~~~~~~~~~~~~~~~~~~~~~~~~~~~~~~~~~~~~~~~~
\!\left|
\begin{array}{cccc}
\!\!\psi_2 &  \!\!\cdots &  \!\!\! \psi_{k} & \!\!\!1 \\
\!\!\vdots & & \!\!\!\vdots  & \!\!\!\vdots \\
\!\!\psi_2\Lambda_2^{k-1}  &  \!\!\cdots &  \!\!\! \psi_{k}\Lambda_{k}^{k-1} & \!\!\! \fbox{$0$}
\end{array}
\!\right|
\nonumber \\
&\!\!\!\!=\!\!\!\!&
\widetilde{\sigma}_{[k-1]}^{-1}\widetilde{\sigma}_{[k]} \in U(2).
\label{identity_psi-sigma}
\end{eqnarray}
Note that the second equality from the bottom is obtained by using the homological relation { \eqref{homological}} and { \eqref{Rmulti}}.
By \eqref{Darboux transf}, \eqref{Recurrece_psi} and \eqref{identity_psi-sigma}, we can conclude that
\begin{eqnarray}
\sigma_{[k+1]}=
(\widetilde{\sigma}_{[k]} - \sigma_{[k]})
\widetilde{\sigma}_{[k-1]}^{-1}\widetilde{\sigma}_{[k]}
(\widetilde{\sigma}_{[k]} - \sigma_{[k]})^{-1}
\sigma_{[k]} \in U(2).
\end{eqnarray}
We remark that in the case of $\mathbb{E}$ the condition $\vert \lambda_j\vert=1$ comes from the condition that $\Lambda_j\Lambda_j^\dagger$ is a scalar matrix together with the reality condition $\mu_j=-1/\overline{\lambda}_j$. In the case of $\mathbb{U}_2$, the condition that $\Lambda_j\Lambda_j^\dagger$ is a scalar matrix and the reality condition $\mu_j=+1/\overline{\lambda}_j$ leads to trivial solutions because $\Lambda_j$ becomes a scalar matrix and hence the Darboux transformation becomes trivial. 

\newpage

\section{Miscellaneous Formula}

\subsection{Flip Symmetry of $n$-Soliton Solutions}

\vspace{2mm}
\noindent
{\bf Proposition D.1} 
\smallskip \\
The data of the $n$-soliton solutions has the following symmetry:
\begin{eqnarray}
\label{Symmetry of Delta, Delta_ij}
\left( \Delta, \Delta_{11}, \Delta_{12}, \Delta_{21}, \Delta_{22}  \right) 
\bigg|_{L_j \rightarrow -L_j}
&=&
\left( \Delta, \Delta_{22}, \Delta_{21}, \Delta_{12}, \Delta_{11}  \right),
\\
\partial_{\mu}\left(\Delta, \Delta_{11}, \Delta_{12}, \Delta_{21}, \Delta_{22}  \right) 
\bigg|_{L_j \rightarrow -L_j}
&=& \partial_{\mu}\left( \Delta, \Delta_{22}, \Delta_{21}, \Delta_{12}, \Delta_{11}  \right) .
\end{eqnarray}
\noindent
(Proof) 
Let $\widetilde{\psi}_j$ be defined by $\widetilde{\psi}_j:=\psi_j\bigg|_{L_j \rightarrow -L_j}
=   
\left(
\!\!\begin{array}{cc}
e^{-L_j} & e^{\overline{L}_j} \\
-e^{L_j} & e^{-\overline{L}_j}
\end{array}
\!\!\right)$ 
which satisfies 
\begin{eqnarray*}
\widetilde{\psi}_j \Lambda_j^{k}
=
E (\psi_j\Lambda_j^{k}) F, ~~
E:=
\left(
\begin{array}{cc}
0 & 1 \\
1 & 0
\end{array}
\right),~~
F:=
\left(
\begin{array}{cc}
-1 & 0 \\
 0 & 1
\end{array}
\right).
\end{eqnarray*}
Then
\begin{eqnarray*}
\sigma\bigg|_{L_j \rightarrow -L_j}
&\!\!\!=\!\!\!&
\left|
\begin{array}{cccccccc}
\widetilde{\psi}_1 & \widetilde{\psi}_2 & \cdots & \widetilde{\psi}_n & 1 \\
\widetilde{\psi}_1 \Lambda_1 & \widetilde{\psi}_2 \Lambda_2 & \cdots & \widetilde{\psi}_n \Lambda_{n} & 0 \\
\vdots & \vdots & \ddots & \vdots & \vdots \\
\widetilde{\psi}_1 \Lambda_1^{n} & \widetilde{\psi}_2\Lambda_2^{n} & \cdots & \widetilde{\psi}_n \Lambda_{n}^{n} & $\fbox{0}$
\end{array}
\right|
\\
&\!\!\!=\!\!\!&
\left|
\begin{array}{cccccccc}
E (\psi_1) F & E (\psi_2) F & \cdots & E (\psi_n)  F & E 1 E \\
E (\psi_1 \Lambda_1) F & E (\psi_2 \Lambda_2) F & \cdots & E (\psi_n \Lambda_{n}) F & E0E \\
\vdots & \vdots & \ddots & \vdots & \vdots \\
E (\psi_1 \Lambda_1^{n}) F & E (\psi_2\Lambda_2^{n}) F & \cdots & E (\psi_n \Lambda_{n}^{n}) F & $\fbox{$E0E$}$
\end{array}
\right|
\\
&\!\!\!=\!\!\!&
E\left|
\begin{array}{cccccccc}
\psi_1 & \psi_2 & \cdots & \psi_n & 1 \\
\psi_1 \Lambda_1 & \psi_2 \Lambda_2 & \cdots & \psi_n \Lambda_{n} & 0 \\
\vdots & \vdots & \ddots & \vdots & \vdots \\
\psi_1 \Lambda_1^{n} & \psi_2\Lambda_2^{n} & \cdots & \psi_n \Lambda_{n}^{n} & $\fbox{0}$
\end{array}
\right|E
=E \sigma E
\end{eqnarray*}
by the multiplicative rule { \eqref{Rmulti} and \eqref{Lmulti}} of the quasideterminant. (Common multiplier of rows is E, common multipliers { of} columns are F and E.)
Now we have
\begin{eqnarray*}
\sigma\bigg|_{L_j \rightarrow -L_j}
=
\left(
\begin{array}{cc}
0 & 1 \\
1 & 0
\end{array}
\right)
\frac{-1}{\Delta}
\left(
\begin{array}{cc}
\Delta_{11} & \Delta_{12} \\
\Delta_{21} & \Delta_{22}
\end{array}
\right)
\left(
\begin{array}{cc}
0 & 1 \\
1 & 0
\end{array}
\right)
=
\frac{-1}{\Delta}
\left(
\begin{array}{cc}
\Delta_{22} & \Delta_{21} \\
\Delta_{12} & \Delta_{11}
\end{array}
\right).
\end{eqnarray*}
Therefore, $\left( \Delta, \Delta_{11}, \Delta_{12}, \Delta_{21}, \Delta_{22}  \right) 
\bigg|_{L_j \rightarrow -L_j}
=
\left( \Delta, \Delta_{22}, \Delta_{21}, \Delta_{12}, \Delta_{11}  \right)$.
%&&\mbox{(Cf: \eqref{Data of J_2}, \eqref{Data of J_3}.)}  \nonumber 

By the fact that 
$
\dfrac{df(-x)}{dx}=-\dfrac{df(x)}{dx}\bigg|_{x \rightarrow -x},
$
we have
\begin{eqnarray*}
&&\left( \partial_{\mu}\Delta, \partial_{\mu}\Delta_{11}, \partial_{\mu}\Delta_{12}, \partial_{\mu}\Delta_{21}, \partial_{\mu}\Delta_{22}  \right) 
\bigg|_{L_j \rightarrow -L_j}
\\
&\!\!\!\!\!\!=\!\!\!\!\!\!&
-\partial_{\mu}\left(
\Delta\big|_{L_j \rightarrow -L_j}, \Delta_{11}\big|_{L_j \rightarrow -L_j}, 
\Delta_{12}\big|_{L_j \rightarrow -L_j}, \Delta_{21}\big|_{L_j \rightarrow -L_j}, 
\Delta_{22}\big|_{L_j \rightarrow -L_j}  \right) 
\\
&\!\!\!\!\!\!=\!\!\!\!\!\!&
-\partial_{\mu}\left( \Delta, \Delta_{22}, \Delta_{21}, \Delta_{12}, \Delta_{11}  \right)
~~\mbox{by $\eqref{Symmetry of Delta, Delta_ij}$}
\\
&\!\!\!\!\!\!=\!\!\!\!\!\!&
-\left( \partial_{\mu}\Delta, \partial_{\mu}\Delta_{22}, \partial_{\mu}\Delta_{21}, \partial_{\mu}\Delta_{12}, \partial_{\mu}\Delta_{11}  \right) .
\end{eqnarray*}

\vspace{2mm}
\noindent
{\bf Corollary D.2} 
\smallskip \\
Proposition D.1 implies:

\noindent
(1) $\mbox{Tr}\left[(\partial_{\mu}\sigma)\sigma^{-1}(\partial_{\nu}\sigma)\sigma^{-1}\right]
$
is an even function with respect to $L_j$.

\noindent
(2) $\mbox{Tr}\left[(\partial_{\mu}\sigma)\sigma^{-1}(\partial_{\nu}\sigma)\sigma^{-1}(\partial_{\rho}\sigma)\sigma^{-1}\right]$
is an odd function with respect to $L_j$.
\medskip \\
%\noindent
(Proof) 
\smallskip \\
(1) \vspace{-10mm}
\begin{eqnarray*}
&&\mbox{Tr}\left[(\partial_{\mu}\sigma)\sigma^{-1}(\partial_{\nu}\sigma)\sigma^{-1}\right]
\bigg|_{L_j \rightarrow -L_j}  
\\
&\!\!\!\!\!=\!\!\!\!\!&
\frac{-1}{|\sigma|\Delta^{2}}
\left\{
\left|
\begin{array}{cc}
\partial_{\mu} \Delta_{11} & \partial_{\mu} \Delta_{12} \\
\partial_{\nu} \Delta_{21} & \partial_{\nu} \Delta_{22}
\end{array}
\right|
+
\left|
\begin{array}{cc}
\partial_{\nu} \Delta_{11} & \partial_{\nu} \Delta_{12} \\
\partial_{\mu} \Delta_{21} & \partial_{\mu} \Delta_{22}
\end{array}
\right|
-
2|\sigma|(\partial_{\mu} \Delta)(\partial_{\nu} \Delta)
\right\}
\bigg|_{L_j \rightarrow -L_j}
\\
&\!\!\!\!\!\!=\!\!\!\!\!\!&
\frac{-1}{|\sigma|\Delta^{2}}
\left\{
\left|
\begin{array}{cc}
-\partial_{\mu} \Delta_{22} & -\partial_{\mu} \Delta_{21} \\
-\partial_{\nu} \Delta_{12} & -\partial_{\nu} \Delta_{11}
\end{array}
\right|
+
\left|
\begin{array}{cc}
-\partial_{\nu} \Delta_{22} & -\partial_{\nu} \Delta_{21} \\
-\partial_{\mu} \Delta_{12} & -\partial_{\mu} \Delta_{11}
\end{array}
\right|
-
2|\sigma|(-\partial_{\mu} \Delta)(-\partial_{\nu} \Delta)
\right\}  ~~~~~~~~
\\
&\!\!\!\!\!=\!\!\!\!\!&
\frac{-1}{|\sigma|\Delta^{2}}
\left\{
\left|
\begin{array}{cc}
\partial_{\nu} \Delta_{11} & \partial_{\nu} \Delta_{12} \\
\partial_{\mu} \Delta_{21} & \partial_{\mu} \Delta_{22}
\end{array}
\right|
+
\left|
\begin{array}{cc}
\partial_{\mu} \Delta_{11} & \partial_{\mu} \Delta_{12} \\
\partial_{\nu} \Delta_{21} & \partial_{\nu} \Delta_{22}
\end{array}
\right|
-
2|\sigma|(\partial_{\mu} \Delta)(\partial_{\nu} \Delta)
\right\}
\\
&\!\!\!\!\!=\!\!\!\!\!&
\mbox{Tr}\left[(\partial_{\mu}\sigma)\sigma^{-1}(\partial_{\nu}\sigma)\sigma^{-1}\right].
\end{eqnarray*}
(2)
\vspace{-10mm}
\begin{eqnarray*}
&&\mbox{Tr}\left[(\partial_{\mu}\sigma)\sigma^{-1}(\partial_{\nu}\sigma)\sigma^{-1}(\partial_{\rho}\sigma)\sigma^{-1}\right]
\bigg|_{L_j \rightarrow -L_j }   \\
&\!\!\!\!\!\! = \!\!\!\!\!\!&
\frac{1}{2|\sigma|^2\Delta^{4}}
\left|
\begin{array}{cccc}
\Delta_{11} & \Delta_{12} & \Delta_{21} & \Delta_{22} \\
\partial_{\mu}\Delta_{11} & \partial_{\mu} \Delta_{12} & \partial_{\mu} \Delta_{21} & \partial_{\mu} \Delta_{22} \\
\partial_{\nu} \Delta_{11} & \partial_{\nu} \Delta_{12} & \partial_{\nu} \Delta_{21} & \partial_{\nu} \Delta_{22} \\
\partial_{\rho} \Delta_{11} & \partial_{\rho} \Delta_{12} & \partial_{\rho} \Delta_{21} & \partial_{\rho} \Delta_{22}
\end{array}
\right|_{L_j \rightarrow -L_j}   
\\
&\!\!\!\!\!\!=\!\!\!\!\!\!&
\frac{1}{2|\sigma|^2\Delta^{4}}
\left|
\begin{array}{cccc}
\Delta_{22} & \Delta_{21} & \Delta_{12} & \Delta_{11} \\
-\partial_{\mu}\Delta_{22} & -\partial_{\mu} \Delta_{21} & -\partial_{\mu} \Delta_{12} & -\partial_{\mu} \Delta_{11} \\
-\partial_{\nu} \Delta_{22} & -\partial_{\nu} \Delta_{21} & -\partial_{\nu} \Delta_{12} & -\partial_{\nu} \Delta_{11} \\
-\partial_{\rho} \Delta_{22} & -\partial_{\rho} \Delta_{21} & -\partial_{\rho} \Delta_{12} & -\partial_{\rho} \Delta_{11}
\end{array}
\right|
\\
&\!\!\!\!\!\!=\!\!\!\!\!\!&
\frac{(-1)^3}{2|\sigma|^2\Delta^{4}}
\left|
\begin{array}{cccc}
\Delta_{22} & \Delta_{21} & \Delta_{12} & \Delta_{11} \\
\partial_{\mu}\Delta_{22} & \partial_{\mu} \Delta_{21} & \partial_{\mu} \Delta_{12} & \partial_{\mu} \Delta_{11} \\
\partial_{\nu} \Delta_{22} & \partial_{\nu} \Delta_{21} & \partial_{\nu} \Delta_{12} & \partial_{\nu} \Delta_{11} \\
\partial_{\rho} \Delta_{22} & \partial_{\rho} \Delta_{21} & \partial_{\rho} \Delta_{12} & \partial_{\rho} \Delta_{11}
\end{array}
\right|
=
%\\
%&\!\!\!\!\!\!=\!\!\!\!\!\!&
\frac{-1}{2|\sigma|^2\Delta^{4}}
\left|
\begin{array}{cccc}
\Delta_{11} & \Delta_{12} & \Delta_{21} & \Delta_{22} \\
\partial_{\mu}\Delta_{11} & \partial_{\mu} \Delta_{12} & \partial_{\mu} \Delta_{21} & \partial_{\mu} \Delta_{22} \\
\partial_{\nu} \Delta_{11} & \partial_{\nu} \Delta_{12} & \partial_{\nu} \Delta_{21} & \partial_{\nu} \Delta_{22} \\
\partial_{\rho} \Delta_{11} & \partial_{\rho} \Delta_{12} & \partial_{\rho} \Delta_{21} & \partial_{\rho} \Delta_{22}
\end{array}
\right|
\\
&\!\!\!\!\!\!=\!\!\!\!\!\!&
-\mbox{Tr}\left[(\partial_{\mu}\sigma)\sigma^{-1}(\partial_{\nu}\sigma)\sigma^{-1}(\partial_{\rho}\sigma)\sigma^{-1}\right].
\end{eqnarray*}
%which is an odd function with respect to $L_j$. Therefore, the integrands \eqref{WZ term_real spaces} of the Wess-Zumino term are clearly even functions with respect to $L_i$.

\subsection{Data of $n$-Soliton Solutions}

In this subsection we present the data of $n$-soliton solutions in terms of $X_j, \Theta_{j}$. 
We introduce the convention $\varepsilon_{j}$ which takes values in $\left\{ \pm 1, \pm i \right\}$ and define an informal symbol $\mathcal{P}\{ (\varepsilon_1, \cdots ,\varepsilon_n) \}$ to denote the set of all permutations of $(\varepsilon_1, \cdots ,\varepsilon_n)$.
The data of $n$-soliton solutions can be expressed formally as the following $\Delta$, $\Delta_{ij}$ with some undetermined coefficients $a(\boldsymbol{\varepsilon}):=a(\varepsilon_1, \cdots, \varepsilon_n)$ and $A(\boldsymbol{\varepsilon}):=A(\varepsilon_1, \cdots, \varepsilon_n)$:
\begin{eqnarray*}
\Delta  
&\!\!\!=\!\!\!&
\displaystyle{\sum_{\footnotesize{\begin{array}{c} 1 \leq j \leq n, \\ \varepsilon = \pm 1  \end{array}}} \!\!\!\!\!a(\boldsymbol{\varepsilon})\exp{\left(\sum_{j=1}^{n} \varepsilon_{j}X_{j}\right)}}
\\
&&\!\! + 
\!\displaystyle{
	\sum_{\footnotesize{\begin{array}{c} 1 \leq k \textless \ell \leq n, \\ 
	\varepsilon_j = \pm 1, j \neq k, \ell  
			\\ 
	 (\varepsilon_{k}, \varepsilon_{\ell}) \in \mathcal{P}\{(i, -i)\} \end{array} }} \!\!\!\!\!\!\!\!\!\!\!\!\!\!\!\!a(\boldsymbol{\varepsilon})\exp({\sum_{ j \neq k, \ell}^{n} \varepsilon_{j}X_{j} + \varepsilon_{k}\Theta_{k} + \varepsilon_{\ell}\Theta_{\ell}})
} 
+ ~ \mbox{terms involving more $i\Theta_j$'s} \nonumber  \\
&\!\!\!\! = \!\!\!\!&
\frac{1}{2}
\displaystyle{\sum_{\footnotesize{\begin{array}{c} 1 \leq j \leq n, \\ \varepsilon_j = \pm 1 \end{array}}} 
	\left[
	\begin{array}{l}
	\displaystyle{a(\boldsymbol{\varepsilon})\exp{(\sum_{j=1}^{n} \varepsilon_{j}X_{j})}
	}
+\displaystyle{a(-\boldsymbol{\varepsilon})\exp{(-\sum_{j=1}^{n} \varepsilon_{j}X_{j})}}
\end{array}
\right]
}
\\
&&\!\! + 
\underbrace{
\frac{1}{2}
\!\!\!\!\!\!\!\!\displaystyle{
	\sum_{\footnotesize{\begin{array}{c} 1 \leq k \textless \ell \leq n, \\ 
			\varepsilon_j = \pm 1, j \neq k, \ell  
			\\ 
			(\varepsilon_{k}, \varepsilon_{\ell}) \in \mathcal{P}\{(i, -i)\} \end{array} }}
	 \!\!\!\!\left[
	 \begin{array}{l}
	~~\displaystyle{a(\boldsymbol{\varepsilon})\exp({\sum_{ j \neq k, \ell}^{n} \varepsilon_{j}X_{j} + \varepsilon_{k}\Theta_{k} + \varepsilon_{\ell}\Theta_{\ell}})}
	 \\
	 +\displaystyle{a(-\boldsymbol{\varepsilon})\exp(-{\sum_{ j \neq k, \ell }^{n} \varepsilon_{j}X_{j} - \varepsilon_{k}\Theta_{k} - \varepsilon_{\ell}\Theta_{\ell}})
	}
	 \end{array}
	 \right] + \cdots
}
}
\nonumber \\
&& ~~~~~~~~~~~~~~~~~~~~ \sum_{\footnotesize{\begin{array}{l} 1\leq k \textless \ell \leq n, \\ \varepsilon_{j, j \neq k, \ell} = \pm 1 \end{array}} }\!\!\!\!\!\!\mathcal{O}\left(\cosh(\sum_{j=1, j \neq k, \ell}^{n}\varepsilon_jX_j)\right)
\end{eqnarray*}
By symmetry \eqref{Symmetry of Delta, Delta_ij (2)}, $\Delta\Big|_{(X_j, \Theta_j) \rightarrow (-X_j, \Theta_j)} = \Delta$  implies
$
a(\boldsymbol{\varepsilon}) = a(-\boldsymbol{\varepsilon})
$,
and hence
\begin{eqnarray}
\Delta   \label{Data of n soliton_Delta}
=
\!\!\!\!\displaystyle{\sum_{\footnotesize{\begin{array}{c} 1 \leq j \leq n, \\ \varepsilon_j = \pm 1 \end{array}}}} 
	\!\!\!\!a(\boldsymbol{\varepsilon})	\cosh{\left(\sum_{j=1}^{n} \varepsilon_{j}X_{j}\right)}
+
\!\!\!\!\sum_{\footnotesize{\begin{array}{l} 1\leq k \textless \ell \leq n, \\ \varepsilon_{j, j \neq k, \ell} = \pm 1 \end{array}} }\!\!\!\!\!\!\mathcal{O}\left(\cosh(\sum_{j=1, j \neq \ell, m}^{n}\!\!\!\!\varepsilon_jX_j)\right).  ~~
\end{eqnarray}
Similarly, we have
\begin{eqnarray}
\Delta_{11}
&\!\!\!\!=\!\!\!\!&
\displaystyle{\sum_{\footnotesize{\begin{array}{c} 1 \leq j \leq n, \\ \varepsilon_j = \pm 1 \end{array}}} \!\!\!\!\!\!\!\!A(\boldsymbol{\varepsilon})\exp{\left(\sum_{j=1}^{n} \varepsilon_{j}X_{j}\right)}}
\nonumber
\\
&&\!\!\!\! + 
\!\!\!\!\displaystyle{
	\sum_{\footnotesize{\begin{array}{c} 1 \leq k \textless \ell \leq n, \\ 
			\varepsilon_j = \pm 1, j \neq k, \ell  
			\\ 
			(\varepsilon_{k}, \varepsilon_{\ell}) \in \mathcal{P}\{(i, -i)\} \end{array} }} \!\!\!\!\!\!\!\!\!\!\!\!\!\!\!\!\!\!\!\!A(\boldsymbol{\varepsilon})\exp({\sum_{ j \neq k, \ell}^{n} \varepsilon_{j}X_{j} + \varepsilon_{k}\Theta_{k} + \varepsilon_{\ell}\Theta_{\ell}})
}
+  \mbox{terms involving more $i\Theta_j$'s} \nonumber  \\
&\!\!\!\! = \!\!\!\!&
\frac{1}{2}
\displaystyle{\sum_{\footnotesize{\begin{array}{c} 1 \leq j \leq n, \\ \varepsilon_j = \pm 1 \end{array}}} 
	\!\!\left[
	\begin{array}{l}
	\displaystyle{A(\boldsymbol{\varepsilon})\exp{(\sum_{j=1}^{n} \varepsilon_{j}X_{j})}
	}
	+\displaystyle{A(-\boldsymbol{\varepsilon})\exp{(-\sum_{j=1}^{n} \varepsilon_{j}X_{j})}}
	\end{array}
	\right]    \label{Data of n soliton_Delta_11}
}
\nonumber
\\
&&\!\!\!\! + 
\underbrace{
	\frac{1}{2}
	\!\!\!\!\!\!\displaystyle{
		\sum_{\footnotesize{\begin{array}{c} 1 \leq k \textless \ell \leq n, \\ 
				\varepsilon_j = \pm 1, j \neq k, \ell  
				\\ 
				(\varepsilon_{k}, \varepsilon_{\ell}) \in \mathcal{P}\{(i, -i)\} \end{array} }}
		\!\!\!\!\!\left[
		\begin{array}{l}
		~~\displaystyle{A(\boldsymbol{\varepsilon})\exp({\sum_{ j \neq k, \ell }^{n} \varepsilon_{j}X_{j} + \varepsilon_{k}\Theta_{k} + \varepsilon_{\ell}\Theta_{\ell}})}
		\\
		+\displaystyle{A(-\boldsymbol{\varepsilon})\exp(-{\sum_{ j \neq k, \ell}^{n} \varepsilon_{j}X_{j} - \varepsilon_{k}\Theta_{k} - \varepsilon_{\ell}\Theta_{\ell}})
		}
		\end{array}
		\!\!\right] + \cdots
	}
}
\nonumber \\
&& ~~~~~~~~~~~~~~~~~~~~ \sum_{\footnotesize{\begin{array}{l} 1\leq k \textless \ell \leq n, \\ \varepsilon_{j, j \neq k, \ell} = \pm 1 \end{array}} }\!\!\!\!\!\!\mathcal{O}\left(\cosh(\sum_{j=1, j \neq k, \ell}^{n}\varepsilon_jX_j)\right)
\end{eqnarray}
By symmetry \eqref{Symmetry of Delta, Delta_ij (2)}, $\Delta_{22}=\Delta_{11}\Big|_{(X_j, \Theta_j) \rightarrow (-X_j, \Theta_j)}$ implies that
%\begin{eqnarray}
\label{Data of n soliton_Delta_22}
$\Delta_{22} = \Delta_{11}\Big|_{A(\boldsymbol{\varepsilon}) \rightarrow A(-\boldsymbol{\varepsilon})}$.   
%\end{eqnarray}
\begin{eqnarray}
\Delta_{12}  \label{Data of n soliton_Delta_12}
&\!\!\!\!=\!\!\!\!&
\displaystyle{
	\sum_{\footnotesize{\begin{array}{c} 1 \leq k \leq n, \\ 
			\varepsilon_j = \left\{
			\begin{array}{l} \pm 1, j \neq k  
			\\ 
			+i, j = k \end{array}\right. \end{array}}} \!\!\!\!\!\!\!\!\!\!\!\!\!\!\!\!\!\!\!\!A(\boldsymbol{\varepsilon})\exp({\sum_{ j \neq k}^{n} \varepsilon_{j}X_{j} + \varepsilon_{k}\Theta_{k}})
}
\nonumber
\\
&& \!\!\!\!+
\left\{
\underbrace{
\begin{array}{l}
\displaystyle{
\!\!\!\!\sum_{\footnotesize{\begin{array}{c} 1 \leq k \textless \ell  \textless m \leq n, \\ 
			\varepsilon_j = \pm 1, j \neq k, \ell, m  
			\\ 
			(\varepsilon_{k}, \varepsilon_{\ell}, \varepsilon_{m}) \in \mathcal{P}\{(i, i ,-i)\} \end{array} }} \!\!\!\!\!\!\!\!\!\!\!\!\!\!\!\!\!\!\!\!A(\boldsymbol{\varepsilon})\exp({\sum_{ j \neq k, \ell, m}^{n} \varepsilon_{j}X_{j} + \varepsilon_{k}\Theta_{k} + \varepsilon_{\ell}\Theta_{\ell} + \varepsilon_{m}\Theta_{m}})
}
\\
+  ~\mbox{terms involving more $i\Theta_j$'s}  
\end{array}
}
\right\}
\nonumber \\
&& ~~~~~~~~~~~~~~~~~~~~ \sum_{\footnotesize{\begin{array}{l} 1\leq k \textless \ell \textless m \leq n, \\ \varepsilon_{j, j \neq k, \ell, m} = \pm 1 \end{array}} }\!\!\!\!\!\!\mathcal{O}\left(\cosh(\sum_{j=1, j \neq k, \ell, m}^{n}\varepsilon_jX_j)\right)
\end{eqnarray}
By symmetry \eqref{Symmetry of Delta, Delta_ij (2)}, 
%\begin{eqnarray}
\label{Data of n soliton_Delta_21}
$\Delta_{21}=\Delta_{12}\Big|_{(X_j, \Theta_j) \rightarrow (-X_j, \Theta_j)}$.
%\end{eqnarray}
Concrete examples are mentioned as follows:
\medskip \\
$\bullet$~{\bf The data of one-soliton solutions:}
\begin{eqnarray}
\label{Data of J_2}
\begin{array}{l}
\Delta=a(1)\left(e^{X_1} + e^{-X_1}\right),~  
\smallskip \\ 
\Delta_{11}=A(1)e^{X_1} + A(-1)e^{-X_1},~~~
\Delta_{22}=A(-1) e^{X_1} + A(1) e^{-X_1},~ 
\smallskip \\
\Delta_{12}=A(i)e^{i\Theta_1},~~~~~~~~~~~~~~~~~~~~~
\Delta_{21}=A(i)e^{-i\Theta_1},
\end{array} 
\end{eqnarray}
where
$%\begin{eqnarray}
a(1)=1, ~~ A(1)=\lambda_1, ~~ A(-1)=\mu_1, ~~ A(i)=-(\lambda_1 - \mu_1).
$
%\end{eqnarray}
\medskip \\
$\bullet$~ {\bf The data of two-soliton solutions ($\Theta_{jk}:=\Theta_j - \Theta_k, ~\! j,k =1,2$) :}  
\begin{eqnarray}
\label{Data of J_3}
\begin{array}{l}
\Delta=
\left\{
\begin{array}{l}
~~a(1,1)\left(e^{X_1 + X_2} + e^{-(X_1 + X_2} \right)  
+a(1,-1)\left(e^{X_1 - X_2} + e^{-(X_1 - X_2} \right) 
\smallskip \\
+a(i,-i)\left(e^{i\Theta_{12}} + e^{-i\Theta_{12}} \right)
\end{array} 
\right\},  
\medskip \\
\Delta_{11}
=
\left\{
\begin{array}{l}
~~\left[A(1,1)e^{X_1 + X_2} + A(-1,-1)e^{-(X_1 + X_2)}\right]
\smallskip \\
+\left[A(1,-1)e^{X_1 - X_2} + A(-1,1)e^{-(X_1 - X_2)}\right]  
\smallskip \\
+\left[A(i,-i)e^{i\Theta_{12}} + A(-i,i)e^{-i\Theta_{12}}\right]
\end{array}
\right\},   
\medskip \\
\Delta_{22}=\Delta_{11}\Big|_{(X_j, \Theta_j) \rightarrow (-X_j, -\Theta_j)}
=\Delta_{11}\Big|_{A(\varepsilon_1, \varepsilon_2) \rightarrow A(-\varepsilon_1, -\varepsilon_2)},
\bigskip \\
\Delta_{12}
=
\left\{
\begin{array}{l}
~~A(1,i)e^{X_1 + i\Theta_2}  \smallskip +A(-1,i)e^{-X_1 + i\Theta_2}  \smallskip \\
+A(i,1)e^{X_2 + i\Theta_1} + A(i,-1)e^{-X_2 + i\Theta_1} 
\end{array}
\right\}, 
\medskip \\
\Delta_{21}=\Delta_{12}\Big|_{(X_j, \Theta_j) \rightarrow (-X_j, -\Theta_j)}.
\end{array} ~~
\end{eqnarray}
where
\begin{eqnarray}
\label{Coefficients of Delta, Delta_ij}
\begin{array}{l}
\begin{array}{l}
a(1,1)=(\lambda_1-\lambda_2)(\mu_1-\mu_2),~~
a(1,-1)=(\lambda_1-\mu_2)(\mu_1-\lambda_2),
\smallskip \\
a(i,-i)=(\lambda_1-\mu_1)(\lambda_2-\mu_2)
\end{array} 
\medskip \\
\begin{array}{l}
A(1,1)=-(\lambda_1-\lambda_2)(\mu_1-\mu_2)\lambda_1\lambda_2,~~ A(-1,-1)=-(\lambda_1-\lambda_2)(\mu_1-\mu_2)\mu_1\mu_2,
\smallskip \\
A(1,-1)=-(\lambda_1-\mu_2)(\mu_1-\lambda_2)\lambda_1\mu_2,~~
A(-1,1)=-(\lambda_1-\mu_2)(\mu_1-\lambda_2)\mu_1\lambda_2,
\smallskip \\
A(i,-i)=-(\lambda_1-\mu_1)(\lambda_2-\mu_2)\lambda_1\mu_1,~~ 
A(-i,i)=-(\lambda_1-\mu_1)(\lambda_2-\mu_2)\lambda_2\mu_2,
\end{array}
\medskip \\
\begin{array}{l}
A(1,i)=(\lambda_1-\lambda_2)(\lambda_1-\mu_2)(\lambda_2-\mu_2)\mu_1,
\smallskip \\
A(-1,i)=(\mu_1-\mu_2)(\mu_1-\lambda_2)(\lambda_2-\mu_2)\lambda_1,
\smallskip \\
A(i,1)=(\lambda_1-\lambda_2)(\lambda_1-\mu_1)(\mu_1-\lambda_2)\mu_2,
\smallskip \\
A(i,-1)=(\mu_1-\mu_2)(\lambda_1-\mu_1)(\lambda_1-\mu_2)\lambda_2.
\end{array}
\end{array}
\end{eqnarray}

\subsection{Exact Calculation of NL$\sigma$M Term (Two-Soliton)}
\label{Exact calculation of the NL Sigma Model Term (2-Soliton)}
For preparation, we introduce some symmetries between the coefficients of the soliton data  before our calculation.
Our observation is as the following Remark 1 $\sim$ 3 which can be checked simply from  \eqref{Data of J_3} and \eqref{Coefficients of Delta, Delta_ij}.
\bigskip \\
{\bf Remark 1}
\begin{eqnarray}
\label{Relations of coefficients_0}
\begin{array}{l}
A(1,1)=-\lambda_1\lambda_2 ~\!a(1, 1),~~ A(-1,-1)=-\mu_1\mu_2 ~\!a(1, 1),
\smallskip \\
A(1,-1)=-\lambda_1\mu_2~\!a(1, -1),~~
A(-1,1)=-\mu_1\lambda_2 ~\!a(1, -1).
\end{array}
\end{eqnarray} 
{\bf Remark 2}  
\begin{eqnarray}
\label{Relations of coefficients_1}
\begin{array}{l}
A(1, i)A(-1, i)
=\lambda_1\mu_1(\lambda_2 - \mu_2)^2 ~\! a(1, 1)a(1, -1),~
\smallskip \\
A(i, 1)A(i, -1)
=\lambda_2\mu_2(\lambda_1 - \mu_1)^2 ~\! a(1, 1)a(1, -1),~
\smallskip \\
A(-1, i)A(i, 1)
=\lambda_1\mu_2(\mu_1 - \lambda_2)^2 ~\! a(1, 1)a(i, -i),~
\smallskip \\
A(1, i)A(i, -1)
=\mu_1\lambda_2(\lambda_1 - \mu_2)^2 ~\! a(1, 1)a(i, -i),~
\smallskip \\
A(-1, i)A(i, -1)
=\lambda_1\lambda_2 (\mu_1 - \mu_2)^2~\!a(1, -1)a(i, -i),~
\smallskip \\
A(1, i)A(i, 1)
=\mu_1\mu_2 (\lambda_1 - \lambda_2)^2~\!a(1, -1)a(i, -i).
\end{array}
\end{eqnarray}
{\bf Remark 3}
\begin{eqnarray}
\label{Relations of coefficients_2}
\begin{array}{l}
A(1, i) - A(-1, i) = A(i, -1) - A(i, 1) = (\lambda_1\mu_1 - \lambda_2\mu_2)~\!a(i, -i), 
\smallskip \\
A(i, 1) - A(-1, i) = A(i, -1) - A(1, i) = (\lambda_1\mu_2 - \mu_1\lambda_2)~\!a(1, -1),
\smallskip \\
A(1, i) + A(i,1) = A(-1, i) + A(i, -1) = (\lambda_1\lambda_2 - \mu_1\mu_2)~\!a(1, 1).
\end{array}
\end{eqnarray}
which implies the relation
\begin{eqnarray}
\label{Relations of coefficients_3}
A(1, i) - A(-1, i) + A(i, 1) - A(i, -1) =0
\end{eqnarray}
directly by taking some simple addition and subtraction over \eqref{Relations of coefficients_2}. On the other hand, by taking sum of squares over $L. H. S.$ and $R. H. S.$ of \eqref{Relations of coefficients_2} and using the relation \eqref{Relations of coefficients_3}, we get the following nontrivial identity.
\medskip \\
{\bf Remark 4}
\begin{eqnarray}
\label{Relations of coefficients_4}
&&\!\!\!\! (\lambda_1\lambda_2 - \mu_1\mu_2)^2a^2(1, 1)
+(\lambda_1\mu_2 - \mu_1\lambda_2)^2a^2(1, -1)
+(\lambda_1\mu_1 - \lambda_2\mu_2)^2a^2(i, -i)  ~~~~ \nonumber \\
&\!\!\!\!=\!\!\!\!&
A^2(1, i) + A^2(-1, i) + A^2(i, 1) + A^2(i, -1). 
\end{eqnarray}

Now let us start our main calculation of  the NL sigma model term for two-soliton.
By using the soliton data \eqref{Data of J_3} and after a little bit tedious calculation, we can conclude that
\begin{eqnarray}
&&\!\!\!\! \partial_{\mu}\Delta_{11} \partial_{\mu}\Delta_{22} - \partial_{\mu}\Delta_{12} \partial_{\mu}\Delta_{21} - |\sigma|(\partial_{\mu}\Delta)^2    \nonumber \\
&\!\!\!\!=\!\!\!\!&
\begin{array}{l}
~(r_{\mu}^{(1)} + r_{\mu}^{(2)})^2
\left[ A(1, 1)A(-1, -1) - |\sigma|a^2(1,1) \right]
\left[
e^{2(X_1 + X_2)} + e^{-2(X_1 + X_2)}
\right]
\medskip  \\
\!\!+(r_{\mu}^{(1)} - r_{\mu}^{(2)})^2
\left[ A(1, -1)A(-1, 1) - |\sigma|a^2(1, -1) \right]
\left[
e^{2(X_1 - X_2)} + e^{-2(X_1 - X_2)}
\right]
\medskip \\
\!\!+(s_{\mu}^{(1)} - s_{\mu}^{(2)})^2
\left[ A(i, -i)A(-i, i) - |\sigma|a^2(i, -i) \right]
\left(
e^{2\Theta_{12}} + e^{-2\Theta_{12}}
\right)
\end{array}
\nonumber \\
&&+\!\left\{
\begin{array}{l}
~(r_{\mu}^{(1)} \!+ r_{\mu}^{(2)})(r_{\mu}^{(1)} \!- r_{\mu}^{(2)})
\!\left[
\begin{array}{l}
A(1, 1)A(-1, 1) + A(1, -1)A(-1, -1) 
\medskip \\
\!- 2|\sigma|a(1, 1)a(1, -1)
\end{array}
\!\!\right]
\medskip \\
\!\!-(r_{\mu}^{(1)} \!+ s_{\mu}^{(2)})(r_{\mu}^{(1)} \!- s_{\mu}^{(2)})
A(1, i)A(-1, i)
\end{array}
\!\!\!\right\}
\!\!\left[ e^{2X_1} + e^{-2X_1} \right] 
\nonumber \\
&&-\!\left\{
\begin{array}{l}
~(r_{\mu}^{(1)} \!+ r_{\mu}^{(2)})(r_{\mu}^{(1)} \!- r_{\mu}^{(2)})
\!\left[
\begin{array}{l}
A(1, 1)A(1, -1) + A(-1, 1)A(-1, -1) 
\medskip \\
\!- 2|\sigma|a(1, 1)a(1, -1)
\end{array}
\!\!\right]
\medskip \\
\!\!-(s_{\mu}^{(1)} \!+ r_{\mu}^{(2)})(s_{\mu}^{(1)} \!- r_{\mu}^{(2)})
A(i, 1)A(i, -1)
\end{array}
\!\!\!\right\}
\!\!\left[ e^{2X_2} + e^{-2X_2} \right]
\nonumber  \\
&&+\!\left\{
\begin{array}{l}
(r_{\mu}^{(1)} \!+ \!r_{\mu}^{(2)})(s_{\mu}^{(1)} \!- \!s_{\mu}^{(2)})
\!\left[
\begin{array}{l}
A(1, 1)A(-i, i) + A(-1, -1)A(i, -i) 
\medskip \\
\!- 2|\sigma|a(1, 1)a(i, -i)
\end{array}
\!\!\right]
\medskip   \\
\!\!\!\!-(r_{\mu}^{(1)} \!- \!s_{\mu}^{(2)})(s_{\mu}^{(1)} \!+ \!r_{\mu}^{(2)})
A(-1, i)A(i, 1)
\end{array}
\!\!\!\right\}
\!\!\left[
\begin{array}{l}
~\!e^{X_1 - X_2 + i\Theta_{12}} 
\medskip \\
\!\!\!+ e^{-(X_1 - X_2 + i\Theta_{12})} 
\end{array}
\!\!\right] 
\nonumber  \\
&&-\!\left\{
\begin{array}{l}
(r_{\mu}^{(1)} \!+ \!r_{\mu}^{(2)})(s_{\mu}^{(1)} \!- \!s_{\mu}^{(2)})
\!\left[
\begin{array}{l}
A(1, 1)A(i, -i) + A(-1, -1)A(-i, i) 
\medskip \\
\!- 2|\sigma|a(1, 1)a(i, -i)
\end{array}
\!\!\right]
\medskip  \\
\!\!\!\!-(r_{\mu}^{(1)} \!+ \!s_{\mu}^{(2)})(s_{\mu}^{(1)} \!- \!r_{\mu}^{(2)})
A(1, i)A(i, -1)
\end{array}
\!\!\!\right\}
\!\!\left[
\begin{array}{l}
~\!e^{X_1 + X_2 - i\Theta_{12}} 
\medskip \\
\!\!\!+ e^{-(X_1 + X_2 - i\Theta_{12})} 
\end{array}
\!\!\right] 
\nonumber  \\
&&+\!\left\{
\begin{array}{l}
(r_{\mu}^{(1)} \!- \!r_{\mu}^{(2)})(s_{\mu}^{(1)} \!- \!s_{\mu}^{(2)})
\!\left[
\begin{array}{l}
A(1, -1)A(-i, i) + A(-1, 1)A(i, -i) 
\medskip \\
\!- 2|\sigma|a(1, -1)a(i, -i)
\end{array}
\!\!\right]
\medskip  \\
\!\!\!\!-(r_{\mu}^{(1)} \!-\! s_{\mu}^{(2)})(s_{\mu}^{(1)} \!-\! r_{\mu}^{(2)})
A(-1, i)A(i, -1)
\end{array}
\!\!\!\right\}
\!\!\left[
\begin{array}{l}
~\!e^{X_1 - X_2 + i\Theta_{12}} 
\medskip \\
\!\!\!+ e^{-(X_1 - X_2 + i\Theta_{12})} 
\end{array}
\!\!\right] 
\nonumber  \\
&&-\!\left\{
\begin{array}{l}
(r_{\mu}^{(1)} \!- \!r_{\mu}^{(2)})(s_{\mu}^{(1)} \!- \!s_{\mu}^{(2)})
\!\left[
\begin{array}{l}
A(1, -1)A(i, -i) + A(-1, 1)A(-i, i) 
\medskip \\
\!- 2|\sigma|a(1, -1)a(i, -i)
\end{array}
\!\!\right]
\medskip  \\
\!\!\!\!-(r_{\mu}^{(1)} \!+ \!s_{\mu}^{(2)})(s_{\mu}^{(1)} \!+ \!r_{\mu}^{(2)})
A(1, i)A(i, 1)
\end{array}
\!\!\!\right\}
\!\!\left[
\begin{array}{l}
~\!e^{X_1 - X_2 - i\Theta_{12}} 
\medskip \\
\!\!\!+ e^{-(X_1 - X_2 - i\Theta_{12})} 
\end{array}
\!\!\right] 
\nonumber \\
%\end{eqnarray}
&&
%\begin{eqnarray}
-\left\{
\begin{array}{l}
~(r_{\mu}^{(1)} + r_{\mu}^{(2)})^2
\left[ A^2(1, 1) + A^2(-1, -1) -2|\sigma|a^2(1, 1) \right]
\medskip \\
\!\!+(r_{\mu}^{(1)} - r_{\mu}^{(2)})^2
\left[ A^2(1, -1) + A^2(-1, 1) -2|\sigma|a^2(1, -1) \right]
\medskip \\
\!\!+(s_{\mu}^{(1)} - s_{\mu}^{(2)})^2
\left[ A^2(i, -i) + A^2(-i, i) -2|\sigma|a^2(i, -i) \right]
\medskip \\
\!\!-(r_{\mu}^{(1)} + s_{\mu}^{(2)})^2 A^2(1, i)
-(r_{\mu}^{(1)} - s_{\mu}^{(2)})^2 A^2(-1, i)
\medskip \\
\!\!-(s_{\mu}^{(1)} + r_{\mu}^{(2)})^2 A^2(i, 1)
-(s_{\mu}^{(1)} - r_{\mu}^{(2)})^2 A^2(i, -1)
\end{array}
\right\} ~~~~~~~~~~~~~~~~~~~~
\nonumber 
\end{eqnarray}
%\newpage
By \eqref{Relations of coefficients_0} and \eqref{Relations of coefficients_1}, we find that the coefficients of the leading terms $\exp(\pm 2(X_1 \pm X_2))$ are identical to zero and the remaining terms can be rewritten as
\begin{eqnarray}
\label{Calculation of NL Sigma Term_2-soliton_1 }
&&\!\!\!\! \partial_{\mu}\Delta_{11} \partial_{\mu}\Delta_{22} - \partial_{\mu}\Delta_{12} \partial_{\mu}\Delta_{21} - |\sigma|(\partial_{\mu}\Delta)^2    \nonumber \\
&\!\!\!\!=\!\!\!\!&
\begin{array}{l}
\left\{
\begin{array}{l}
-(r_{\mu}^{(2)} + s_{\mu}^{(2)})(r_{\mu}^{(2)} - s_{\mu}^{(2)})A(1, i)A(-1, i)
\left(
e^{2X_1} + e^{-2X_1}
\right)
\medskip \\
-(r_{\mu}^{(1)} + s_{\mu}^{(1)})(r_{\mu}^{(1)} - s_{\mu}^{(1)})A(i, 1)A(i, -1)
\left(
e^{2X_2} + e^{-2X_2}
\right)
\medskip \\
-(r_{\mu}^{(1)} - s_{\mu}^{(1)})(r_{\mu}^{(2)} + s_{\mu}^{(2)})A(-1, i)A(i, 1)
\left(
e^{X_1 + X_2 + i\Theta_{12}} + e^{-(X_1 + X_2 + i\Theta_{12})}
\right)
\medskip \\
-(r_{\mu}^{(1)} + s_{\mu}^{(1)})(r_{\mu}^{(2)} - s_{\mu}^{(2)})A(1, i)A(i, -1)
\left(
e^{X_1 + X_2 - i\Theta_{12}} + e^{-(X_1 + X_2 - i\Theta_{12})}
\right)
\medskip \\
+(r_{\mu}^{(1)} - s_{\mu}^{(1)})(r_{\mu}^{(2)} - s_{\mu}^{(2)})A(-1, i)A(i, -1)
\left(
e^{X_1 - X_2 + i\Theta_{12}} + e^{-(X_1 - X_2 + i\Theta_{12})}
\right)
\medskip \\
+(r_{\mu}^{(1)} + s_{\mu}^{(1)})(r_{\mu}^{(2)} + s_{\mu}^{(2)})A(1, i)A(i, 1)
\left(
e^{X_1 - X_2 - i\Theta_{12}} + e^{-(X_1 - X_2 - i\Theta_{12})}
\right)
\end{array}
\!\!\right\}
\medskip \\
\begin{array}{l}
-\left\{
\underbrace{
\begin{array}{l}
~(r_{\mu}^{(1)} + r_{\mu}^{(2)})^2
(\lambda_1\lambda_2 - \mu_1\mu_2)^2~\!a^2(1, 1)
\medskip \\
\!\!+(r_{\mu}^{(1)} - r_{\mu}^{(2)})^2
(\lambda_1\mu_2 - \mu_1\lambda_2)^2~\!a^2(1, -1)
\medskip \\
\!\!+(s_{\mu}^{(1)} - s_{\mu}^{(2)})^2
(\lambda_1\mu_1 - \lambda_2\mu_2)^2~\!a^2(i, -i)
\medskip \\
\!\!-(r_{\mu}^{(1)} + s_{\mu}^{(2)})^2 A^2(1, i)
-(r_{\mu}^{(1)} - s_{\mu}^{(2)})^2 A^2(-1, i)
\medskip \\
\!\!-(s_{\mu}^{(1)} + r_{\mu}^{(2)})^2 A^2(i, 1)
-(s_{\mu}^{(1)} - r_{\mu}^{(2)})^2 A^2(i, -1)
\end{array}
}
\right\}
\end{array}
\end{array} ~~~~~~
\\
&&  ~~~~~~~~~~~~~~~~~~~~~~~~~~~~~~~~~~~ =: \Xi   \nonumber 
\end{eqnarray}
Next, we want to show that the constant term $\Xi$ above can be absorbed completely into the non-constant terms.
By the definition of $r_{\mu}^{(j)}, s_{\mu}^{(j)}$ above (\ref{Derivative of exponential functions_1}), we can replace $r_{\mu}^{(j)}, s_{\mu}^{(j)}$ by $\ell_{\mu}^{(j)}$ and obtain
\begin{eqnarray}
%&\!\!\!\!=\!\!\!\!&
%\begin{array}{l}
%-4\left\{
%\begin{array}{l}
%~\ell_{\mu}^{(2)}\overline{\ell}_{\mu}^{(2)}A(1, i)A(-1, i)
%\left(
%e^{2X_1} + e^{-2X_1}
%\right)
%\medskip \\
%\!\!+\ell_{\mu}^{(1)}\overline{\ell}_{\mu}^{(1)}A(i, 1)A(i, -1)
%\left(
%e^{2X_2} + e^{-2X_2}
%\right)
%\medskip \\
%\!\!+\overline{\ell}_{\mu}^{(1)}\ell_{\mu}^{(2)}A(-1, i)A(i, 1)
%\left(
%e^{X_1 + X_2 + i\Theta_{12}} + e^{-(X_1 + X_2 + i\Theta_{12})}
%\right)
%\medskip \\
%\!\!+\ell_{\mu}^{(1)} \overline{\ell}_{\mu}^{(2)}A(1, i)A(i, -1)
%\left(
%e^{X_1 + X_2 - i\Theta_{12}} + e^{-(X_1 + X_2 - i\Theta_{12})}
%\right)
%\medskip \\
%\!\!-\overline{\ell}_{\mu}^{(1)}\overline{\ell}_{\mu}^{(2)}A(-1, i)A(i, -1)
%\left(
%e^{X_1 - X_2 + i\Theta_{12}} + e^{-(X_1 - X_2 + i\Theta_{12})}
%\right)
%\medskip \\
%\!\!-\ell_{\mu}^{(1)}\ell_{\mu}^{(2)}A(1, i)A(i, 1)
%\left(
%e^{X_1 - X_2 - i\Theta_{12}} + e^{-(X_1 - X_2 - i\Theta_{12})}
%\right)
%\end{array}
%\right\}
%\medskip  \\
%\!\!\!\!\!\!\!\!\begin{array}{l}
\Xi&\!\!\!\!=\!\!\!\!&\left\{
%\underbrace{
	\begin{array}{l}
	~\left(\ell_{\mu}^{(1)} + \overline{\ell}_{\mu}^{(1)} + \ell_{\mu}^{(2)} + \overline{\ell}_{\mu}^{(2)}\right)^2
	(\lambda_1\lambda_2 - \mu_1\mu_2)^2~\!a^2(1, 1)
	\medskip \\
	\!\!+\left(\ell_{\mu}^{(1)} + \overline{\ell}_{\mu}^{(1)} - \ell_{\mu}^{(2)} - \overline{\ell}_{\mu}^{(2)}\right)^2
	(\lambda_1\mu_2 - \mu_1\lambda_2)^2~\!a^2(1, -1)
	\medskip \\
	\!\!+\left(\ell_{\mu}^{(1)} - \overline{\ell}_{\mu}^{(1)} - \ell_{\mu}^{(2)} + \overline{\ell}_{\mu}^{(2)}\right)^2
	(\lambda_1\mu_1 - \lambda_2\mu_2)^2~\!a^2(i, -i)
	\medskip \\
	\!\!-\left(\ell_{\mu}^{(1)} + \overline{\ell}_{\mu}^{(1)} + \ell_{\mu}^{(2)} - \overline{\ell}_{\mu}^{(2)}\right)^2 \!\!A^2(1, i)
	-\left(\ell_{\mu}^{(1)} + \overline{\ell}_{\mu}^{(1)} - \ell_{\mu}^{(2)} + \overline{\ell}_{\mu}^{(2)}\right)^2 \!\!A^2(-1, i)
	\medskip \\
	\!\!-\left(\ell_{\mu}^{(1)} - \overline{\ell}_{\mu}^{(1)} + \ell_{\mu}^{(2)} + \overline{\ell}_{\mu}^{(2)}\right)^2 \!\!A^2(i, 1)
	-\left(\ell_{\mu}^{(1)} - \overline{\ell}_{\mu}^{(1)} - \ell_{\mu}^{(2)} - \overline{\ell}_{\mu}^{(2)}\right)^2 \!\!A^2(i, -1)
	\end{array}
%}
\right\}
%\end{array}
%\end{array} ~~~~ 
%\\
%&& ~~~~~~~~~~~~~~~~~~~~~~~~~~~~~~~~~~~~~~~~~~~ =: \Xi   \nonumber 
%\end{eqnarray}
%Next, we want to show that the constant term $\Xi$ above can be absorbed completely into the non-constant terms.
%In fact, the constant term $\Xi$ can be decomposed  into the following form:
%\begin{eqnarray}
%\Xi
\nonumber 
\end{eqnarray}
By \eqref{Relations of coefficients_4} and \eqref{Relations of coefficients_2},
\begin{eqnarray}
\Xi &\!\!\!\!=\!\!\!\!&
-4\left\{
\begin{array}{l}
~\ell_{\mu}^{(1)}\overline{\ell}_{\mu}^{(1)}
\left[
-(\lambda_1\mu_1 - \lambda_2\mu_2)^2 a^2(i, -i) + A^2(i, 1) + A^2(i, -1)
\right]
\medskip \\
\!\!+\ell_{\mu}^{(2)}\overline{\ell}_{\mu}^{(2)}
\left[
-(\lambda_1\mu_1 - \lambda_2\mu_2)^2 a^2(i, -i) + A^2(1, i) + A^2(-1, i)
\right]
\medskip \\
\!\!+\ell_{\mu}^{(1)}\overline{\ell}_{\mu}^{(2)}
\left[
-(\lambda_1\mu_2 - \mu_1\lambda_2)^2 a^2(1, -1) + A^2(1, i) + A^2(i, -1)
\right]
\medskip \\
\!\!+\overline{\ell}_{\mu}^{(1)}\ell_{\mu}^{(2)}
\left[
-(\lambda_1\mu_2 - \mu_1\lambda_2)^2 a^2(1, -1) + A^2(-1, i) + A^2(i, 1)
\right]
\medskip \\
\!\!-\ell_{\mu}^{(1)}\ell_{\mu}^{(2)}
\left[
-(\lambda_1\lambda_2 - \mu_1\mu_2)^2 a^2(1, 1) + A^2(1, i) + A^2(i, 1)
\right]
\medskip \\
\!\!-\overline{\ell}_{\mu}^{(1)}\overline{\ell}_{\mu}^{(2)}
\left[
-(\lambda_1\lambda_2 - \mu_1\mu_2)^2 a^2(1, 1) + A^2(-1, i) + A^2(i, -1)
\right]
\end{array}
\right\}
\nonumber \\
&\!\!\!\!=\!\!\!\!&
-8\left\{
\begin{array}{l}
~\ell_{\mu}^{(1)}\overline{\ell}_{\mu}^{(1)} A(i, 1)A(i, -1)
+\ell_{\mu}^{(2)}\overline{\ell}_{\mu}^{(2)} A(1, i)A(-1, i)
\medskip \\
\!\!+\ell_{\mu}^{(1)}\overline{\ell}_{\mu}^{(2)} A(1, i)A(i, -1)
+\overline{\ell}_{\mu}^{(1)} \ell_{\mu}^{(2)} A(-1, i)A(i, 1)
\medskip \\
\!\!+\ell_{\mu}^{(1)}\ell_{\mu}^{(2)} A(1, i)A(i, 1)
+\overline{\ell}_{\mu}^{(1)} \overline{\ell}_{\mu}^{(2)} A(-1, i)A(i, -1)
\end{array}
\right\}.
\nonumber 
\end{eqnarray}
Comparing with \eqref{Calculation of NL Sigma Term_2-soliton_1 }, we have
\begin{eqnarray}
&&\!\!\!\! \partial_{\mu}\Delta_{11} \partial_{\mu}\Delta_{22} - \partial_{\mu}\Delta_{12} \partial_{\mu}\Delta_{21} - |\sigma|(\partial_{\mu}\Delta)^2    \nonumber \\
&\!\!\!\!=\!\!\!\!&
\begin{array}{l}
-4\left\{
\begin{array}{l}
~\ell_{\mu}^{(1)}\overline{\ell}_{\mu}^{(1)}A(i, 1)A(i, -1)
\left(
e^{X_2} + e^{-X_2}
\right)^2
\medskip \\
\!\!+\ell_{\mu}^{(2)}\overline{\ell}_{\mu}^{(2)}A(1, i)A(-1, i)
\left(
e^{X_1} + e^{-X_1}
\right)^2
\medskip \\
\!\!+\ell_{\mu}^{(1)} \overline{\ell}_{\mu}^{(2)}A(1, i)A(i, -1)
\left(
e^{\frac{X_1 + X_2 - i\Theta_{12}}{2}} + e^{-\frac{(X_1 + X_2 - i\Theta_{12})}{2}}
\right)^2
\medskip \\
\!\!+\overline{\ell}_{\mu}^{(1)}\ell_{\mu}^{(2)}A(-1, i)A(i, 1)
\left(
e^{\frac{X_1 + X_2 + i\Theta_{12}}{2}} + e^{-\frac{(X_1 + X_2 + i\Theta_{12})}{2}}
\right)^2
\medskip \\
\!\!-\ell_{\mu}^{(1)}\ell_{\mu}^{(2)}A(1, i)A(i, 1)
\left(
e^{\frac{X_1 - X_2 - i\Theta_{12}}{2}} - e^{-\frac{(X_1 - X_2 - i\Theta_{12})}{2}}
\right)^2
\medskip \\
\!\!-\overline{\ell}_{\mu}^{(1)}\overline{\ell}_{\mu}^{(2)}A(-1, i)A(i, -1)
\left(
e^{\frac{X_1 - X_2 + i\Theta_{12}}{2}} - e^{-\frac{(X_1 - X_2 + i\Theta_{12})}{2}}
\right)^2
\end{array}
\right\}
\end{array}
\nonumber 
\end{eqnarray}
By \eqref{Tr(A_m A_n)_2} and \eqref{Relations of coefficients_1}, we can conclude that
\begin{eqnarray}
%&&
\mbox{Tr}\left[(\partial_{\mu}\sigma)\sigma^{-1} \right]^2   
\nonumber 
%\\
%&\!\!\!\!=\!\!\!\!&
%\frac{2}{|\sigma|\Delta^{2}}
%\left\{
%\left|
%\begin{array}{cc}
%\partial_{\mu} \Delta_{11} & \partial_{\mu} \Delta_{12} \\
%\partial_{\mu} \Delta_{21} & \partial_{\mu} \Delta_{22}
%\end{array}
%\right|
%-
%|\sigma|(\partial_{\mu} \Delta)^2
%\right\}
%\nonumber
%&\!\!\!\!=\!\!\!\!& 
%\frac{
%	-8\left\{
%	\begin{array}{l}
%	~\ell_{\mu}^{(1)}\overline{\ell}_{\mu}^{(1)}A(i, 1)A(i, -1)
%	\cosh^2 X_2
%	+\ell_{\mu}^{(2)}\overline{\ell}_{\mu}^{(2)}A(1, i)A(-1, i)
%	\cosh^2 X_1
%	\medskip \\
%	\!\!+\ell_{\mu}^{(1)} \overline{\ell}_{\mu}^{(2)}A(1, i)A(i, -1)
%	\displaystyle{\cosh^2 \left(\frac{X_1 + X_2 - i\Theta_{12}}{2} \right)}
%	\medskip \\
%	\!\!+\overline{\ell}_{\mu}^{(1)}\ell_{\mu}^{(2)}A(-1, i)A(i, 1)
%	\displaystyle{\cosh^2 \left(\frac{X_1 + X_2 + i\Theta_{12}}{2} \right)}
%	\medskip \\
%	\!\!-\ell_{\mu}^{(1)}\ell_{\mu}^{(2)}A(1, i)A(i, 1)
%	\displaystyle{\sinh^2 \left(\frac{X_1 - X_2 - i\Theta_{12}}{2}
%		\right)}
%	\medskip \\
%	\!\!-\overline{\ell}_{\mu}^{(1)}\overline{\ell}_{\mu}^{(2)}A(-1, i)A(i, -1)
%	\displaystyle{\sinh^2 \left(\frac{X_1 - X_2 + i\Theta_{12}}{2} \right)}
%	\end{array}
%	\right\}
%}
%{
%	\displaystyle{
%		|\sigma|\left[
%		\begin{array}{l}
%		a(1, 1)\cosh(X_1 + X_2)
%		+
%		a(1, -1)\cosh(X_1 - X_2)
%		+
%		a(i, -i)\cos \Theta_{12}
%		\end{array}
%		\right]^2
%	}
%} ~~~~~~
%\\
&\!\!\!\!=\!\!\!\!&
\frac{
	-8\left\{
	\begin{array}{l}
	~a(1, 1)a(1, -1)
	\left[
	\ell_{\mu}^{(1)}\overline{\ell}_{\mu}^{(1)}\Lambda_{11}
	\cosh^2 X_2
	+\ell_{\mu}^{(2)}\overline{\ell}_{\mu}^{(2)}\Lambda_{22}
	\cosh^2 X_1
	\right]
	\medskip \\
	\!\!+ a(1, 1)a(i, -i)\left[
	\begin{array}{l}
	~\ell_{\mu}^{(1)} \overline{\ell}_{\mu}^{(2)}\Lambda_{12}
	\displaystyle{\cosh^2 \left(\frac{X_1 + X_2 - i\Theta_{12}}{2} \right)}
	\medskip \\
	\!\!+ \overline{\ell}_{\mu}^{(1)}\ell_{\mu}^{(2)}\Lambda_{21}~\!
	\displaystyle{\cosh^2 \left(\frac{X_1 + X_2 + i\Theta_{12}}{2} \right)}
	\end{array}
	\right]
	\medskip \\
	\!\!- a(1, -1)a(i, -i)\left[
	\begin{array}{l}
	~\ell_{\mu}^{(1)}\ell_{\mu}^{(2)}\Omega_{12}~\!
	\displaystyle{\sinh^2 \left(\frac{X_1 - X_2 - i\Theta_{12}}{2}
		\right)}
	\medskip \\
	\!\!+\overline{\ell}_{\mu}^{(1)}\overline{\ell}_{\mu}^{(2)}\widetilde{\Omega}_{12}~\!
	\displaystyle{\sinh^2 \left(\frac{X_1 - X_2 + i\Theta_{12}}{2} \right)}
	\end{array}
	\right]
	\end{array}
	\right\}
}
{
	\displaystyle{
		\left[
		\begin{array}{l}
		a(1, 1)\cosh(X_1 + X_2)
		+
		a(1, -1)\cosh(X_1 - X_2)
		+
		a(i, -i)\cos \Theta_{12}
		\end{array}
		\right]^2
	}
}, 
\nonumber 
\end{eqnarray}
where
\begin{eqnarray*}
\displaystyle{
	\Lambda_{jk}
	:=
	\frac{(\lambda_j-\mu_k)^2}{\lambda_j\mu_k},~~ 
	\Omega_{jk}
	:=
	\frac{(\lambda_j-\lambda_k)^2}{\lambda_j\lambda_k}, ~~
	\widetilde{\Omega}_{jk}
	:=\frac{(\mu_j - \mu_k)^2}{\mu_j\mu_k}.
}
\end{eqnarray*}

\subsection{Exact Calculation of Wess-Zumino Action Density (Two-Soliton)}
\label{WZ2}

By substituting \eqref{Determinant of n-soliton solution}, \eqref{Data of J_3} and \eqref{Derivative of exponential functions_1} into \eqref{Tr(A_m A_n A_p)_2} for $(m, n, p) = (\mu, \nu, \rho)$, we have
\begin{eqnarray}
\label{Density of the WZ Term_decomposition App}
\!\!\!\!\!\!\!\!&&\mbox{Tr}\left[(\partial_{\mu}\sigma)\sigma^{-1}(\partial_{\nu}\sigma)\sigma^{-1}(\partial_{\rho}\sigma)\sigma^{-1}\right]
\smallskip 
=
\displaystyle{\frac{1}{2}}
%\left\{
%\begin{array}{l}
(B_{\mu\nu\rho}+B_{\nu\rho\mu}+B_{\rho\mu\nu}),\\
\!\!\!\!\!\!\!\!&&
B_{\mu\nu\rho}:=
\displaystyle{\frac{1}{|\sigma|^2\Delta^4}}
\left(
\left|
\!\!\begin{array}{cc}
\Delta_{11}\!\! & \Delta_{22}\!\! \\
\partial_{\mu} \Delta_{11}\!\! & \partial_{\mu} \Delta_{22}\!\!
\end{array}
\right|
%\cdot
\left|
\!\!\begin{array}{cc}
\partial_{\nu} \Delta_{12}\!\! & \partial_{\nu} \Delta_{21}\!\! \\
\partial_{\rho} \Delta_{12}\!\! & \partial_{\rho} \Delta_{21}\!\!
\end{array}
\right|
+
\left|
\!\!\begin{array}{cc}
\Delta_{12}\!\! & \Delta_{21}\!\! \\
\partial_{\mu} \Delta_{12}\!\! & \partial_{\mu} \Delta_{21}\!\!
\end{array}
\right|
%\cdot
\left|
\!\!\begin{array}{cc}
\partial_{\nu} \Delta_{11}\!\! & \partial_{\nu} \Delta_{22}\!\! \\
\partial_{\rho} \Delta_{11}\!\! & \partial_{\rho} \Delta_{22}\!\!
\end{array}
\right|\right). 
\nonumber
%\end{array}
%\right\}\!\!+
%{\mbox{CycPerm}}(\mu, \nu, \rho)
%\!\!. ~~~~~~~~
\end{eqnarray}
where
\begin{eqnarray}
\label{WZ term_part1}
&&\displaystyle{\frac{1}{|\sigma|\Delta^2}}
\left|
\!\!\begin{array}{cc}
\Delta_{11}\!\! & \Delta_{22}\!\! \\
\partial_{\mu} \Delta_{11}\!\! & \partial_{\mu} \Delta_{22}\!\!
\end{array}
\right|
\nonumber \\
&\!\!\!\!=\!\!\!\!&
\frac{
-\left\{
\begin{array}{l}
 ~~~2r_{\mu}^{(1)}
ab\mathcal{D}_{11}~\!\mathrm{cosh}(2X_2)
+2r_{\mu}^{(2)}
ab\mathcal{D}_{22}~\!\mathrm{cosh}(2X_1)
\smallskip \\
+\left[ (r_{\mu}^{(1)} + r_{\mu}^{(2)}) + (s_{\mu}^{(1)} - s_{\mu}^{(2)}) \right]
ac\mathcal{D}_{12}~\!\mathrm{cosh}(X_1 + X_2 -i\Theta_{12})
\smallskip \\
+\left[ (r_{\mu}^{(1)} + r_{\mu}^{(2)}) - (s_{\mu}^{(1)} - s_{\mu}^{(2)}) \right]
ac\mathcal{D}_{21}~\!\mathrm{cosh}(X_1 + X_2 +i\Theta_{12})
\smallskip \\
+\left[ (r_{\mu}^{(1)} - r_{\mu}^{(2)}) + (s_{\mu}^{(1)} - s_{\mu}^{(2)}) \right]
bc\mathcal{E}_{12}~\!\mathrm{cosh}(X_1 - X_2 -i\Theta_{12})
\smallskip \\
-\left[ (r_{\mu}^{(1)} - r_{\mu}^{(2)}) - (s_{\mu}^{(1)} - s_{\mu}^{(2)}) \right]
bc\widetilde{\mathcal{E}}_{12}~\!\mathrm{cosh}(X_1 - X_2 +i\Theta_{12}) 
\end{array}
\right\} -F
} 
{
	2\left[
	\begin{array}{l}
	a~\!\mbox{cosh}(X_1 + X_2)
	+
	b~\!\mbox{cosh}(X_1 - X_2)
	+
	c~\!\mbox{cos}\Theta_{12}
	\end{array}
	\right]^2   
}~,~~~~~~~
\end{eqnarray}
\begin{eqnarray}
\label{WZ term_part3}
&&\!\!\!\!\displaystyle{\frac{1}{|\sigma|\Delta^2}}
\left|
\!\!\begin{array}{cc}
\Delta_{12}\!\! & \Delta_{21}\!\! \\
\partial_{\mu} \Delta_{12}\!\! & \partial_{\mu} \Delta_{21}\!\!
\end{array}
\right| 
\nonumber \\
&\!\!\!\!=\!\!\!\!&
\frac{
-\left\{
\begin{array}{l}
~~~2s_{\mu}^{(1)}
\!ab\mathcal{d}_{11}~\!\mathrm{cosh}(2X_2)
+2s_{\mu}^{(2)}
\!ab\mathcal{d}_{22}~\!\mathrm{cosh}(2X_1)
\smallskip \\
+\left[ (r_{\mu}^{(1)} - r_{\mu}^{(2)}) + (s_{\mu}^{(1)} + s_{\mu}^{(2)}) \right]
\!ac\mathcal{d}_{12}~\!\mathrm{cosh}(X_1 + X_2 -i\Theta_{12})
\smallskip \\
-\left[ (r_{\mu}^{(1)} - r_{\mu}^{(2)}) - (s_{\mu}^{(1)} + s_{\mu}^{(2)}) \right]
\!ac\mathcal{d}_{21}~\!\mathrm{cosh}(X_1 + X_2 +i\Theta_{12})
\smallskip \\
+\left[ (r_{\mu}^{(1)} + r_{\mu}^{(2)}) + (s_{\mu}^{(1)} + s_{\mu}^{(2)}) \right]
\!bc\mathcal{e}_{12}~\!\mathrm{cosh}(X_1 - X_2 -i\Theta_{12})
\smallskip \\
-\left[ (r_{\mu}^{(1)} + r_{\mu}^{(2)}) - (s_{\mu}^{(1)} + s_{\mu}^{(2)}) \right]
\!bc\widetilde{\mathcal{e}}_{12}~\!\mathrm{cosh}(X_1 - X_2 +i\Theta_{12})
\end{array}
\right\} + f
}
{
	2\left[
	\begin{array}{l}
	a~\!\mbox{cosh}(X_1 + X_2)
	+
	b~\!\mbox{cosh}(X_1 - X_2)
	+
	c~\!\mbox{cos}\Theta_{12}
	\end{array}
	\right]^2   
}~, ~~~~~~
\end{eqnarray}
\begin{eqnarray}
\label{WZ term_part4}
&&\!\!\!\!\displaystyle{\frac{1}{|\sigma|\Delta^2}}
\left|
\!\!\begin{array}{cc}
\partial_{\nu} \Delta_{11}\!\! & \partial_{\nu} \Delta_{22}\!\! \\
\partial_{\rho} \Delta_{11}\!\! & \partial_{\rho} \Delta_{22}\!\!
\end{array}
\right|
\nonumber \\
&\!\!\!\!=\!\!\!\!&
\frac{
\left\{
\begin{array}{l}
~2\left[
r_{\nu}^{(1)}r_{\rho}^{(2)} -  r_{\rho}^{(1)}r_{\nu}^{(2)} \right]
ab\left[
\mathcal{D}_{11}\sinh{(2X_2)}
-\mathcal{D}_{22}\sinh{(2X_1)}
\right]
\medskip \\
-\left[
(r_{\nu}^{(1)} + r_{\nu}^{(2)})(s_{\rho}^{(1)} - s_{\rho}^{(2)}) -  
(r_{\rho}^{(1)} + r_{\rho}^{(2)})(s_{\nu}^{(1)} - s_{\nu}^{(2)}) 
\right] \cdot
\smallskip \\
~~~ac\left[
\mathcal{D}_{12}~\!\mathrm{sinh}(X_1 + X_2 -i\Theta_{12})
-\mathcal{D}_{21}~\!\mathrm{sinh}(X_1 + X_2 +i\Theta_{12})
\right]
\medskip \\
-\left[
(r_{\nu}^{(1)} - r_{\nu}^{(2)})(s_{\rho}^{(1)} - s_{\rho}^{(2)}) -  
(r_{\rho}^{(1)} - r_{\rho}^{(2)})(s_{\nu}^{(1)} - s_{\nu}^{(2)}) 
\right]\cdot
\smallskip \\
~~~bc\left[
\mathcal{E}_{12}~\!\mathrm{sinh}(X_1 - X_2 -i\Theta_{12})
+ \widetilde{\mathcal{E}}_{12}~\!\mathrm{sinh}(X_1 - X_2 +i\Theta_{12})
\right]
\end{array}
\right\}
}
{
2\left[
\begin{array}{l}
a~\!\mbox{cosh}(X_1 + X_2)
+
b~\!\mbox{cosh}(X_1 - X_2)
+
c~\!\mbox{cos}\Theta_{12}
\end{array}
\right]^2   
}, ~~~~~~
\end{eqnarray}
\begin{eqnarray}
\label{WZ term_part2}
&&\!\!\!\!\displaystyle{\frac{1}{|\sigma|\Delta^2}}
\left|
\!\!\begin{array}{cc}
\partial_{\nu} \Delta_{12}\!\! & \partial_{\nu} \Delta_{21}\!\! \\
\partial_{\rho} \Delta_{12}\!\! & \partial_{\rho} \Delta_{21}\!\!
\end{array}
\right|
\nonumber \\
&\!\!\!\!=\!\!\!\!&
\frac{
\left\{
\begin{array}{l}
~2\left[ 
\begin{array}{l}
~~(s_{\nu}^{(1)}r_{\rho}^{(2)} -  s_{\rho}^{(1)}r_{\nu}^{(2)})
ab\mathcal{d}_{11}\sinh{(2X_2)}
\smallskip \\
-(r_{\nu}^{(1)}s_{\rho}^{(2)} -  r_{\rho}^{(1)}s_{\nu}^{(2)}) 
ab\mathcal{d}_{22} \sinh{(2X_1)}
\end{array}
\right]
\medskip \\
-\left[
\begin{array}{l}
~~(r_{\nu}^{(1)} + s_{\nu}^{(2)})(s_{\rho}^{(1)} - r_{\rho}^{(2)}) 
\smallskip \\
-(r_{\rho}^{(1)} + s_{\rho}^{(2)})(s_{\nu}^{(1)} - r_{\nu}^{(2)}) 
\end{array}
\right]
\!ac\mathcal{d}_{12}~\!\mathrm{sinh}(X_1 + X_2 -i\Theta_{12})
\medskip \\
-\left[
\begin{array}{l}
~~(r_{\nu}^{(1)} - s_{\nu}^{(2)})(s_{\rho}^{(1)} + r_{\rho}^{(2)}) 
\smallskip \\
-(r_{\rho}^{(1)} - s_{\rho}^{(2)})(s_{\nu}^{(1)} + r_{\nu}^{(2)}) 
\end{array}
\right]
\!ac\mathcal{d}_{21}~\!\mathrm{sinh}(X_1 + X_2 +i\Theta_{12})
\medskip \\
-\left[
\begin{array}{l}
~~(r_{\nu}^{(1)} + s_{\nu}^{(2)})(s_{\rho}^{(1)} + r_{\rho}^{(2)}) 
\smallskip \\
- (r_{\rho}^{(1)} + s_{\rho}^{(2)})(s_{\nu}^{(1)} + r_{\nu}^{(2)}) 
\end{array}
\right]
\!bc\mathcal{e}_{12}~\!\mathrm{sinh}(X_1 - X_2 -i\Theta_{12})
\medskip \\
-\left[
\begin{array}{l}
~~(r_{\nu}^{(1)} - s_{\nu}^{(2)})(s_{\rho}^{(1)} - r_{\rho}^{(2)}) 
\smallskip \\
-(r_{\rho}^{(1)} - s_{\rho}^{(2)})(s_{\nu}^{(1)} - r_{\nu}^{(2)}) 
\end{array}
\right]
\!bc\widetilde{\mathcal{e}}_{12}~\!\mathrm{sinh}(X_1 - X_2 +i\Theta_{12})
\end{array}
\right\}
}
{
	2\left[
	\begin{array}{l}
	a~\!\mbox{cosh}(X_1 + X_2)
	+
	b~\!\mbox{cosh}(X_1 + X_2)
	+
	c~\!\mbox{cos}\Theta_{12}
	\end{array}
	\right]^2   
}~. ~~~~
\end{eqnarray}
$a, b, c$ are defined in Table \ref{Table_1} and $\mathcal{D}_{jk}, \mathcal{d}_{jk}, \mathcal{E}_{jk}, \mathcal{\widetilde{E}}_{jk}, \mathcal{e}_{jk}$ are defined as the following Table \ref{Table_2} for each real space respectively. $F$ and $f$ are some constants. Note that 
\begin{eqnarray}
(\mathcal{\overline{D}}_{jk} , ~\!\mathcal{\overline{E}}_{jk}) 
&\!\!\!\!=\!\!\!\!& 
\left\{
\begin{array}{l}
(-\mathcal{D}_{kj}, ~\!\mathcal{\widetilde{E}}_{jk}) ~~~~\mbox{on $\mathbb{U}_1$}
\smallskip \\
(\mathcal{D}_{kj}, ~\!-\mathcal{\widetilde{E}}_{jk}) ~~~\mbox{on $\mathbb{E}$}
\end{array},
\right.
\\
(\mathcal{\overline{d}}_{jk} , ~\!\mathcal{\overline{e}}_{jk}) 
&\!\!\!\!=\!\!\!\!& 
(\mathcal{d}_{kj}, ~\!\mathcal{\widetilde{e}}_{jk}) ~~~~~~~~~~~~\mbox{on $\mathbb{U}_1, \mathbb{E}$},
\end{eqnarray}
which implies \eqref{WZ term_part1} $\sim$ \eqref{WZ term_part2} are all pure imaginary functions on $\mathbb{U}_1$ and hence \eqref{Density of the WZ Term_decomposition App} is real-valued on $\mathbb{U}_1$. By \eqref{WZ term_real spaces}, the Wess-Zumino term is real-valued on $\mathbb{U}_1$.
On the other hand, we find that \eqref{WZ term_part1}, \eqref{WZ term_part4} are real functions and \eqref{WZ term_part3}, \eqref{WZ term_part2} are pure imaginary functions on $\mathbb{E}$. Therefore, \eqref{Density of the WZ Term_decomposition App} is a pure imaginary function on $\mathbb{E}$. 
This implies the Wess-Zumino term is real-valued on $\mathbb{E}$ because of  (Cf: \eqref{WZ term_real spaces})  
\begin{eqnarray}
{\cal{L}}_{\scriptsize{\mbox{WZ}}}
&\!\!\!\stackrel{\scriptsize{\mathbb{E}}}{=}\!\!\!\!&
-\frac{i}{8\pi}\left(
\mathrm{Tr}\left( \theta_1 \theta_3 \theta_4 \right)x^{1}
+\mathrm{Tr}\left( \theta_2 \theta_3 \theta_4 \right)x^{2}
-\mathrm{Tr}\left( \theta_3 \theta_1 \theta_2 \right)x^{3}
-\mathrm{Tr}\left( \theta_4 \theta_1 \theta_2 \right)x^{4}\right) ~~~~
\end{eqnarray}
\begin{table}[h]
	\!\!\!\!\caption{Summary of Coefficients}
	\label{Table_2}
	\medskip
\begin{center}
	\begin{tabular}{|c|c|c|}
		%\hspace{-1cm}
		\hline
		Space & $\mathbb{U}_1$ &  $\mathbb{E}$   \\ 
		(Metric) & $(+,+,-,-)$ & $(+,+,+,+)$   \\
		\hline
		\hline
		$\mathcal{D}_{jk}$  & $\underline{
			( \lambda_j - \overline{\lambda}_k )( \lambda_j + \overline{\lambda}_k )}$ &  $\underline{-( \lambda_j\overline{\lambda}_k - 1)( \lambda_j\overline{\lambda}_k + 1 )}$  \\ 
		& $\lambda_j\overline{\lambda}_k$ & $\lambda_j\overline{\lambda}_k$ 
		 \\ 
		\hline
		$\mathcal{E}_{jk}$ &  $\underline{( \lambda_j - \lambda_k)( \lambda_j + \lambda_k)}$ & $\underline{( \lambda_j - \lambda_k)( \lambda_j + \lambda_k)}$ 
		\\
		& $\lambda_j\lambda_k$ & $\lambda_j\lambda_k$  
		\\
		\hline 
		$\mathcal{\widetilde{E}}_{jk}$ &  $\underline{( \overline{\lambda}_j - \overline{\lambda}_k)( \overline{\lambda}_j + \overline{\lambda}_k)}$ & $\underline{-( \overline{\lambda}_j - \overline{\lambda}_k)( \overline{\lambda}_j + \overline{\lambda}_k)}$ 
		\\
		& $\overline{\lambda}_j\overline{\lambda}_k$ & $\overline{\lambda}_j\overline{\lambda}_k$  
		\\
		\hline 
		$\mathcal{d}_{jk}$  & $\underline{
			( \lambda_j - \overline{\lambda}_k )^2}$ &  $\underline{
			-( \lambda_j\overline{\lambda}_k + 1 )^2}$  \\ 
		& $\lambda_j\overline{\lambda}_k$ & $\lambda_j\overline{\lambda}_k$ 
		\\ 
		\hline
		$\mathcal{e}_{jk}$ &  $\underline{( \lambda_j - \lambda_k)^2}$ & $\underline{( \lambda_j - \lambda_k)^2}$ 
		\\
		& $\lambda_j\lambda_k$ & $\lambda_j\lambda_k$  
		\\
		\hline 
		$\mathcal{\widetilde{e}}_{jk}$ &  $\underline{( \overline{\lambda}_j - \overline{\lambda}_k)^2}$ & $\underline{( \overline{\lambda}_j - \overline{\lambda}_k)^2}$ 
		\\
		& $\overline{\lambda}_j\overline{\lambda}_k$ & $\overline{\lambda}_j\overline{\lambda}_k$  
		\\
		\hline 
	\end{tabular}
\end{center}
\end{table}

For convenience to discuss the asymptotic behavior of \eqref{Density of the WZ Term_decomposition App}, we consider the asymptotic limit
\begin{eqnarray}
\left\{
\begin{array}{l}
X_1 ~ \mbox{is a finite real number}
\smallskip \\
|X_2|  \gg  |X_1|
\end{array}.
\right.
\end{eqnarray}
By the same calculation as mentioned in section 4.2, we have
\begin{eqnarray*}
\label{WZ term_part1_X_i}
%&&\!\!\!\!
\displaystyle{\frac{1}{|\sigma|\Delta^2}}
\left|
\!\!\begin{array}{cc}
\Delta_{11}\!\! & \Delta_{22}\!\! \\
\partial_{\mu} \Delta_{11}\!\! & \partial_{\mu} \Delta_{22}\!\!
\end{array}
\right|_{|X_2| \gg |X_1|}
%\nonumber \\
&\!\!\!\!=\!\!\!\!&
\frac{-2r_{\mu}^{(1)}ab\mathcal{D}_{11}\cosh^2{X_2} + \mathcal{O}(\cosh{X_2})}
{\left[
	\begin{array}{l}
	a~\!\mbox{cosh}(X_1 + X_2)
	+
	b~\!\mbox{cosh}(X_1 + X_2)
	+
	\mathcal{O}(1)
	\end{array}
	\right]^2 
}
\\
&\!\!\!\!=\!\!\!\!&
\frac{
-2r_{\mu}^{(1)}ab\mathcal{D}_{11} + \mathcal{O}(\mathrm{sech}X_2)
}
{\left[
	(a+b)~\!\mbox{cosh}X_1 + (a-b)~\!\mbox{sinh}X_1\mbox{tanh}X_2 + \mathcal{O}(\mathrm{sech}X_2)
	\right]^2},
\end{eqnarray*}
\begin{eqnarray*}
\label{WZ term_part3_X_i}
%&&\!\!\!\!
\displaystyle{\frac{1}{|\sigma|\Delta^2}}
\left|
\!\!\begin{array}{cc}
\Delta_{12}\!\! & \Delta_{21}\!\! \\
\partial_{\mu} \Delta_{12}\!\! & \partial_{\mu} \Delta_{21}\!\!
\end{array}
\right|_{|X_2| \gg |X_1|}
%\nonumber \\
&\!\!\!\!=\!\!\!\!&
\frac{
	-2s_{\mu}^{(1)}ab\mathcal{d}_{11}\cosh^2{X_2} + \mathcal{O}(\cosh{X_2})
}
{\left[
	\begin{array}{l}
	a~\!\mbox{cosh}(X_1 + X_2)
	+
	b~\!\mbox{cosh}(X_1 + X_2)
	+
	\mathcal{O}(1)
	\end{array}
	\right]^2 
}
\\
&\!\!\!\!=\!\!\!\!&
\frac{
	-2s_{\mu}^{(1)}ab\mathcal{d}_{11} + \mathcal{O}(\mathrm{sech}X_2)
}
{\left[
	(a+b)~\!\mbox{cosh}X_1 + (a-b)~\!\mbox{sinh}X_1\mbox{tanh}X_2 + \mathcal{O}(\mathrm{sech}X_2)
	\right]^2}, 
\end{eqnarray*}
\begin{eqnarray*}
\label{WZ term_part4_X_i}
%&&\!\!\!\!
\displaystyle{\frac{1}{|\sigma|\Delta^2}}
\left|
\!\!\begin{array}{cc}
\partial_{\nu} \Delta_{11}\!\! & \partial_{\nu} \Delta_{22}\!\! \\
\partial_{\rho} \Delta_{11}\!\! & \partial_{\rho} \Delta_{22}\!\!
\end{array}
\right|_{|X_2| \gg |X_1|}
%\nonumber \\
&\!\!\!\!=\!\!\!\!&
\frac{2\left(
	r_{\nu}^{(1)}r_{\rho}^{(2)} -  r_{\rho}^{(1)}r_{\nu}^{(2)} \right) ab
	\mathcal{D}_{11}\cosh{X_2}\sinh{X_2} + \mathcal{O}(\cosh{X_2}) }
{
\left[
\begin{array}{l}
a~\!\mbox{cosh}(X_1 + X_2)
+
b~\!\mbox{cosh}(X_1 + X_2)
+
\mathcal{O}(1)
\end{array}
\right]^2 
}
\nonumber \\
&\!\!\!\!=\!\!\!\!&
\frac{
2\left(
r_{\nu}^{(1)}r_{\rho}^{(2)} -  r_{\rho}^{(1)}r_{\nu}^{(2)} \right) ab
\mathcal{D}_{11}~\!\mathrm{tanh}X_2 + \mathcal{O}(\mathrm{sech}X_2)
}
{
	\left[
	(a+b)~\!\mbox{cosh}X_1 + (a-b)~\!\mbox{sinh}X_1\mbox{tanh}X_2 + \mathcal{O}(\mathrm{sech}X_2)
	\right]^2
},
\end{eqnarray*}
\begin{eqnarray*}
\label{WZ term_part2_X_i}
%&&\!\!\!\!
\displaystyle{\frac{1}{|\sigma|\Delta^2}}
\left|
\!\!\begin{array}{cc}
\partial_{\nu} \Delta_{12}\!\! & \partial_{\nu} \Delta_{21}\!\! \\
\partial_{\rho} \Delta_{12}\!\! & \partial_{\rho} \Delta_{21}\!\!
\end{array}
\right|_{|X_2| \gg |X_1|}
%\nonumber \\
&\!\!\!\!=\!\!\!\!&
\frac{2\left(
s_{\nu}^{(1)}r_{\rho}^{(2)} -  s_{\rho}^{(1)}r_{\nu}^{(2)} 
\right)ab
\mathcal{d}_{11}\cosh{X_2}\sinh{X_2} + \mathcal{O}(\cosh{X_2})}
{
\left[
\begin{array}{l}
a~\!\mbox{cosh}(X_1 + X_2)
+
b~\!\mbox{cosh}(X_1 + X_2)
+
\mathcal{O}(1)
\end{array}
\right]^2 
}
\nonumber \\
&\!\!\!\!=\!\!\!\!&
\frac{
	2\left(
	s_{\nu}^{(1)}r_{\rho}^{(2)} -  s_{\rho}^{(1)}r_{\nu}^{(2)} 
	\right)ab
	\mathcal{d}_{11}~\!\mathrm{tanh}X_2 + \mathcal{O}(\mathrm{sech}X_2)
}
{
	\left[
	(a+b)~\!\mbox{cosh}X_1 + (a-b)~\!\mbox{sinh}X_1\mbox{tanh}X_2 + \mathcal{O}(\mathrm{sech}X_2)
	\right]^2
} . 
\end{eqnarray*}
Therefore, for fixed $X_1$ and $|X_2| \gg |X_1|$ we can conclude that 
\begin{eqnarray*}
B_{\mu\nu\rho}
&\!\!\!\!\stackrel{|X_2| \gg |X_1|}{\simeq}\!\!\!\!&
\displaystyle{
	\frac{
		-4a^2b^2C_{\mu\nu\rho}\mathcal{D}_{11}\mathcal{d}_{11}\mathrm{tanh}X_2 + \mathcal{O}(\mathrm{sech}X_2)	
}
	{	\left[
		(a+b)~\!\mbox{cosh}X_1 + (a-b)~\!\mbox{sinh}X_1\mbox{tanh}X_2 + \mathcal{O}(\mathrm{sech}X_2)
		\right]^4
	}
} \\
&\!\!\!\! \stackrel{X_2 \rightarrow \pm\infty}{\longrightarrow} \!\!\!\!&
~~\displaystyle{
	\frac{
		\mp 32a^2b^2C_{\mu\nu\rho}\mathcal{D}_{11}\mathcal{d}_{11}}
	{(ae^{X_1} + be^{X_1})^4}}
= \mp 2C_{\mu\nu\rho}D_{11}d_{11} \mathrm{sech}^4(X_1 + \delta_1)
\mbox{ ~~for fixed $ X_1$},
\end{eqnarray*}
where the phase shift factor is 
$\displaystyle{\delta_1:=(1/2)~\!\mathrm{log}(a/b)}$. 
and 
$C_{\mu\nu\rho}:=
\left(r_{\mu}^{(1)}s_{\nu}^{(1)} + s_{\mu}^{(1)}r_{\nu}^{(1)}\right)r_{\rho}^{(2)}
-
\left(r_{\mu}^{(1)}s_{\rho}^{(1)} + s_{\mu}^{(1)}r_{\rho}^{(1)}\right)r_{\nu}^{(2)}$. 
%We remark that the asymptotic limit of \eqref{Density of the WZ term_one-third part} behaves as 1-soliton and anti-1-soliton which suggests that \eqref{Density of the WZ Term_decomposition} is like to be an odd function.

%By \eqref{Density of the WZ Term_decomposition} and the fact that
%\begin{eqnarray}\left\{
%\begin{array}{l}~~\left(r_{\mu}^{(1)}s_{\nu}^{(1)} + s_{\mu}^{(1)}r_{\nu}^{(1)}\right)r_{\rho}^{(2)}-\left(r_{\mu}^{(1)}s_{\rho}^{(1)} + s_{\mu}^{(1)}r_{\rho}^{(1)}\right)r_{\nu}^{(2)}
%\smallskip \\
%+ ~(\mu, \nu, \rho) \longrightarrow (\nu, \rho, \mu), ~(\rho, \mu, \nu)
%\end{array}
%\right\} =0,  
%\end{eqnarray} 
%we have 
%\begin{eqnarray}
%\mbox{Tr}\{(\partial_{\mu}\sigma)\sigma^{-1}(\partial_{\nu}\sigma)\sigma^{-1}(\partial_{\rho}\sigma)\sigma^{-1}\}\longrightarrow 0
%\end{eqnarray}
%in the asymptotic region. 
%Therefore, the Wess-Zumino term is asymptotic to zero for 2-soliton case

\subsection{Asymptotic Form of WZW$_4$ Action Density ($n$-Soliton)}
\label{Asymptotic Form of WZW_4 Action Density (n-Soliton)}
Without loss of generality, we consider one of the asymptotic regions of type $\mathscr{R}_K$ which is labeled by (the other cases are equivalent to this one.)
\begin{eqnarray}
\label{specialized epsilon_j}
\varepsilon_j
=\left\{
\begin{array}{l}
+1, ~ j=1, \cdots, K-1 
\\
-1, ~ j= K+1, \cdots, n
%\\
%\pm 1,~ j=K
\end{array}
\right.
\end{eqnarray}
and define
\begin{eqnarray*}
	\boldsymbol{X}_{\!\widehat{K}}
	:=\sum_{j=1, j \neq K}^{n}\varepsilon_j X_j
	\stackrel{\eqref{specialized epsilon_j}}{=}
	X_1 + \cdots + X_{K-1} - X_{K+1} + \cdots + X_{n}.
\end{eqnarray*}
By \eqref{Data of n soliton_Delta}, \eqref{Data of n soliton_Delta_11}, and \eqref{Data of n soliton_Delta_12}, we can conclude that 
\begin{eqnarray}
\label{The data of n-soliton solution (asymptotics) }
\begin{array}{l}
\Delta      
\stackrel{\mathscr{R}_K}{\simeq}2\left\{
\begin{array}{l}
~a(\boldsymbol{1}, ~\!\varepsilon_{K}=+1, ~\! \boldsymbol{-1})\cosh(\boldsymbol{X}_{\!\widehat{K}} + X_{K} ) 
\smallskip \\
\!\!+a(\boldsymbol{1}, ~\!\varepsilon_{K}=-1, ~\! \boldsymbol{-1}) 
\cosh(\boldsymbol{X}_{\!\widehat{K}} - X_{K} ) 
\end{array}
\!\!\right\}
+
\!\!\!\displaystyle{\sum_{j=1, j \neq K}^{n}}\!\!\mathcal{O}\left( \cosh (\boldsymbol{X}_{\!\widehat{K}} -\varepsilon_j X_j) \right), ~~~~
\smallskip \\
\Delta_{11}
\stackrel{\mathscr{R}_K}{\simeq}
\!\!\left\{
\begin{array}{l}
~A(\boldsymbol{1}, ~\!\varepsilon_{K}=+1, ~\! \boldsymbol{-1})
\exp(\boldsymbol{X}_{\!\widehat{K}} + X_{K} )
\smallskip \\
\!\!+A(\boldsymbol{1}, ~\!\varepsilon_{K}=-1, ~\! \boldsymbol{-1})
\exp(\boldsymbol{X}_{\!\widehat{K}} - X_{K} )
\smallskip \\
\!\!+A(\boldsymbol{-1}, ~\!\varepsilon_{K}=+1, ~\! \boldsymbol{1})
\exp(-\boldsymbol{X}_{\!\widehat{K}} + X_{K} )
\smallskip \\
\!\!+A(\boldsymbol{-1}, ~\!\varepsilon_{K}=-1, ~\! \boldsymbol{1})
\exp(-\boldsymbol{X}_{\!\widehat{K}} - X_{K} )
\end{array}
\!\!\right\}
\!+\!\!\!\displaystyle{\sum_{j=1, j \neq J}^{n}}\!\!\mathcal{O}\left( \cosh (\boldsymbol{X}_{\!\widehat{K}} - \varepsilon_{j} X_j) \right), ~~~~
\smallskip \\
\Delta_{22}\left( \mathscr{R}_K \right)
=\Delta_{11}\left( \mathscr{R}_K \right)\Big|_{A(\varepsilon_1, \cdots, \varepsilon_{n}) ~\rightarrow~ A(-\varepsilon_1, \cdots, -\varepsilon_{n})},
\end{array}
\end{eqnarray}
%\medskip \\
\begin{eqnarray}
\begin{array}{l}
\Delta_{12}
\stackrel{\mathscr{R}_K}{\simeq}
\!\!\left\{
\begin{array}{l}
~A(\boldsymbol{1}, ~\!\varepsilon_{K}=+i, ~\! \boldsymbol{-1})\exp(\boldsymbol{X}_{\!\widehat{K}} + i\Theta_{K})
\smallskip \\
\!\!+A(\boldsymbol{-1}, ~\!\varepsilon_{K}=+i, ~\! \boldsymbol{1})\exp(-\boldsymbol{X}_{\!\widehat{K}} + i\Theta_{K})
\end{array}
\!\!\right\}
\!+\!\!\! \displaystyle{\sum_{j=1, j \neq K}^{n}}\!\!\mathcal{O}\left( \cosh (\boldsymbol{X}_{\!\widehat{K}} - \varepsilon_{j} X_j) \right), ~~~~
\smallskip \\
\Delta_{21}\left( \mathscr{R}_K \right)
=\Delta_{12}\left( \mathscr{R}_K \right)\Big|_{(X_j, \Theta_{K}) ~\rightarrow~(-X_j, -\Theta_{K})}. 
\end{array}
\end{eqnarray}
%\subsubsection{}
By \eqref{The data of n-soliton solution (asymptotics) }, \eqref{Derivative of exponential functions_1}, and direct calculation, we have
the asymptotic form of the NL$\sigma$M action density:
\begin{eqnarray*}
&&\!\!\!\!\!\! \mbox{Tr}\left[(\partial_{\mu}\sigma)\sigma^{-1}(\partial^{\mu}\sigma)\sigma^{-1} \right]
\nonumber \\
&\!\!\!\!\stackrel{\mathscr{R}_K}{\simeq}\!\!\!\!&
\frac{
	\left[
	\begin{array}{l}
	%\left[
	%\begin{array}{l}
	~4d_{KK}~\!a(\boldsymbol{1}, \varepsilon_{K}=+1, \boldsymbol{-1})
	%\smallskip \\
	a(\boldsymbol{1}, \varepsilon_{K}=-1, \boldsymbol{-1}) 
	%\end{array}
	%\right]
	\cosh(2\boldsymbol{X}_{\!\widehat{K}})
	\smallskip \\
	+\displaystyle{\sum_{j=1, j \neq K}^{n}\mathcal{O}\left( \cosh (2\boldsymbol{X}_{\!\widehat{K}} - \varepsilon_{j} X_j)\right)}
	\end{array}
	\right]
}
{
	\left\{
	\left[
	\begin{array}{l}
	~~a(\boldsymbol{1}, ~\!\varepsilon_{K}=+1, ~\! \boldsymbol{-1})\cosh(\boldsymbol{X}_{\!\widehat{K}} + X_{K} ) 
	\smallskip \\
	+a(\boldsymbol{1}, ~\!\varepsilon_{K}=-1, ~\! \boldsymbol{-1}) 
	\cosh(\boldsymbol{X}_{\!\widehat{K}} - X_{K} ) 
	\end{array}
	\right]
	\!+\!\!\!\displaystyle{\sum_{j=1, j \neq K}^{n}\!\!\mathcal{O}\left( \cosh (\boldsymbol{X}_{\!\widehat{K}} - \varepsilon_{j} X_j) \right) }
	\right\}^2
} ~~~~~~~~~  \\
&\!\!\!\! = \!\!\!\!&
\frac{
	%\left\{
	%\begin{array}{l}
	8d_{KK}~\!
	%(r_{\mu}^{(J)} + s_{\mu}^{(J)})(r_{\mu}^{(J)} - s_{\mu}^{(J)}) \displaystyle{\frac{(\lambda_{J} - \mu_{J})^2}{\lambda_J \mu_J}}
	a(\boldsymbol{1}, \varepsilon_{K}=+1, \boldsymbol{-1})a(\boldsymbol{1}, \varepsilon_{K}=-1, \boldsymbol{-1}) 
	%\medskip \\
	+\displaystyle{\sum_{j=1, j \neq K}^{n}\mathcal{O}\left( \mbox{sech} X_j \right)}
	%\end{array}
	%\right\}
}
{
	\left\{
	\begin{array}{l}
	~~\left[
	a(\boldsymbol{1}, ~\!\varepsilon_{K}=+1, ~\! \boldsymbol{-1})
	+
	a(\boldsymbol{1}, ~\!\varepsilon_{K}=-1, ~\! \boldsymbol{-1})\right]\cosh X_K
	
	\smallskip \\
	+\left[
	a(\boldsymbol{1}, ~\!\varepsilon_{K}=+1, ~\! \boldsymbol{-1})
	-
	a(\boldsymbol{1}, ~\!\varepsilon_{K}=-1, ~\! \boldsymbol{-1})
	\right] \tanh \boldsymbol{X}_{\!\widehat{K}} ~\! \sinh X_K
	\smallskip \\
	+ \!\! \displaystyle{\sum_{j=1, j \neq K}^{n}\!\!\mathcal{O}\left( \mbox{sech} X_j \right) }
	\end{array}
	\right\}^2
}
\nonumber \\
&\!\!\!\!\stackrel{\boldsymbol{X}_{\!\widehat{K}} \rightarrow \pm\infty}{\longrightarrow}\!\!\!\!&
2d_{KK}\mathrm{sech}^2(X_K \pm \delta_K) ~~\mbox{for fixed real number
$ X_K$},
\end{eqnarray*}
where the phase shift factor is 
$\displaystyle{\delta_K:=\frac{1}{2}~\!\mathrm{log}\left[\frac{a(\boldsymbol{1}, ~\!\varepsilon_{K}=+1, ~\! \boldsymbol{-1})}{a(\boldsymbol{1}, ~\!\varepsilon_{K}=-1, ~\! \boldsymbol{-1})}\right]}$. 
and  $d_{KK}$ is defined in Table \ref{Table_1}.
%\newpage

%\subsubsection{Asymptotic Form of Wess-Zumino Action Density}
By \eqref{The data of n-soliton solution (asymptotics) }, \eqref{Derivative of exponential functions_1}, and direct calculation, we have 
the asymptotic form of the Wess-Zumino action density:
\begin{eqnarray*}
\label{Density of the WZ term_n-soliton} 
B_{\mu\nu\rho}
\stackrel{\mathscr{R}_K}{\simeq}
\displaystyle{
	\frac{
		%\left\{
		%\begin{array}{l}
		-4\mathcal{A\widetilde{A}} C_{\mu\nu\rho}^{(K)} \mathrm{tanh}\boldsymbol{X}_{\!\widehat{K}} 
		%\smallskip \\
		+ \displaystyle{\sum_{j=1, j \neq K}^{n}\mathcal{O}\left( \mbox{sech} X_j \right)}
		%\end{array}	
		%\right\}
	}
	{	|\sigma|^2\left\{
		\begin{array}{l}
		~~\left[
		a(\boldsymbol{1}, ~\!\varepsilon_{K}=+1, ~\! \boldsymbol{-1})
		+
		a(\boldsymbol{1}, ~\!\varepsilon_{K}=-1, ~\! \boldsymbol{-1})\right]\cosh X_K
		
		\smallskip \\
		+\left[
		a(\boldsymbol{1}, ~\!\varepsilon_{K}=+1, ~\! \boldsymbol{-1})
		-
		a(\boldsymbol{1}, ~\!\varepsilon_{K}=-1, ~\! \boldsymbol{-1})
		\right] \tanh \boldsymbol{X}_{\!\widehat{K}}~\! \sinh X_K
		\smallskip \\
		+ \!\! \displaystyle{\sum_{j=1, j \neq K}^{n}\!\!\mathcal{O}\left( \mbox{sech} X_j \right) }
		\end{array}
		\right\}^4
	}
}, ~~~~~~~~
\end{eqnarray*}
where 
\begin{eqnarray*}
C_{\mu\nu\rho}^{(K)}:=
\left(r_{\mu}^{(K)}s_{\nu}^{(K)} + s_{\mu}^{(K)}r_{\nu}^{(K)}\right)\boldsymbol{r}_{\rho}^{(\widehat{K})}
-
\left(r_{\mu}^{(K)}s_{\rho}^{(K)} + s_{\mu}^{(K)}r_{\rho}^{(K)}\right)\boldsymbol{r}_{\nu}^{(\widehat{K})}), 
~
r_{\rho}^{(K)}:=\!\!\!\!\displaystyle{\sum_{ j=1, j \neq K}^{n}}\!\varepsilon_{j}r_{\rho}^{(j)}, 
\end{eqnarray*}
and
\begin{eqnarray*}
	\mathcal{A}
	&\!\!\!\!:=\!\!\!\!&
	\left[
	\begin{array}{l}
		~~\!A({\bf 1}, ~\varepsilon_{K}=+1, ~{\bf-1})A({\bf -1}, ~\varepsilon_{K}=+1, ~{\bf 1})
		\\
		\!\!-A({\bf 1}, ~\varepsilon_{K}=-1, ~{\bf -1})A({\bf -1}, ~\varepsilon_{K}=-1, ~{\bf 1})
	\end{array}
	\right],
	\\
	\mathcal{\widetilde{A}}
	&\!\!\!\!:=\!\!\!\!&
	A({\bf 1}, ~\varepsilon_{K}=+i, ~{\bf-1}) A({\bf -1}, ~\varepsilon_{K}=+i, ~{\bf 1}).
\end{eqnarray*}
By \eqref{Density of the WZ Term_decomposition} and the fact that
\begin{eqnarray*}
C_{\mu\nu\rho}^{(K)}+
C_{\nu\rho\mu}^{(K)}+
C_{\rho\mu\nu}^{(K)}=0,
\end{eqnarray*} 
we have 
\begin{eqnarray*}
\mbox{Tr}\left[(\partial_{\mu}\sigma)\sigma^{-1}(\partial_{\nu}\sigma)\sigma^{-1}(\partial_{\rho}\sigma)\sigma^{-1}\right]
\longrightarrow 0
\end{eqnarray*}
in the asymptotic region. 
Therefore, the Wess-Zumino term is asymptotic to zero for $n$-soliton case.

\end{appendix}

%\vspace{-4mm}

\end{document}